\documentclass[a4paper,11pts,noapacite]{article}
\usepackage[T1]{fontenc}
\usepackage{dsfont}

\usepackage{xr}
\makeatletter
\newcommand*{\addFileDependency}[1]{% argument=file name and extension
\typeout{(#1)}% latexmk will find this if $recorder=0
\@addtofilelist{#1}
\IfFileExists{#1}{}{\typeout{No file #1.}}
}\makeatother

\newcommand*{\myexternaldocument}[1]{%
\externaldocument{#1}%
\addFileDependency{#1.tex}%
\addFileDependency{#1.aux}%
}

\newcommand\sdot[1][.4]{\mathbin{\vcenter{\hbox{\scalebox{#1}{$\,\bullet\,$}}}}}

\myexternaldocument{Supplement}

\usepackage{enumerate}
\usepackage{enumitem}

\usepackage{graphicx}
\usepackage[labelfont=bf,textfont=it]{caption}
\usepackage{subcaption}
\usepackage{array}
\graphicspath{{./Figures/}}
\usepackage[section]{placeins}

\usepackage{float}
\usepackage{multirow}
\usepackage{threeparttable}
\usepackage{chngcntr}

\usepackage{geometry}
\usepackage{lscape}

\usepackage{amsfonts}
\usepackage{amsmath}
\usepackage{amsthm}
\usepackage{amssymb}
\usepackage{mathtools}
\usepackage[mathscr]{eucal}
\usepackage{mathtools}
\usepackage{centernot}
\usepackage{bbm}
\usepackage{tensor}

\usepackage{hyperref}
\hypersetup{
	colorlinks=true,
	linkcolor=blue,
	filecolor=magenta,      
	urlcolor=cyan,
	citecolor=blue,
}

\usepackage{listings}
\usepackage[dvipsnames]{xcolor}
\usepackage[framed]{matlab-prettifier}
\usepackage{colortbl}

\usepackage{xstring}

\usepackage{natbib}
\usepackage{authblk}

\usepackage{tikz}
\usetikzlibrary{bayesnet}
\usepackage{tikz-network}

\usetikzlibrary{shapes.geometric, arrows}
\tikzstyle{startstop} = [rectangle, rounded corners, minimum width=1cm, minimum height=0.5cm, text centered, draw=black]

\tikzstyle{io} = [trapezium, trapezium left angle=70, trapezium right angle=110, minimum width=1cm, minimum height=0.5cm, text centered, draw=black]

\tikzstyle{process} = [rectangle, minimum width=1cm, minimum height=0.5cm, text centered, draw=black]

\tikzstyle{decision} = [diamond, minimum width=1cm, minimum height=0.5cm, text centered, draw=black]

\tikzstyle{arrow} = [thick,->,>=stealth]

\usepackage{xcolor}
\usepackage{listings}
\definecolor{light-gray}{gray}{0.95}
\theoremstyle{definition}
\newtheorem{definition}{Definition}
\theoremstyle{corollary}
\newtheorem{corollary}{Corollary}

\newtheorem{proposition}{Proposition}%[section]

\makeatletter
\newcommand{\ssymbol}[1]{^{\@fnsymbol{#1}}}
\makeatother

\usepackage{comment}

\title{A Dynamic Stochastic Block Model for Multidimensional Networks\thanks{Authors acknowledge financial support from Italian Ministry MIUR under the PRIN projects `\textit{Hi-Di NET - Econometric Analysis of High Dimensional Models with Network Structures in Macroeconomics and Finance}' (grant 2017TA7TYC) and `\textit{Discrete random structures for Bayesian learning and prediction}' (grant 2022CLTYP4), and from the EU under the Next Generation EU Project `\textit{GRINS - Growing Resilient, INclusive and Sustainable}'; National Recovery and Resilience Plan (NRRP). This research used the SCSCF and HPC multiprocessor cluster system provided by the Venice Centre for Risk Analytics (VERA) at Ca' Foscari University of Venice. The views and opinions expressed are only those of the authors and do not necessarily reflect those of the European Union or the European Commission. Neither the European Union nor the European Commission can be held responsible for them.}}%: Granger causality between community structures
\author[$\ssymbol{3}$]{Ovielt Baltodano L\'opez}
\author[$\ssymbol{3}$]{Roberto Casarin\footnote{Corresponding author \\
		\indent \hspace{0.1cm} Email addresses: \href{mailto:ovielt.baltodano@unive.it}{ovielt.baltodano@unive.it} (Ovielt Baltodano L\'opez), \href{mailto:r.casarin@unive.it}{r.casarin@unive.it} (Roberto Casarin)}}
\affil[$\ssymbol{3}$]{\small Department of Economics, Ca' Foscari University of Venice, San Giobbe 873/b 30121, Italy}
\date{July, 2025}

\begin{document}
	
	%extent the Granger--causality definition to the latent network structures inducing clustering in edges. At this purpose, we model the multiplex network dependence by combining a dynamic stochastic block model in each layer with a dependent HMC expressing the transition probabilities as a multinomial logit model,
	
	%Moreover, 
	
	\maketitle	
	\begin{abstract}	
		The availability of relational data can offer new insights into the functioning of the economy. Nevertheless, modeling the dynamics in network data with multiple types of relationships is still a challenging issue. Stochastic block models provide a parsimonious and flexible approach to network analysis. We propose a new stochastic block model for multidimensional networks, where layer--specific hidden Markov--chain processes drive the changes in community formation. The changes in the block membership of a node in a given layer may be influenced by its own past membership in other layers. This allows for clustering overlap, clustering decoupling, or more complex relationships between layers, including settings of unidirectional, or bidirectional, non--linear  Granger block causality. We address the overparameterization issue of a saturated specification by assuming a Multi--Laplacian prior distribution within a Bayesian framework. Data augmentation and Gibbs sampling are used to make the inference problem more tractable. Through simulations, we show that standard linear models and the pairwise approach are unable to detect block causality in most scenarios. In contrast, our model can recover the true Granger causality structure. As an application to international trade, we show that our model offers a unified framework, encompassing community detection and Gravity equation modeling. We found new evidence of block Granger causality of trade agreements and flows and core--periphery structure in both layers on a large sample of countries. 
		\\
		
		%\noindent \textbf{Keywords:} stochastic block models; hidden Markov chain; multidimensional networks; edge clustering; Granger causality; group LASSO; FTAs effectiveness; Gravity equation. 
		
		
	\end{abstract}
	
	\begin{comment}
		Main contribution an extension to Granger--causality  in the network clustering framework.
		
		For the DSBM literature, we contribute by: modeling multilayer dependence, we add covariates for controlling observed heterogeneity, implicitly we introduce time--varying transition matrix. 
		
		For the Markov Switching literature:  we provide s more flexible approach for Panel Markov Chain dependence compared with the Dirichlet representation, and we provide a generalization for more than two states.
		
		In the contingency table literature: we use of Bayesian group--lasso for the Polya--Gamma representation of multinomial logit regression. In the context of HMC, some entries of the transition matrix may not have enough observation and the Multi--Laplacian prior over--perform normal prior.
		
		[we do not emphasize this last point because with the second paper we cover better this part?]
	\end{comment}

%  and an alternative to extreme reductionist approaches \citep[e.g., see][]{Erdos1959, Watts1998, Barabasi1999}.\footnote{See \citet{Bollobas01} for an introduction to random graphs and \citet{newman2010} for an introduction to network analysis.} They 
\newpage
	\section{Introduction} \label{sec:intro}
	Network models are a tool for studying real-world complex systems and have become a convenient framework for describing social and economic interactions \citep[see, for example][]{de2017econometrics}. In terms of higher--order network properties, such as clustering, several stochastic models have been proposed apart from exponential random graphs, e.g., latent space models and Stochastic Block Models (SBMs) \citep{kim2018review,hoff2018}. However, in systems with multiple types of interactions between nodes, a single graph does not fully describe the connectivity structure. Thus, graphs with multiple types of edges (layers) have been introduced. See \citet{Kiv14} for an introduction to multidimensional networks. To the best of our knowledge, relatively few works deal with dynamic clustering models for multidimensional graphs \citep[e.g.,][]{lee2019review,paul2020random,lei2022bias}. The objective of the present work is to extend the Dynamic SBM (DSBM) to accommodate multiple edge types. \textcolor{black}{Our approach can be easily extended to a general multi--layer framework where edges between layers are allowed.}  
	
	In SBMs, edge clustering is driven by a probabilistic classification of the nodes into different communities. \textcolor{black}{In this paper, the terms group, cluster, block, and community are used interchangeably to describe a set of nodes sharing similar connectivity characteristics. These characteristics may include edge probability, strength, and variance of the strength.} An alternative dynamic specification involves using Hidden Markov Chains (HMCs) to capture temporal changes in the node's membership. See \citet{fruhwirth2006finite} for an introduction to Hidden  Markov models. \citet{yang2011detecting} work with (un)directed and unweighted dynamic networks and \citet{matias2017statistical} generalize their model to include weighted networks, discussing the identification issues that arise when block--dependent connectivity parameters are time--varying. Other extensions of SBMs address mixed--membership and further edge formation heterogeneity \citep[e.g.,][]{airoldi2008mixed,zhao2012consistency}.  
	
	In these previous works, the DSBM has accounted for one type of edge. However, in social and economic relationships, there is usually more than one type of tie, such as multiple goods traded between regions or several assets exchanged by financial institutions. Therefore, we propose a DSBM for multidimensional networks (DSBMM) where each type of relationship is represented by a different layer, with no inter--layer edges and a time--invariant and layer--invariant node set \textcolor{black}{and with a fixed number of communities, potentially different between layers}. 
	
	To the best of our knowledge, extensions of the SBM in a multidimensional setting are static and correlational\textcolor{black}{---undirected dependence relationships}. \citet{jovanovski2019bayesian} identify communities per layer (local clustering), and a consensus of all layers (global clustering) with unweighted edges. \citet{stanley2016clustering} in a static framework, assume layer clustering, in such a way that the layers in the same group can share a common SBM. Other studies with no interest in community detection use multivariate distributions to make inference on the dependence between layers, but this restricts all layers to be of the same type, that is, either weighted or unweighted and directed or undirected \citep{salter2017latent}.  \textcolor{black}{Alternative models to the SBM family have been proposed to capture clustering in a dynamic setting, such as the latent space model proposed by  \citet{durante2017bayesian}, where layer--specific and shared latent positions evolve smoothly following Gaussian processes. Although this model is flexible enough to adapt to different network structures, connectivity dynamics, and to exploit information across layers, there are significant differences with the DSBMM. First, the community structure provides a clear interpretation in terms of node classification. In contrast, the latent position coordinates may require additional knowledge of the field of application to address identifiability issues, as in factor models.  Second, although the shared latent positions allow for layer dependence, it does not provide information on the direction of the relationship between each pair of layers, only an undirected global dependence measure between all layers. Moreover, this dynamic latent position model assumes that all layers are unweighted and undirected, whereas the DSBMM encompasses layers with different relationship types simultaneously.} Although these alternative multidimensional models can provide a measure of association or clustering overlap between layers, it is not possible to infer if the nature of the relationship is unidirectional or bidirectional.
	
	In this paper, we introduce a concept of nonlinear Granger block--causality (\textcolor{black}{NGB causality}) that identifies the directed dependence between layers' community structure. For instance, Layer $\ell$ may be \textcolor{black}{NGB} causing Layer $\mathfrak{m}$, but not vice--versa. Specifically, in the DSBMM, the nodes' membership dynamics in each layer is influenced by their respective membership in the other layers. By using a saturated multinomial specification to model the transition matrix of each layer, it is possible to test for Granger causality between layers. The approach also allows for layers of different types, i.e. (un)directed or (un)weighted, and a set of layer--specific covariates to control for the observed node or dyad heterogeneity.
	
	We propose a Bayesian inference, based on a Multi--Laplacian prior, to induce a group LASSO penalty and address the overparameterization issue \citep{raman2009bayesian}. A Pólya Gamma representation is introduced to make the multinomial model tractable \citep{polson2013bayesian}. Our simulation results show that the Multi--Laplacian prior performs better than a Normal prior in retrieving the underlying causal structure of the DGP (\textcolor{black}{NGB causality}). This difference in performance is due to the twofold shrinkage of the Multi--Laplacian prior. In a Normal prior setting, information on the parameter grouping is excluded, thus the correlation between \textcolor{black}{lagged HMC variables} is not exploited, whereas in a Multi--Laplacian setting, different shrinkage effects, within groups of parameters and between groups, are allowed. A comparison between DSBMM and a benchmark model, that is a Bayesian Vector Autoregression (BVAR), shows that the latter is not able to detect the \textcolor{black}{NGB causality} under the great majority of scenarios.
	
	We apply our DSBMM model to trade flows and free trade agreements (FTA) and contribute to the debate on the effectiveness of FTA as a policy instrument for global trade integration \citep{baier2007free,baier2019widely}. The wide range of empirical results regarding the effects of FTAs on trade, combined with the complexity of overlapping FTAs between countries, suggests that considering the network structure can provide a complementary view of the FTA's influence on the topological properties of trade flows. Specifically, we test if FTA clustering helps predict the international trade community membership after controlling for country and dyad observed heterogeneity. Moreover, our model--based approach for community detection accounts for uncertainty in the parameters and community membership. Community detection in international trade flows has been analyzed mainly using modularity algorithms, which are descriptive and do not provide probabilistic statements. For example, \citet{bartesaghi2020communicability} focuses on aggregate trade flows to infer groups of countries with dense connectivity, and \citet{barigozzi2011identifying} uses commodity-specific layers to identify communities of countries per product. Furthermore, by allowing covariates to affect trade flows, our DSBMM integrates community detection into Gravity equation models, which are extensively used in international trade \citep[e.g.,][]{piermartini2016estimating}. 
 
 The paper is organized as follows. Section \ref{sec:DSBMM} introduces the DSBMM and presents some numerical illustrations of the model properties. Section \ref{sec:postapp} addresses the inference procedure, including the choice of the prior distribution, the details of the Gibbs sampler, and a comparison with panel BVAR models {\color{black}and other dynamic network models} using synthetic data. Section \ref{sec:tradeFTA} provides an application to the Trade Flows--FTAs multidimensional network. Section \ref{sec:con} summarizes the main conclusions.

\section{A Dynamic SBM for multidimensional networks}
	\label{sec:DSBMM}

 The community detection task within a dynamic framework is defined in this section, and the multidimensional dependence is presented in detail, accompanied by examples. 
	\subsection{Multidimensional DSBM}\label{sec:multDSBM}
	
A graph can be defined as the ordered triplet $\mathcal{G}=\{\mathcal{V},\mathcal{E},Y\}$, where the node set  $\mathcal{V}=\{1,\cdots,N\}$ is the same across the layer set $\mathcal{L}=\{1,\ldots,L\}$ and the edge set collection $\mathcal{E}=\{\mathcal{E}^{(1)},\ldots,\mathcal{E}^{(L)}\}$, $\mathcal{E}^{(\ell)}\subseteq \mathcal{V}\times \mathcal{V}$ allows only for intra--layer edges. The set $Y=\{Y^{(1)},\cdots, Y^{(L)}\}$ includes the layer--specific adjacency matrices with $Y^{(\ell)}\in \mathbb{R}^{N}\times\mathbb{R}^{N}$ where the $(i,j)$--th element of $Y^{(\ell)}$ is $Y^{(\ell)}_{ij}=0$ if $(i,j)\not\in \mathcal{E}^{(\ell)}$ and $Y^{(\ell)}_{ij}=a\in \mathbb{R}\backslash\{0\}$ if $(i,j)\in \mathcal{E}^{(\ell)}$. The multidimensional may jointly consider (un)directed and (un)weighted layers, resulting in (symmetric) asymmetric and (binary) real-valued adjacency matrices.
	
Considering that higher-order network structures, such as clubs or closed communities, are common features in real-world graphs \citep[e.g.,][]{csermely2013structure}, an SBM can be used in each layer to model network topology through node grouping. We define the set $\mathfrak{V}=\{\mathfrak{V}^{(1)},\dots,\mathfrak{V}^{(L)}\}$, where each element is a layer--specific partition of the node set into $Q^{(\ell)}$ subsets, that is, $\mathfrak{V}^{(\ell)}=\{\mathcal{V}^{(\ell)}_{1},\ldots,\mathcal{V}^{(\ell)}_{Q^{(\ell)}}\}$ and each $\mathcal{V}^{(\ell)}_{q}\subseteq \mathcal{V}$ is called block or community in layer $\ell$, satisfying two properties: $\mathcal{V}^{(\ell)}_{q}\cap \mathcal{V}^{(\ell)}_{r}=\emptyset$, $q\neq r$, and $\cup_{q\in \mathcal{Q}^{(\ell)}}\mathcal{V}^{(l)}_{q}=\mathcal{V},\ \mathcal{Q}^{(\ell)}=\{1,\dots,Q^{(\ell)}\}$. \textcolor{black}{The number of blocks $Q^{(\ell)}$ is fixed.}

It is assumed that the multidimensional graph evolves over time, $\mathcal{G}_{1:T}=\{\mathcal{G}_{t}\}_{t\in \mathcal{T}}$, with $\mathcal{G}_{t}=\{\mathcal{V},\mathcal{E}_t,Y_t\}$, $\mathcal{T}=\{1,\ldots,T\}$, and a latent sequence of partitions $\mathfrak{V}_{1:T}=\{\mathfrak{V}_t\}_{t\in \mathcal{T}}$ drives its topology. For a layer $\ell$, the block membership of a node is indicated by the latent variable $Z^{(\ell)}_{it}=q\in \mathcal{Q}^{(\ell)}$, for $t\in \mathcal{T}$ \textcolor{black}{and $i\in \mathcal{V}$}. Notice that the partition sequence and the memberships are intrinsically related $\mathcal{V}_{qt}^{(\ell)}=\{i\in\mathcal{V}|Z^{(\ell)}_{it}=q\}$ \textcolor{black}{for $q\in \mathcal{Q}^{(\ell)}$, $t\in \mathcal{T}$ and $\ell \in \mathcal{L}$}. Then, the membership dynamics $Z^{(\ell)}_{{\color{black}i\sdot}}=\{Z^{(\ell)}_{it}\}_{t\in \mathcal{T}}$, follows a hidden Markov--chain process \textcolor{black}{for $i \in \mathcal{V}$}. Specifically, for the initial membership $Z^{(\ell)}_{i1}$, the probabilities of belonging to a block are collected into the vector $\alpha^{(\ell)}=(\alpha^{(\ell)}_1,\dots,\alpha^{(\ell)}_{Q^{(\ell)}})'$, where $\sum_{q\in \mathcal{Q}^{(\ell)}} \alpha^{(\ell)}_q =1$. In other words, $Z_{i1}^{(\ell)}\sim\operatorname{Ca}(\alpha^{(\ell)})$, where $\operatorname{Ca}(p)$ denotes a categorical (or multinoulli) distribution with probability vector $p$. The membership changes for $t\in \mathcal{T}\backslash\{1\}$ are determined by the transition matrix $P_{it}^{(\ell)}$ \textcolor{black}{for $i \in \mathcal{V} $}, where \textcolor{black}{$P_{i t}^{(\ell)}=\left(P_{i t, r q}^{(\ell)}\right)_{r, q \in \mathcal{Q}^{(\ell)}}$,} $P^{(\ell)}_{it,qr}\in (0,1)$ for $q,r \in \mathcal{Q}^{(\ell)}$ and $\sum_{r\in\mathcal{Q}^{(\ell)}} P^{(\ell)}_{it,qr}=1$. The transition probabilities vary across nodes and time. Thus, $Z_{it}^{(\ell)}|Z_{it-1}^{(1:L)},\sim\operatorname{Ca}(P_{it,Z^{(\ell)}_{it-1}\sdot}^{(\ell)})$, where $P_{it,Z^{(\ell)}_{it-1}\sdot}^{(\ell)}=(P_{it,Z^{(\ell)}_{it-1}1}^{(\ell)},\ldots,P_{it,Z^{(\ell)}_{it-1}Q^{(\ell)}}^{(\ell)})'$. 
	
The contemporaneous network  $Y^{(\ell)}_t=\{Y^{(\ell)}_{ijt}\}_{i,j\in \mathcal{V}}$ only depends on $Z^{(\ell)}_{{\color{black}\sdot t}}=\{Z^{(\ell)}_{it}\}_{i\in\mathcal{V}}$ and layer--specific covariates $X^{(\ell)}_{1:T}=\{X_{ijt}^{(\ell)}\}_{i,j\in \mathcal{V},\ t\in \mathcal{T}}$, i.e. given $Z^{(\ell)}_{1:T}\textcolor{black}{=\{Z^{(\ell)}_{{\color{black}\sdot t}}\}_{t\in \mathcal{T}}}$ and $X^{(\ell)}_{1:T}$, $Y^{(\ell)}_{1:T}=\textcolor{black}{\{Y^{(\ell)}_t\}_{t\in \mathcal{T}}}$ components are independent \textcolor{black}{across time and edges}. Each entry of the $\ell$--th adjacency matrix follows the mixed distribution,
	\begin{equation} 
		\label{eq:1}	Y_{ijt}^{(\ell)}\left|X^{(\ell)}_{ijt},Z^{(\ell)}_{it}=q, Z^{(\ell)}_{jt}=r, \vartheta^{(\ell)}\right.\sim (1-\nu^{(\ell)}_{qr})\delta(y)+\nu^{(\ell)}_{qr} f^{(\ell)}\left(y|\theta^{(\ell)}_{qr},X^{(\ell)}_{ijt}\right), 
	\end{equation} 
		{\color{black}					where $\vartheta^{(\ell)}=\{\nu^{(\ell)}, \theta^{(\ell)}\}$ are the set of connectivity parameters with the $Q^{(\ell)}$ square matrix $\nu^{(\ell)}=(\nu_{qr}^{(\ell)})_{q,r\in \mathcal{Q}^{(\ell)}}$  and the three--dimensional tensor collecting $(S+1)$ square matrices of dimension $Q^{(\ell)}$ $\theta^{(\ell)}=(\theta_{sqr}^{(\ell)})_{s \in \{0,1,\ldots,S\}\ , q,r\in \mathcal{Q}^{(\ell)}}$. The function $\delta(\cdot)$ denotes the Dirac function at zero, $\nu_{qr}^{(\ell)}$ the probability of having an active edge between two nodes, and $f^{(\ell)}(y|\theta^{(\ell)}_{qr}, X^{(\ell)}_{ijt})$ is a layer--specific probability (mass) density function with parameters $\theta^{(\ell)}_{qr}$. Notice that the DSBM can be interpreted as a dimensionality reduction of the tensor $Y_{1:T}^{(\ell)}$ with dimensions $N^2T$ that is summarized by the set of features in the set $\vartheta^{(\ell)}$ with dimensions $(S+2)Q^{(\ell)2}$. The dynamic network and the set of features are  linked through the block memberships by the following relationships
$$\nu_{Z^{(\ell)}_{it}Z^{(\ell)}_{jt}}^{(\ell)}=\tilde{Z}^{(\ell)\prime}_{it}\nu^{(\ell)}\tilde{Z}^{(\ell)}_{jt}, \quad \quad \theta^{(\ell)}_{Z^{(\ell)}_{it}Z^{(\ell)}_{jt}}=(\tilde{Z}^{(\ell)\prime}_{it}\theta_{0}^{(\ell)}\tilde{Z}^{(\ell)}_{jt},\ldots,\tilde{Z}^{(\ell)\prime}_{it}\theta_{S}^{(\ell)}\tilde{Z}^{(\ell)}_{jt})',$$
where $\theta_{s}^{(\ell)}=(\theta_{sqr}^{(\ell)})_{q,r\in \mathcal{Q}}$ is a slice of the tensor $\theta^{(\ell)}$ for $s\in \{0,\ldots,S\}$, and $\tilde{Z}^{(\ell)}_{it}=(0\ 0\ \ldots 1 \ 0 \ldots 0)'$ is a $Q^{(\ell)}$--vector of the standard basis with one in the $Z^ {(\ell)}_{it}$--th position, $i,j\in \mathcal{V}$, $t\in \mathcal{T}$, $Z^{(\ell)}_{it}\in \mathcal{Q}^{(\ell)}$. For example, if the weights are log--normally distributed, the $\theta^{(\ell)}_{Z^{(\ell)}_{it}Z^{(\ell)}_{jt}}=(\beta^{(\ell)\prime}_{qr},\sigma_{qr}^{(\ell)2})'$ includes: i) a vector $\beta^{(\ell)}_{Z^{(\ell)}_{it}Z^{(\ell)}_{jt}}=(\tilde{Z}^{(\ell)\prime}_{it}\beta_{0}^{(\ell)}\tilde{Z}^{(\ell)}_{jt},\ldots,\tilde{Z}^{(\ell)\prime}_{it}\beta_{S-1}^{(\ell)}\tilde{Z}^{(\ell)}_{jt})'$ of intercept and the coefficients of the covariates $(X^{(\ell)}_{ijt})$; and ii) the variance $\sigma^{(\ell)2}_{Z^{(\ell)}_{it}Z^{(\ell)}_{jt}}=\tilde{Z}^{(\ell)\prime}_{it}\sigma^{(\ell)}\tilde{Z}^{(\ell)}_{jt}$.} In this sense, the DSBMM model can also be interpreted as a regression model with parameters partially pooled across dyads and time. Consequently, (\ref{eq:1}) is more flexible than a standard fixed parameters regression, and more feasible and parsimonious than a time--dyad varying case $\vartheta_{ijt}^{(\ell)}$. Thus, the DSBMM extends the dynamic single--layer SBM \citep{olivella2022dynamic} and the static SBM with observed heterogeneity \citep{mariadassou2010uncovering} to a dynamic multidimensional setup with time--varying node or dyad characteristics, $X^{(\ell)}_{ijt}$. In other words, apart from introducing inter--layer dependence, in the one--layer case $L=1$, the DSBMM in (\ref{eq:1}) differs from \citet{yang2011detecting,matias2017statistical} in accounting for node and dyad observed heterogeneity through $X^{(\ell)}_{ijt}$.
	
	%Note that although in (\ref{eq:1}) all parameters are block dependent, this does not preclude the possibility of having some parameters of $\theta^{(\ell)}_{qr}$ fixed across blocks, it is only a special case.
	
	We introduce a different representation of the model, which offers many advantages. Let $D_{ijt}^{(\ell)}$ be an indicator variable following a Bernoulli distribution, such that $D_{ijt}^{(\ell)}=1$ if $(i,j)\in \mathcal{E}_t^{(\ell)}$ and $D_{ijt}^{(\ell)}=0$ if $(i,j)\not\in \mathcal{E}_t^{(\ell)}$, then (\ref{eq:1}) can be rewritten as
	\begin{eqnarray}
		Y^{(\ell)}_{i j t}\left|X^{(\ell)}_{ijt},D^{(\ell)}_{ijt},Z^{(\ell)}_{it}=q, Z^{(\ell)}_{jt}=r,\vartheta^{(\ell)}\right. &\sim& 			\delta(y) (1-D^{(\ell)}_{ij t})+f^{(\ell)}\left(y| \theta^{(\ell)}_{qr},X^{(\ell)}_{ijt}\right) D^{(\ell)}_{ijt},\label{eq:2}\\
  D^{(\ell)}_{ijt}\left|X^{(\ell)}_{ijt},Z^{(\ell)}_{it}=q, Z^{(\ell)}_{jt}=r,\vartheta^{(\ell)}\right.&\sim& \operatorname{Bern}\left(\nu^{(\ell)}_{qr}\right).\label{eq:3}
	\end{eqnarray}
	
\textcolor{black}{A first advantage is that (\ref{eq:2}) and (\ref{eq:3}) account for both weighted and unweighted layers. In case of unweighted layers, equations (\ref{eq:2}) and (\ref{eq:3}) simplify to (\ref{eq:3}). The second advantage concerns the model's flexibility regarding the edge's existence. If $Y^{(\ell)}_{i j t}=0$ indicates the absence of an edge and not a missing value, then this representation provides a two--part network formation process and the common stochastic membership accounts for possible dependence between the two parts. If all zeros are missing data, $D^{(\ell)}_{ijt}=1$ refers to an observed $Y^{(\ell)}_{i j t}$ and $D^{(\ell)}_{ijt}=0$ denotes missing values. This representation can be interpreted as a random process for the missing values, driven by a common stochastic membership. Assuming dependence between the missing value mechanism and the observable data is common in many applied network areas, from sample selection or common factor approaches, and extending pattern mixture  \citep{helpman2008estimating, little1993pattern,song2004imputation}. The third advantage is that this representation can be used to include covariates in \eqref{eq:3}, i.e. $\tau\left(\nu^{(\ell)}_{qr}\right)= X^{(\ell)'}_{ijt}\theta^{(\ell)}_{qr}$, where $\tau:(0,1)\to\mathbb{R}$ is a link function. Indeed, other latent components, apart from the stochastic membership, can be used as in Heckman's procedure or common factor approaches \citep{van2011bayesian}. }

 %Nevertheless, the model in (\ref{eq:2}) and (\ref{eq:3}) can be easily extended to account for missing values and latent indicator variables following a Heckman's procedure \citep[e.g., see][]{van2011bayesian}.

%Notice that (\ref{eq:1}) implies not only a time dependence through the HMCs, but a specific edge dependence. If $Y_{ijt}^{(\ell)}\mid D^{(\ell)}_{ijt}=1,Z^{(\ell)}_{it}=q, Z^{(\ell)}_{jt}=r,\vartheta^{(\ell)}\sim \operatorname{N}(\beta_{qr0},\sigma_{qr}^2)$, $\beta^{(\ell)}_{qr0}\sim\operatorname{N}(0,\underline{\Sigma}^{(\ell)})$ and $\nu^{(\ell)}_{qr}\sim \operatorname{Beta}(\underline{b}^{(\ell)},\underline{c}^{(\ell)})$, then	\[\!\!\operatorname{cov}\left(Y^{(\ell)}_{i_{1}j_{1}t},Y^{(\ell)}_{i_{2}j_{2}t}\mid Z^{(\ell)}_{i_1t},Z^{(\ell)}_{j_1t},Z^{(\ell)}_{i_2t},Z^{(\ell)}_{j_2t}\right)= \left[\frac{\underline{b}^{(\ell)}\underline{c}^{(\ell)}+(\underline{b}^{(\ell)})^2(\underline{b}^{(\ell)}+\underline{c}^{(\ell)}+1)}{(\underline{b}^{(\ell)}+\underline{c}^{(\ell)})^2(\underline{b}^{(\ell)}+\underline{c}^{(\ell)}+1)}\right]\underline{\Sigma}^{(\ell)}\mathbb{I}_{\{Z_{i_{2}t}\}}(Z_{i_{1}t})\mathbb{I}_{\{Z_{j_{2}t}\}}(Z_{j_{1}t}).\]

{\color{black}
The specification (\ref{eq:1}) implies not only a time dependence through the HMCs, but a cross--sectional edge dependence. In Proposition \ref{prop:modelprop}, this property is exemplified under specific distributional assumptions. The edges within the same community pair interaction have a covariance different from zero, while the edges with different community pair interactions have zero covariance, i.e. a block covariance structure. Moreover, the discrete mixture also provides more flexibility in the relationship between the first and second moments of the weights. In particular, while the log--normal distribution of the weights implies a positive quadratic relationship between mean and variance within community pair interaction, the mixture allows for discontinuities that accommodate monotonic and non--monotonic jumps in the relationship mean--variance. 

\begin{proposition} \label{prop:modelprop}
	Let $Y_{ijt}^{(\ell)}\left|X^{(\ell)}_{ijt},Z^{(\ell)}_{it}=q, Z^{(\ell)}_{jt}=r, \vartheta^{(\ell)}\right.\sim (1-\nu^{(\ell)}_{qr})\delta(y)+\nu^{(\ell)}_{qr} \operatorname{LN}\left(y|\beta^{(\ell)}_{0qr},\sigma_{qr}^{(\ell)2}\right)$, $\beta^{(\ell)}_{0qr}\sim\operatorname{N}(\underline{\beta}^{(\ell)}_{0},\underline{\Sigma}^{(\ell)})$, $\nu^{(\ell)}_{qr}\sim \operatorname{Beta}(\underline{b}^{(\ell)},\underline{c}^{(\ell)})$, and $\sigma_{qr}^{(\ell)2}\sim \operatorname{G}(1,1/\underline{e}^{(\ell)})$, then
	
	\begin{enumerate}[leftmargin=0.2cm]
		\item $\mathbb{E}(Y^{(\ell)}_{ijt}|\vartheta^{(\ell)},Z^{(\ell)}_{it}=q,Z^{(\ell)}_{jt}=r)=\nu^{(\ell)}_{qr}\exp(\beta^{(\ell)}_{0qr}+\sigma_{qr}^{(\ell)2}/2)$;
		\item 				$
		\mathbb{V}(Y^{(\ell)}_{ijt}|\vartheta^{(\ell)},Z^{(\ell)}_{it}=q,Z^{(\ell)}_{jt}=r)=\mathbb{E}(Y^{(\ell)}_{ijt}|\vartheta^{(\ell)},Z^{(\ell)}_{it}=q,Z^{(\ell)}_{jt}=r)^2 \left(\frac{\exp(\sigma_{qr}^{(\ell)2})}{\nu^{(\ell)}_{qr}}-1\right)
		$;
		\item $\mathbb{C}ov(Y^{(\ell)}_{i_1j_1t},Y^{(\ell)}_{i_2j_2t}|Z^{(\ell)}_{i_1t},Z^{(\ell)}_{j_1t},Z^{(\ell)}_{i_2t},Z^{(\ell)}_{j_2t})=$\newline \hspace*{3.3cm} $ \left(\frac{\underline{b}^{(\ell)}(\underline{b}^{(\ell)}+1)\exp(2(\underline{\beta}^{(\ell)}_{0}+\underline{\Sigma}^{(\ell)}))\underline{e}^{(\ell)}}{(\underline{b}^{(\ell)}+\underline{c}^{(\ell)})(\underline{b}^{(\ell)}+\underline{c}^{(\ell)}+1)(\underline{e}^{(\ell)}-1)}-\frac{4\underline{e}^{(\ell)2}\underline{b}^{(\ell)2}\exp(2\underline{\beta}^{(\ell)}_{0}+\underline{\Sigma}^{(\ell)})}{(\underline{b}^{(\ell)}+\underline{c}^{(\ell)})^2(2\underline{e}^{(\ell)}-1)^2}\right)\mathbb{I}_{\{Z_{i_{2}t}\}}(Z_{i_{1}t})\mathbb{I}_{\{Z_{j_{2}t}\}}(Z_{j_{1}t})$.
	\end{enumerate}
	
	\end{proposition}
The conditional expected weights follow a Pareto--log--normal distribution, a heavy--tail distribution used in network analysis to describe the degree distribution and also in international trade theory to capture the productivity heterogeneity of the firms at a granular level \citep[e.g.,][]{nigai2017tale,fang2011double}. As in the Pareto distribution, the Pareto--log--normal distribution's tail properties would be preserved under convolutions, meaning that the behavior of the extreme values in the network strength would inherit similar characteristics. The variance--to--mean ratio (VM) measures the overdispersion features and is given in the following. 
\begin{corollary} \label{col:1} Based on Proposition \ref{prop:modelprop}, the VM ratio is given by     $$VM=\frac{\exp(\underline{\beta}^{(\ell)}_0+\underline{\Sigma}^{(\ell)}/2)(2\underline{e}^{(\ell)}-1)}{2}\left(\frac{\exp(\underline{\Sigma}^{(\ell)})}{\underline{e}^{(\ell)}-2}-\frac{4\underline{e}^{(\ell)}\underline{b}^{(\ell)}}{(\underline{b}^{(\ell)}+\underline{c}^{(\ell)})(2\underline{e}^{(\ell)}-1)^2}\right), \ \underline{e}^{(\ell)}>2. $$
\end{corollary}
As $\underline{e}^{(\ell)}$ goes to infinity, the moments and VM ratio gets closer to the log--normal case, while as $\underline{e}^{(\ell)}\to2^+$ the VM ratio explodes due to the heavy tails, with tail index of $1/\underline{e}^{(\ell)}$.  These properties can be used to calibrate the hyperparameters of the prior in applications, where prior information on the tails of the strength is available.

	\textcolor{black}{Notice that by specifying inter--layer edge conditional distribution given the layer blocks, the model framework and properties can be easily extended to a general multi--layer networks, where layers are node aligned and edges between layers are allowed. Since this specification is not relevant to our application, we leave it for further research.}
}
	
	\subsection{Layer Dependence}
\label{sec:laydep}

A statistical model for multidimensional networks should account for edge redundancy, that is, the persistence of edges between the same pair of nodes across network layers, and more structurally, clustering redundancy \citep{han2015consistent,stanley2016clustering,jovanovski2019bayesian}. For instance, if a set of nodes is densely connected in one layer, a similar structure is highly likely to be present in the second layer. Nevertheless, existing approaches are static and correlational and only account for clustering overlap. To capture dependence between the node partitions in the different layers, including non--overlap dependence, and identify a \textcolor{black}{NGB} causal structure, we can exploit the time dimension and assume dependent Markov Chain processes \citep{otranto2005multi,agudze2021markov}.
	
We are interested in a Granger non--causality relationship across layers. In this respect, the transition probabilities for each layer only depend on the collection of memberships $\mathfrak{V}_{t-1}$, that is $Z_{it-1}=\textcolor{black}{(Z^{(\ell)}_{it-1})_{\ell \in \mathcal{L}}}$. This assumption simplifies as follows the joint probability $\mathbb{P}(Z_{it}|Z_{it-1})$, $i\in \mathcal{V}$,   
	\begin{equation}
		\mathbb{P}(Z_{it}|Z_{it-1})=\mathbb{P}\left(Z^{(1)}_{it}|Z_{it-1}\right)\prod_{\ell=2}^{L}\mathbb{P}\left(Z^{(\ell)}_{it}|Z^{(1:\ell-1)}_{it},Z_{it-1}\right)=\prod_{\ell=1}^{L}\mathbb{P}\left(Z^{(\ell)}_{it}|Z_{it-1}\right).\label{eq:membership}
	\end{equation}
	
In contrast to previous studies, which typically cover dependent Markov chains with only two states, the DSBMM requires a more general approach. Therefore, we use a multinomial logit form to express the Markov chain dependence through the transition matrix $\mathbb{P}(Z^{(\ell)}_{it}=r|Z_{it-1})=P^{(\ell)}_{it,Z^{(\ell)}_{it-1}r}, \ r=1,\dots, Q^{(\ell)}$ with $Q^{(\ell)}$ states. Similarly to the log--linear models used for contingency tables, $Z_{it-1}$ is represented by a full set of dummy variables, \textcolor{black}{from now on lagged HMC variables}, $W^{(\ell)}_{it,q}=\mathbb{I}_{\{q\}}\left(Z^{(\ell)}_{it}\right)$, and $W^{(\ell)}_{it}=(W^{(\ell)}_{it,1},\ldots, W^{(\ell)}_{it, Q^{(\ell)}-1})'$, including the interactions and the main effects of the different layers, that is
	\begin{equation}
		\label{eq:dichmod}
		\begin{aligned}
			\log\left(P^{(\ell)}_{it,Z^{(\ell)}_{it-1}r} C^{(\ell)}_{it}\right)&=\kappa^{(\ell)}_{0,r}+\underbrace{\sum_{\mathfrak{m}=1}^L W^{(\mathfrak{m})'}_{it-1}\kappa^{(\ell)}_{\mathfrak{m},r}}_{\text{main effects}}+\underbrace{\sum_{\mathfrak{m}>\mathfrak{n}}\left(W^{(\mathfrak{m})'}_{it-1}\otimes W^{(\mathfrak{n})'}_{it-1}\right)\kappa^{(\ell)}_{\mathfrak{m}\mathfrak{n},r}}_{\text{first order effects}}+\ldots+\underbrace{\bigotimes_{\mathfrak{m}=1}^{L}\left(W^{(\mathfrak{m})'}_{it-1}\right)\kappa^{(\ell)}_{1\ldots L,r}}_{(L-1)\text{-th order effects}},
		\end{aligned}
	\end{equation}
where $\kappa^{(\ell)}_{\mathcal{U},r}=\left(\kappa^{(\ell)}_{\mathcal{U},r,1},\ldots,\kappa^{(\ell)}_{\mathcal{U},r,s(\mathcal{U})}\right)'$ \textcolor{black}{is the collection of parameters measuring the effect of past membership values on the transition to community $r$ in layer $\ell$, and $s(\mathcal{U})=\prod_{\mathfrak{m}\in\mathcal{U}}\left(Q^{(\mathfrak{m})}-1\right)$ is the size of the vector, and $ \mathcal{U}\in \mathcal{P}$ identifies the set of interacting layers with  $\mathcal{P}$ the power set of the layer set $\mathcal{L}$}. \textcolor{black}{For instance, the main effects of the layer 1 on the log odds of changing to block $r$ in the layer $\ell$, i.e. $\mathcal{U}=\{1\}$, is captured by the vector $\kappa^{(\ell)}_{\{1\},r}=\left(\kappa^{(\ell)}_{\{1\},r,1},\ldots,\kappa^{(\ell)}_{\{1\},r,s(\{1\})}\right)'$, where the number of elements is related to the number of communities in layer $s(\{1\})=Q^{(1)}-1$. Following the example, the first--order interaction effects of layer 1 and layer $\ell$ are collected in the vector  $\kappa^{(\ell)}_{\{\ell,1\},r}=\left(\kappa^{(\ell)}_{\{\ell,1\},r,1},\ldots,\kappa^{(\ell)}_{\{\ell,1\},r,s(\{\ell,1\})}\right)'$, where $s(\{\ell,1\})=(Q^{(1)}-1)(Q^{(\ell)}-1)$.} On the left side, $C^{(\ell)}_{it}=\sum_{k\in \mathcal{Q}^{(\ell)} }\exp\left(\widetilde{W}_{it-1}\kappa_k^{(\ell)}\right)$ is the normalizing constant of the multinomial model, where $\widetilde{W}_{it-1}=$ $(1,W^{(\ell)'}_{it-1}$, $\ldots$, $\bigotimes_{\mathfrak{m}=1}^{L}(W^{(\mathfrak{m})'}_{it-1}))$ and $\kappa_{r}^{(\ell)}=\left(\kappa^{(\ell)}_{0,r},\ldots,\kappa^{(\ell)'}_{1\ldots L,r}\right)'$ are a row and a column vector, respectively, with  $p+1$ elements, with $p=\sum_{\mathcal{U}\in \mathcal{P}} s(\mathcal{U})$. Equation (\ref{eq:dichmod}) can be rewritten in an equivalent, more compact form:
	\begin{equation}
		\label{eq:multn1}
		P^{(\ell)}_{it,Z^{(\ell)}_{it-1}r}=\frac{\exp\left(\widetilde{W}_{it-1}\kappa_{r}^{(\ell)}\right)}{\sum_{k=1}^{Q^{(\ell)}}\exp\left(\widetilde{W}_{it-1}\kappa_{k}^{(\ell)}\right)}.
	\end{equation}

For identification, $Z^{(\ell)}_{it}=Q^{(\ell)}$ can be used as reference state/community, i.e. $\kappa^{(\ell)}_{Q^{(\ell)}}=\mathbf{0}$. This multivariate logistic specification is more flexible than the linear specification proposed by \citet{agudze2021markov} for a panel Markov switching model because it does not have to impose any restriction on the parameter space of $\kappa_{r}^{(\ell)}$. Consequently, (\ref{eq:multn1}) captures the special case of clustering overlap, as well as more complex dependence structures, such as the decoupling one. In this modeling framework, it is straightforward to add other exogenous variables that affect memberships and the transition matrix \citep{kaufmann2015k,holsclaw2017bayesian}. This flexibility in (\ref{eq:multn1}) comes at a cost of lower tractability and a larger number of parameters. In this paper, we provide a solution to both issues based on a suitable inference approach.

We extend to the block membership processes the definition of non--causality introduced by \citet{mosconi2006non}. Let $\{Z_{{\color{black}\sdot t}}=(Z^{(1)}_{{\color{black}\sdot t}},\dots,Z^{(L)}_{{\color{black}\sdot t}})', \ t\in \mathcal{T}\}$, or simply $\{Z_{{\color{black}\sdot t}}\}$, be a random sequence in the probability space with the triplet $(\Xi, \mathcal{A},\mathbb{P})$\textcolor{black}{, where $\Xi$ refers to the sample space, that is the potential memberships in all layers, $\mathcal{A}$ the sigma algebra of events, in other words the potential trajectories of the memberships and $\mathbb{P}$ its probability measure}. An information set and a restricted information set are required to define a non--causality property. The available information up to time $t$, in terms of  $\{Z_{{\color{black}\sdot t}}\}$, is referred to as canonical filtration $\left\{\mathcal{F}_t, \ t\in \mathcal{T}\right\}$, which is a sub--$\sigma$--algebra of $\mathcal{A}$. The reduced information sets are  $\mathcal{R}^{(-\ell)}_t=\sigma((Z_{{\color{black}\sdot s}}^{(-\ell)}),\ 1\leq s\leq t)$ and $\mathcal{Z}^{(\ell)}_t=\sigma(Z_{{\color{black}\sdot s}}^{(\ell)},\ 1\leq s\leq t)$, for a layer $\ell\in \mathcal{L}$, so that $\mathcal{Z}^{(\ell)}_t\subseteq\mathcal{R}^{(-\mathfrak{m})}_t\subseteq\mathcal{F}_t$ for all  $\mathfrak{m}\neq \ell$.

	\begin{definition}[Strong one--step--ahead Nonlinear \textcolor{black}{conditional} Granger block non--causality] \label{def1}
		The process $\{Z^{(\mathfrak{m})}_{{\color{black}\sdot t}}\}$ does not strongly \textcolor{black}{NGB} cause $\{Z^{(\ell)}_{{\color{black}\sdot t}}\}$ one--step ahead \textcolor{black}{conditionally given $\mathcal{R}^{(-\mathfrak{m})}_{t-1}$}, and write $Z^{(\mathfrak{m})}\not \stackrel{G}{\Rightarrow} Z^{(\ell)}$ \textcolor{black}{for $\mathfrak{m} \neq \ell$}, if 	
		\[\mathcal{Z}^{(\ell)}_t\bot \ \mathcal{Z}^{(\mathfrak{m})}_{t-1}\left|\mathcal{R}^{(-\mathfrak{m})}_{t-1}\right., \qquad \forall \ t\in \mathcal{T}.\]		
		\textcolor{black}{If $\{Z^{(\ell)}_{{\color{black}\sdot t}}\}$ is a set of first--order Markov chain processes such that a node membership does not depend on other node memberships across all layers, i.e. $\mathbb{P}(Z^{(\ell)}_{it}|Z_{\sdot t-1})=\mathbb{P}(Z^{(\ell)}_{it}|Z_{it-1})$ for $\ell=1,\ldots,L$, as in  \eqref{eq:membership} of our DSBMM,} then $Z^{(\mathfrak{m})}\not \stackrel{G}{\Rightarrow} Z^{(\ell)}$, if 
\[ \mathbb{P}\left(Z^{(\ell)}_{it}|Z_{it-1}\right)=\mathbb{P}\left(Z^{(\ell)}_{it}|Z^{(-\mathfrak{m})}_{it-1}\right),\ \forall \ t\in \{2,\dots,T\}, \ i \in \mathcal{V}.\]
		
	\end{definition}

	\begin{figure}[t]
		\centering
		\caption{Dynamic undirected and unweighted network with two layers over time (different panels), where Layer 1 \textcolor{black}{NGB} causes Layer 2. In each panel, the block structure (grey shades) and node alignment (dashed lines).}
		\label{fig:1}
  		\begin{subfigure}{0.4\textwidth}
			\centering
			\caption{$t=t_1$ ($t_1<t_2$)}
			\SetCoordinates[yAngle=35]
			\begin{tikzpicture}[multilayer=3d,scale=0.6]
				\SetLayerDistance{3}
				\SetVertexStyle[FillColor=white]	
				\begin{Layer}[layer=1]
					%\draw[very thick] (-.5,-.5) rectangle (2.5,2);
					\node at (-.5,-.5)[below right]{Layer 1};
				\end{Layer}
				
				\begin{Layer}[layer=2]
					%\draw[very thick] (-.5,-.5) rectangle (2.5,2);
					\node at (-.5,-.5)[below right]{Layer 2};
				\end{Layer}

				\Plane[x=-.5,y=-.5,width=6,height=4, opacity=0.1,layer=1,NoFill]
				\Plane[x=-.5,y=-.5,width=6,height=4,opacity=0.1,layer=2,NoFill]

				\Vertex[x=0.2,y=0.5,label=1,layer=1,color=Gray]{A}
				\Vertex[x=1.6,y=0.5,label=2,layer=1,color=Gray]{B}
				\Vertex[x=0.8,y=1.5,label=3,layer=1,color=Gray]{C}
				
				\Vertex[x=3,y=3,label=4,layer=1,color=Gray!30]{D}
				\Vertex[x=1.9,y=3,label=5,layer=1,color=Gray!30]{E}
				\Vertex[x=2.5,y=2,label=6,layer=1,color=Gray!30]{F}
				
				\Vertex[x=4,y=1.5,label=7,layer=1,color=Gray!0]{G}
				\Vertex[x=3.4,y=0.5,label=8,layer=1,color=Gray!0]{H}	
				\Vertex[x=4.8,y=0.5,label=9,layer=1,color=Gray!0]{I}
				\Vertex[x=1.3,y=1.5,label=1,layer=2,color=Gray]{A2}
				\Vertex[x=3.2,y=0.3,label=2,layer=2,color=Gray]{B2}
				\Vertex[x=3.6,y=1.5,label=3,layer=2,color=Gray]{C2}
				\Vertex[x=2.5,y=2.5,label=4,layer=2,color=Gray]{D2}
				\Vertex[x=1.7,y=0.3,label=7,layer=2,color=Gray]{G2}
				
				\Vertex[x=0,y=2,label=5,layer=2,color=Gray!0]{E2}
				\Vertex[x=1.1,y=3,label=9,layer=2,color=Gray!0]{I2}
				
				\Vertex[x=5,y=2,label=8,layer=2,color=Gray!30]{H2}	
				\Vertex[x=3.7,y=3,label=6,layer=2,color=Gray!30]{F2}
				
				% Layer 1
				\Edge[bend=-25](A)(B)
				\Edge[bend=25](A)(C)
				\Edge[bend=-25](B)(C)
				
				\Edge[bend=-25](D)(E)
				\Edge[bend=25](D)(F)
				\Edge[bend=-25](E)(F)
				
				\Edge[bend=-25](G)(H)
				\Edge[bend=25](G)(I)
				\Edge[bend=-25](H)(I)
				
				%\Edge[bend=-25](C)(F)
				%\Edge[bend=-25](D)(G)
				
				% Layer 2
				\Edge[bend=25](A2)(D2) \Edge[bend=-25](A2)(G2) \Edge[bend=25](D2)(C2) \Edge[bend=25](C2)(B2) \Edge[bend=25](B2)(G2) \Edge[bend=25](D2)(G2) \Edge[bend=25](A2)(C2)
				
				\Edge[bend=25](E2)(I2) \Edge[bend=-25](H2)(F2)
				
				%Inter-layer
				\Edge[style=dashed,lw=0.5](A)(A2) \Edge[style=dashed,lw=0.5](B)(B2) \Edge[style=dashed,lw=0.5](C)(C2) \Edge[style=dashed,lw=0.5](D)(D2) \Edge[style=dashed,lw=0.5](E)(E2) \Edge[style=dashed,lw=0.5](F)(F2) \Edge[style=dashed,lw=0.5](G)(G2) \Edge[style=dashed,lw=0.5](H)(H2) \Edge[style=dashed,lw=0.5](I)(I2)
				
			\end{tikzpicture}
		\end{subfigure}
		\begin{subfigure}{0.4\textwidth}
			\centering
			\caption{$t=t_2$ ($t_1<t_2$)}
			\SetCoordinates[yAngle=35]
			\begin{tikzpicture}[multilayer=3d,scale=0.6]
				\SetLayerDistance{3}
				\SetVertexStyle[FillColor=white]	
				\begin{Layer}[layer=1]
					%\draw[very thick] (-.5,-.5) rectangle (2.5,2);
					\node at (-.5,-.5)[below right]{Layer 1};
				\end{Layer}
				
				\begin{Layer}[layer=2]
					%\draw[very thick] (-.5,-.5) rectangle (2.5,2);
					\node at (-.5,-.5)[below right]{Layer 2};
				\end{Layer}

				\Plane[x=-.5,y=-.5,width=6,height=4, opacity=0.1,layer=1,NoFill]
				\Plane[x=-.5,y=-.5,width=6,height=4,opacity=0.1,layer=2,NoFill]

				\Vertex[x=0.2,y=0.5,label=1,layer=1,color=Gray]{A}
				\Vertex[x=1.6,y=0.5,label=2,layer=1,color=Gray]{B}
				\Vertex[x=0.8,y=1.5,label=3,layer=1,color=Gray]{C}
				
				\Vertex[x=3,y=3,label=4,layer=1,color=Gray!30]{D}
				\Vertex[x=1.9,y=3,label=5,layer=1,color=Gray!30]{E}
				\Vertex[x=2.5,y=2,label=6,layer=1,color=Gray!30]{F}
				
				\Vertex[x=4,y=1.5,label=7,layer=1,color=Gray!0]{G}
				\Vertex[x=3.4,y=0.5,label=8,layer=1,color=Gray!0]{H}	
				\Vertex[x=4.8,y=0.5,label=9,layer=1,color=Gray!0]{I}
				\Vertex[x=0.2,y=0.5,label=1,layer=2,color=Gray]{A2}
				\Vertex[x=1.6,y=0.5,label=2,layer=2,color=Gray]{B2}
				\Vertex[x=0.8,y=1.5,label=3,layer=2,color=Gray]{C2}
				
				\Vertex[x=3,y=3,label=4,layer=2,color=Gray!30]{D2}
				\Vertex[x=1.9,y=3,label=5,layer=2,color=Gray!30]{E2}
				\Vertex[x=2.5,y=2,label=6,layer=2,color=Gray!30]{F2}
				
				\Vertex[x=4,y=1.5,label=7,layer=2,color=Gray!0]{G2}
				\Vertex[x=3.4,y=0.5,label=8,layer=2,color=Gray!0]{H2}	
				\Vertex[x=4.8,y=0.5,label=9,layer=2,color=Gray!0]{I2}
				
				% Layer 1
				\Edge[bend=-25](A)(B)
				\Edge[bend=25](A)(C)
				\Edge[bend=-25](B)(C)
				
				\Edge[bend=-25](D)(E)
				\Edge[bend=25](D)(F)
				\Edge[bend=-25](E)(F)
				
				\Edge[bend=-25](G)(H)
				\Edge[bend=25](G)(I)
				\Edge[bend=-25](H)(I)
				
				%\Edge[bend=-25](C)(F)
				%\Edge[bend=-25](D)(G)
				
				% Layer 2		
				\Edge[bend=-25](A2)(B2)
				\Edge[bend=25](A2)(C2)
				\Edge[bend=-25](B2)(C2)
				
				\Edge[bend=-25](D2)(E2)
				\Edge[bend=25](D2)(F2)
				\Edge[bend=-25](E2)(F2)
				
				\Edge[bend=-25](G2)(H2)
				\Edge[bend=25](G2)(I2)
				\Edge[bend=-25](H2)(I2)
				
				%\Edge[bend=-25](C2)(F2)
				%\Edge[bend=-25](D2)(G2)
				
				%Inter-layer
				\Edge[style=dashed,lw=0.5](A)(A2) \Edge[style=dashed,lw=0.5](B)(B2) \Edge[style=dashed,lw=0.5](C)(C2) \Edge[style=dashed,lw=0.5](D)(D2) \Edge[style=dashed,lw=0.5](E)(E2) \Edge[style=dashed,lw=0.5](F)(F2) \Edge[style=dashed,lw=0.5](G)(G2) \Edge[style=dashed,lw=0.5](H)(H2) \Edge[style=dashed,lw=0.5](I)(I2)
				
			\end{tikzpicture}
		\end{subfigure}		
	\end{figure}

 	\begin{figure}[t]
		\centering
		\caption{Alluvial plots of the node membership dynamics. In each panel, the plots for Layer 2 (left) and Layer 3 (right). In each plot, the initial (left bar) and final (right bar) block configurations, and the block membership of the nodes (colored curves) over 15 periods (horizontal axis). The height of the bar and the thickness of the curves are proportional to the number of nodes. }
		\label{fig:alluvial_lay2_exmp1}
  \setlength{\tabcolsep}{3pt}
  \renewcommand{\arraystretch}{0.2}
  \begin{tabular}{cccc}
  \multicolumn{2}{c}{\small (a) Unidirectional causality}&  \multicolumn{2}{c}{\small (b) Bidirectional causality}\\
\includegraphics[trim=70mm 20mm 70mm 0,clip,height=170pt, width=100pt]{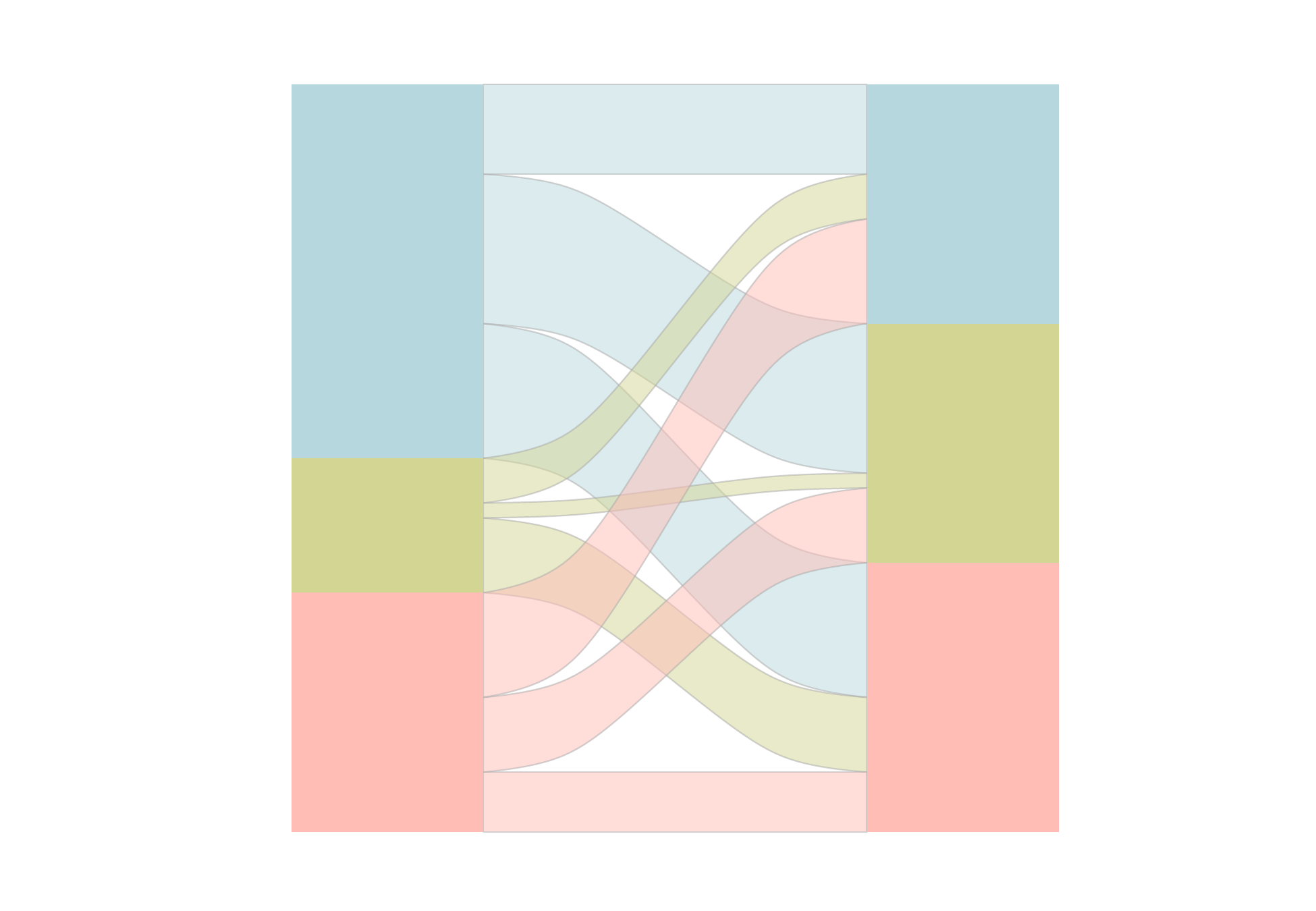}& 
\includegraphics[trim=70mm  20mm 70mm 0,clip,height=170pt, width=100pt]{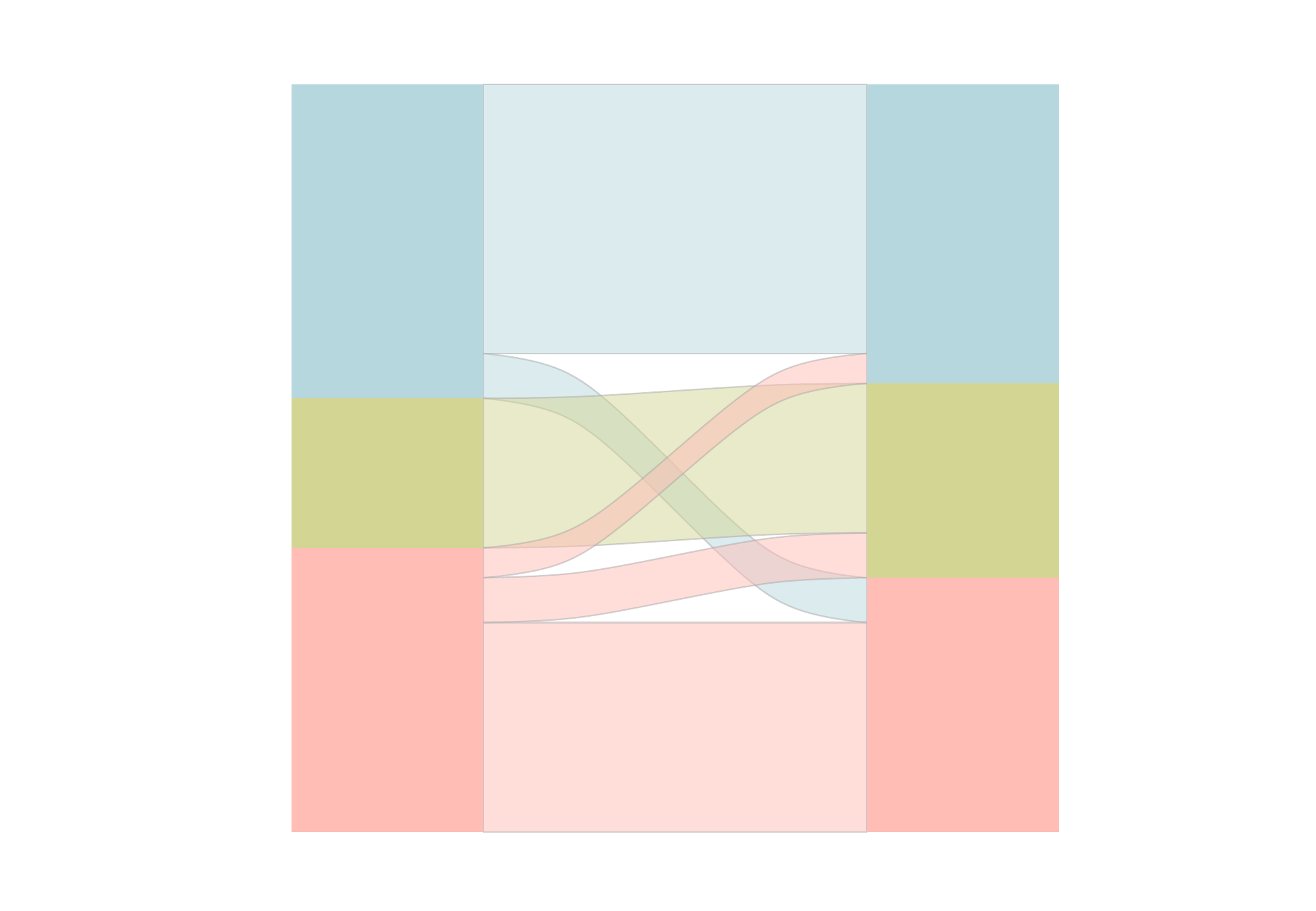}&
\includegraphics[trim=70mm  20mm 70mm 0,clip,height=170pt, width=100pt]{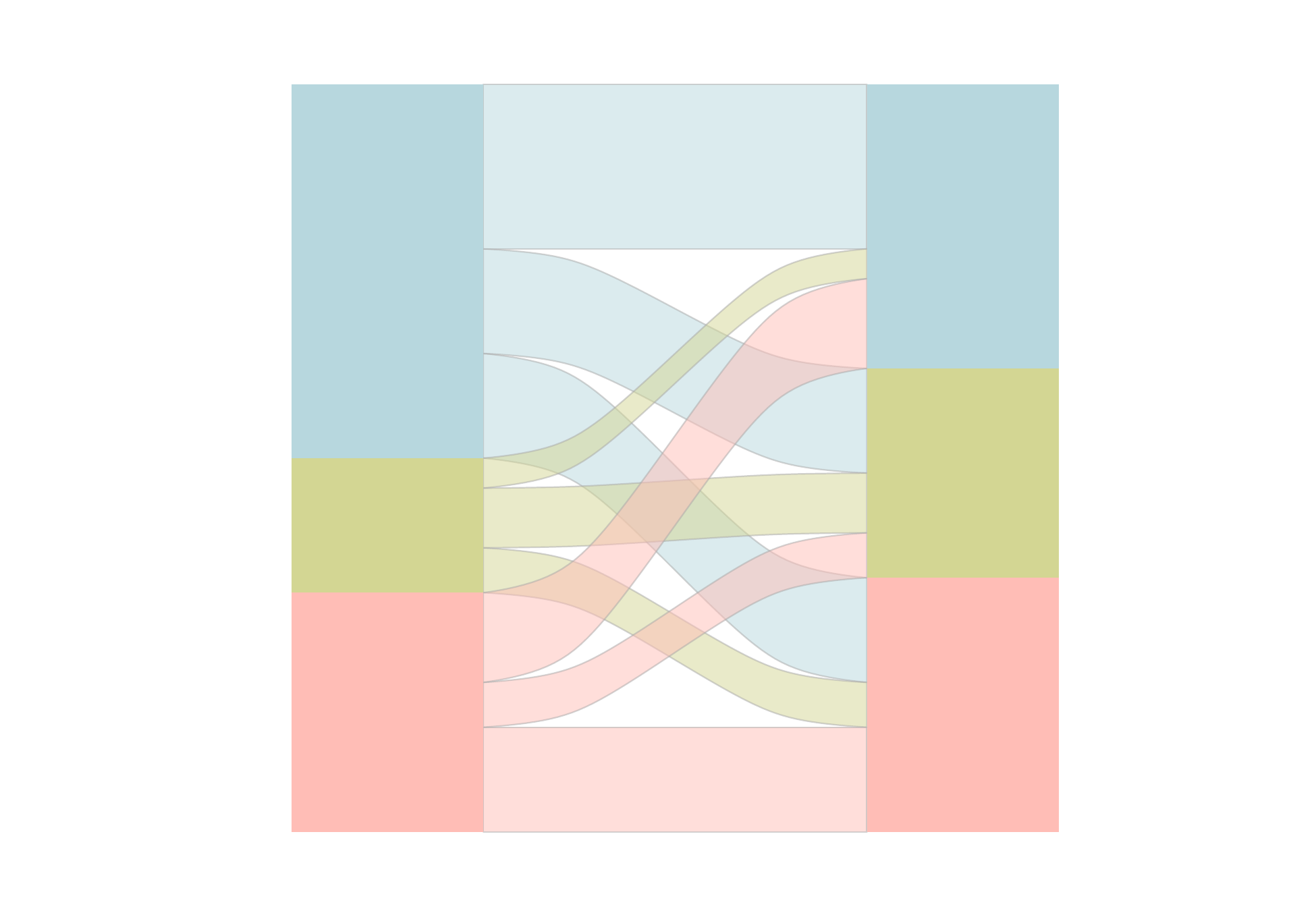}& 
\includegraphics[trim=70mm  20mm 70mm 0,clip,height=170pt, width=100pt]{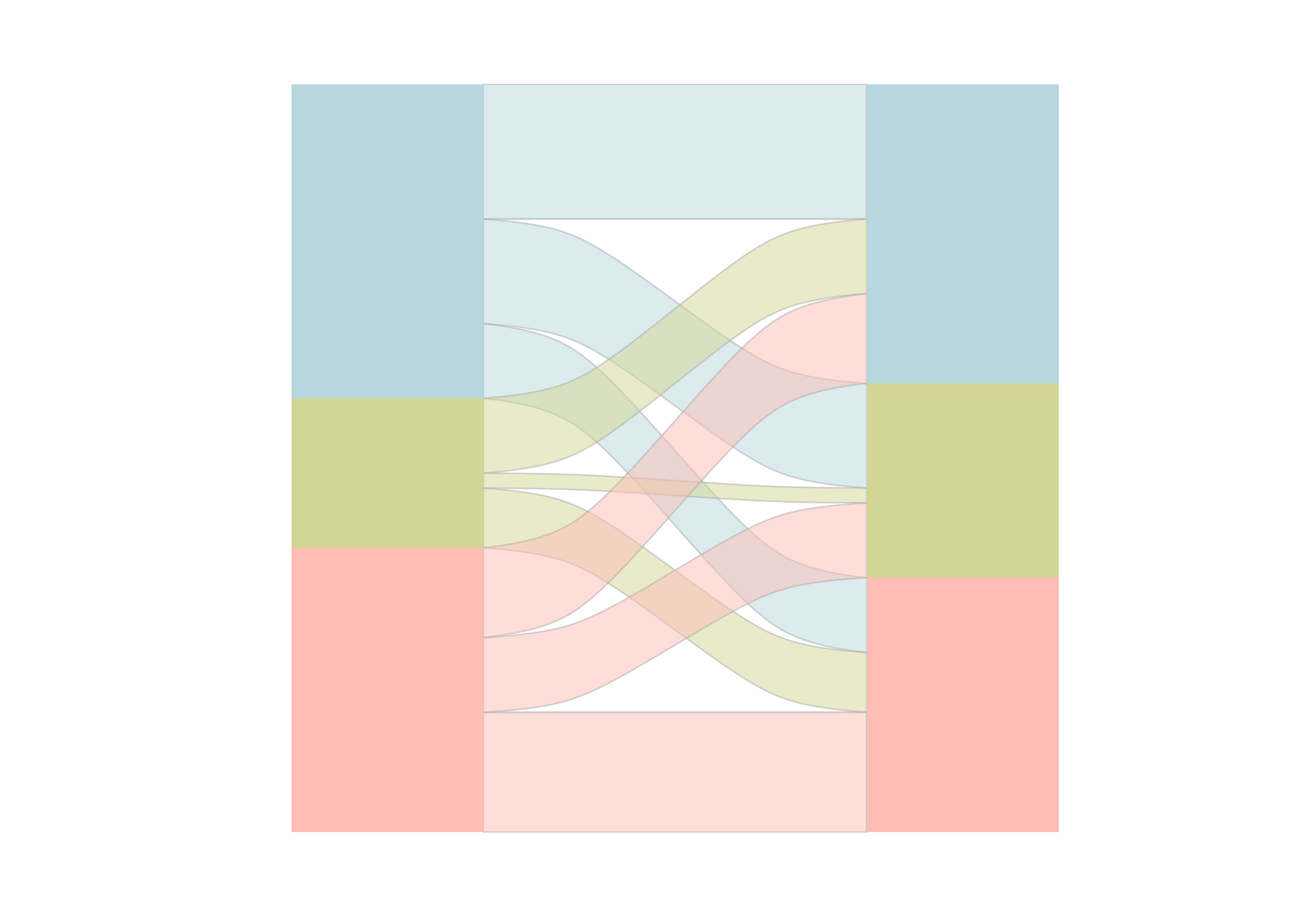}
  \end{tabular}
	\end{figure}	

 Following the specification (\ref{eq:dichmod}), Layer $\mathfrak{m}$ does not \textcolor{black}{NGB} cause Layer $\ell$ if all the main effects and interactions parameters $\kappa^{(\ell)}_{\mathcal{U},r}$ involving $\mathfrak{m}$ in the equation (\ref{eq:dichmod}) are null, otherwise $\mathfrak{m}$ does \textcolor{black}{NGB} cause Layer $\ell$. \textcolor{black}{Granger causality hypothesis testing based on this definition and the parameterization used will be introduced later on in Section \ref{sec:modselcaus}.}
  
{\color{black} Based on Definition \ref{def1}, the potential scenarios of dependence, as in a VAR, are: a unidirectional causality ($Z^{(\mathfrak{m})}\stackrel{G}{\Rightarrow} Z^{(\ell)}$ and $Z^{(\ell)}\not \stackrel{G}{\Rightarrow} Z^{(\mathfrak{m})}$ or vice--versa), bidirectional causality ($Z^{(\mathfrak{m})}\stackrel{G}{\Rightarrow} Z^{(\ell)}$ and $Z^{(\ell)}\stackrel{G}{\Rightarrow} Z^{(\mathfrak{m})}$) or absence of causality ($Z^{(\mathfrak{m})}\not\stackrel{G}{\Rightarrow} Z^{(\ell)}$ and $Z^{(\ell)}\not\stackrel{G}{\Rightarrow} Z^{(\mathfrak{m})}$)}. An illustration of the unidirectional \textcolor{black}{NGB causality} is presented in Figure \ref{fig:1}. It shows a two--layer undirected and unweighted network with nine nodes clustered into three communities, indicated by different gray shades. At time $t_1$ (panel a), the partition of the nodes in Layer 1 differs significantly from that in Layer 2. However, over time the partition in Layer 2 aligns with the one in Layer 1, producing a clustering/community overlap. Based on the definition, Layer 1 is \textcolor{black}{NGB}, causing Layer 2. This membership coupling effect is only an example, as other coupling or decoupling scenarios are possible within our DSBMM (see Section \ref{sec:conn}). 

\textcolor{black}{Our definition of causality shares some similarities with and differs in many aspects from other notions of blockwise Granger causality \citep[e.g.,][]{dufour2010short,hu2015shortcomings}. These latter are linear since they have been proposed in a VAR setup. Moreover, they refer to a relationship between groups of variables (blocks) at different lags. In our DSBMM there are collections of membership variables within each layer and a non--linear relationship between them is assumed. Thus, our NGB for DSBMM is naturally a non--linear causality notion, and regards the relationship between groups of membership variables in different layers (blocks) at different lags. The blocks of membership processes should not be confused with the blocks of nodes, nevertheless one can expect that the NGB in the membership influences the dynamics in the node block structure. As in the conditional Granger causality for linear models, our NGB controls for other blocks (i.e., layers) mediated effects, including in the conditional relationships and in the testing procedure the other--layer node memberships with lags. Finally, in our NGB, unobserved factors influencing node clustering in layer $\mathfrak{m}$ are related to their counterparts in Layer $\ell$. Thus, a substantial difference between NGB and standard Granger causality notions, is that NGB is defined for latent processes, whereas standard notions refer to observable processes.}

%edge clustering is in the other layers induced by unobserved factors summarized by the nodes' membership. Then, studying the relationship between membership in different layers would be equivalent to investigating whether the unobserved factors influencing clustering in layer $\mathfrak{m}$ are related to their counterparts in Layer $\ell$.
	
We illustrate our DSBMM through simulations in two settings: unidirectional and bidirectional \textcolor{black}{NGB causality} (see Settings \ref{ex:1} and \ref{ex:2} in Section \ref{sec:conn} of the Supplementary Material for further details). In these two settings, the network has three layers. Layers 1 and 2 are weighted and directed and have two and three communities, respectively. Layer 3 is unweighted and undirected and has three communities. The number of nodes is $N=50$, and the time horizon is $T=15$. For simplicity no regressors are included and the weighted layers are log--normal distributed with mean $\beta^{(\ell)}_{qr0}$ and variance $\sigma^{(\ell)2}_{qr}$, that is $f^{(\ell)}(y| \theta^{(\ell)}_{qr},X^{(\ell)}_{ijt})$ is equivalent to  $\operatorname{LN}(\beta^{(\ell)}_{qr0},\sigma^{(\ell)2}_{qr})$. 
 
The unidirectional configuration returns the membership dynamics given in Panel (a) of Figure \ref{fig:alluvial_lay2_exmp1}. The alluvial plots show that memberships remain stable in Layer 3, and after 15 periods, the blocks in Layer 2 are almost aligned with those in Layer 3; that is, Layer 3 is \textcolor{black}{NGB}, which causes Layer 2. In the bidirectional case (Panel b) the memberships change significantly in both layers and after 15 periods the blocks in the two layers are aligned.
 
	\section{Bayesian Inference}
	\label{sec:postapp}
	The prior choices are discussed in this section, and the full conditional posterior distributions are derived together with the Markov chain Monte Carlo (MCMC) approximation algorithm. The proofs of the results are given in Appendix \ref{sec:fullcond}
	\subsection{Prior specification}
	\label{sec:prior}
	
	Given the presence of latent variables in the DSBMM, the observed likelihood is not tractable. Moreover, the specification (\ref{eq:dichmod}) introduces all possible interaction effects between layers, causing over--parameterization. A Bayesian approach addresses these issues and provides uncertainty measures of the latent variables and the parameters such as credible intervals \citep{yang2011detecting}. The prior distributions used for the connectivity parameters are 
 \begin{eqnarray}
 \label{eq:priorcon}
     \beta^{(\ell)}_{qr}\sim \operatorname{N}(\underline{\beta}^{(\ell)}_{qr}\underline{\Sigma}^{(\ell)}_{qr}) , \quad \sigma_{qr}^{(\ell)2}\sim\operatorname{IG}(\underline{d}^{(\ell)}_{qr}/2,\underline{e}^{(\ell)}_{qr}/2), \quad \nu^{(\ell)}_{qr}\sim \operatorname{Beta}(\underline{b}^{(\ell)}_{qr},\underline{c}^{(\ell)}_{qr}) , \quad \alpha^{(\ell)}\sim\operatorname{Dir}(\underline{\alpha}^{(\ell)}), 
 \end{eqnarray}
 which belong to the Normal, Inverse--Gamma, Beta, and Dirichlet families, respectively. These distributions are assumed to be independent and are commonly used in the SBM literature, given their convenience as conditional conjugate \citep{yang2011detecting,lee2019review}. %It should be noticed that all parameters are scalar, but $\alpha^{(\ell)}$ and $\beta^{(\ell)}_{qr}$, the size of this latter depending on the number of covariates included. 
 
 Regarding the transition parameters of a Markov chain, the standard choice is a Dirichlet prior. Nevertheless, in our DSBMM, the multinomial representation of the HMC dependence calls for the use of an informative prior over the real line, such as the Normal distribution \citep{held2006bayesian, fruhwirth2010data,billio2016interconnections}. The Normal prior induces a ridge shrinkage, pushing the estimates of the transition probabilities away from zero and one, which is particularly useful for the entries of the transition matrix with few to no observations. However, this prior choice may still be unsatisfactory because it does not cope with the over--parameterization resulting from including all layers and order effects in (\ref{eq:dichmod}). Since the main interest of DSBMM is to select the layers that have a significant influence on transition entries and not to select individual variables, applying a global shrinkage or a global variable selection method would omit important information in considering the correlation between \textcolor{black}{lagged HMC variables}---i.e., variable grouping. We follow a Bayesian group--LASSO, through a Multi--Laplacian prior \citep{raman2009bayesian}.
 %, equivalent to the flexible penalty used by \citet{yuan2006model} in the context of Poisson models for multidimensional contingency tables. 
\begin{definition} 
Let $p$ be the number of covariates and $\mathcal{K}=\{\mathcal{U}_1,\ldots,\mathcal{U}_{m}\}$ a partition of $\{1\ldots,p\}$.
 The Multi--Laplacian distribution prior for $\kappa$ centered at $\mathbf{0}$ is defined as the product of
\[\text{M-Laplace}\left(\kappa_{\mathcal{U}}\mid \mathbf{0}, c(\mathcal{U})^{-1}\right) \propto c(\mathcal{U})^{s(\mathcal{U})/2} \exp \left(-c(\mathcal{U})\left\|\kappa_{\mathcal{U}}\right\|_2\right),\] 
over $\mathcal{U}\in \mathcal{K}$, where $c(\mathcal{U})=[s(\mathcal{U})\rho]^{1/2}>0$ is the precision parameter, $\kappa_{\mathcal{U}}=\left(\kappa_{\mathcal{U},1},\ldots,\kappa_{\mathcal{U},s(\mathcal{U})}\right)'$ and $s(\mathcal{U})$ the number of \textcolor{black}{lagged HMC variables} in $\mathcal{U}$. 
\end{definition}

The Multi--Laplacian distribution selects among groups of covariates and shrinks the parameters of highly correlated variables within each subset $\mathcal{U}$. The precision $c(\mathcal{U})$ governs the sparsity level in the parameters $\kappa_{\mathcal{U}}$.
If all lagged HMC variable groups are singletons, i.e., $s(\mathcal{U})=1$ for all $\mathcal{U} \in \mathcal{K}$, the group LASSO is equivalent to a standard Bayesian LASSO. As in the classical LASSO, the objective is to identify the most relevant variables, but at a group level.

We assume a Multi--Laplacian prior for the parameters of the higher--order effects and of the main effect of other layers. The Normal prior is assumed for the remaining parameters. As stated in the following, our prior distribution for $\kappa^ {(\ell)}$ can be represented as a continuous scale mixture of Normal distributions with a Gamma mixing distribution \citep{park2008bayesian}.

\begin{proposition} \label{prop:multiL}
  Assume a Multi--Laplacian prior for: i) the higher--order effects, that is $\kappa^{(\ell)}_{\mathcal{U},r}$ such that $\mathcal{U}\in \mathcal{P}$, $|\mathcal{U}|\neq 1$, $\ell \in \mathcal{L}$ and $r\in \mathcal{Q}\backslash\{Q\}$; and, ii) the main effects of other layers, that is $\kappa^{(\ell)}_{\mathfrak{m},r}$, such that $\ell\neq \mathfrak{m}$ and $r\in \mathcal{Q}\backslash\{Q\}$. Assume a Normal prior for the intercept and the own layer parameters in the main effects, $\kappa_{0,r}^{(\ell)}$ and $\kappa^{(\ell)}_{\{\ell\},r}$, respectively, for $\ell \in \mathcal{L}$ and $r\in \mathcal{Q}\backslash\{Q\}$. The prior distribution for $\kappa^{(\ell)}$ satisfies  
  	\begin{equation}
		\label{eq:grouplass}
		\begin{aligned}
			&\qquad \qquad  \kappa_{r}^{(\ell)}|\zeta^{(\ell)}_{r}, \zeta^{(\ell)}_{0}\sim \operatorname{N}\left(\underline{\kappa}_r^{(\ell)},\underline{K}_{r}^{(\ell)}\right)\\
			& \qquad \qquad \zeta_{r\mathcal{U}}^{(l)2}|\rho^{(\ell)}\sim \operatorname{G}\left(\frac{s(\mathcal{U})+1}{2},\frac{2}{\rho^{(l)} s(\mathcal{U})}\right),
		\end{aligned}
	\end{equation}
 where 			$\underline{K}_r^{(\ell)}=\operatorname{diag}(\zeta_{0}^{(l)2},\zeta_{r1}^{(l)2}\mathbf{1}_{s(1)}',\ldots,\zeta_{0}^{(l)2}\mathbf{1}_{s(\ell)}',\ldots,\zeta_{r\mathcal{U}}^{(l)2}\mathbf{1}_{s(\mathcal{U})}',\ldots,\zeta_{r1\ldots L}^{(l)2}\mathbf{1}_{s(1\ldots L)}' )$, $\zeta_{0}^{(l)2}$ is a hyperparameter of the Normal prior, $\zeta_{r}^{(\ell)}=[\zeta_{r1}^{(\ell)2},\ldots,\zeta_{rs(1)}^{(\ell)2},\ldots,\zeta_{r\mathcal{U}}^{(\ell)2},\ldots, \zeta_{r1\ldots L}^{(l)2}]$ corresponds to the data augmentation of the Multi--Laplacian prior, $ s(\mathcal{U})$ is the number of \textcolor{black}{lagged HMC variables} in the set of regressors $\mathcal{U}$ (see equation \ref{eq:dichmod}), and $\operatorname{G}(a,b)$ denotes the Gamma distribution with shape parameter $a$ and scale $b$.
\end{proposition}

The shrinking in (\ref{eq:grouplass}) is twofold: shrinking order and layer effects following a $l_1$--penalty to choose the most relevant groups and within--group shrinking following a $l_2$--penalty to deal with highly correlated variables and entries with few to no transitions. The diagonal elements of the prior variance $\underline{K}_r^{(\ell)}$ are groupwise sampled from a Gamma distribution conditionally on $\rho^{(l)}$. Comparing (\ref{eq:grouplass}) with the original proposal by \citet{raman2009bayesian}, two changes are introduced. First, contrary to a Poisson model for contingency tables, a multinomial logit implies $Q^{(\ell)}-1$ regressions, all of them sharing the same parameter $\rho^{(\ell)}$. In this way, variable selection applies even across equations. This cross--equation information improves the estimation of $\rho^{(\ell)}$ by jointly contrasting more groups of variables. Second, not only is the intercept of each of the $Q^{(\ell)}-1$ regression excluded from the Multi--Laplacian prior, but also the main effects of the own layer in order to capture the HMC persistence. As a consequence, the variable selection does not apply to the intercept and the own lagged membership parameters, $Z^{(\ell)}_{it-1}$.

The variable selection can be sensitive to the choice of $\rho^{(\ell)}$ because this latter has a direct relationship with the Lagrange parameter of the LASSO regression, $c^{(\ell)}$. Thus, a Gamma prior is used for $\rho^{(\ell)}$ with hyperparameters $\iota^{(\ell)}_{1}$ and $\iota^{(\ell)}_{2}$
\begin{equation}
\label{eq:rhoprior}
    \rho^{(\ell)}\sim \operatorname{G}\left(\iota^{(\ell)}_{1},\iota^{(\ell)}_{2}\right).
\end{equation}
 The Bayesian interpretation of penalizations as shrinkage priors allows for the estimation of the shrinkage parameter $c^{(\ell)}$ (through $\rho^{(\ell)}$) as a natural hierarchical extension and it provides valid standard errors and other measures of uncertainty \citep{casella2010penalized}. {\color{black} Since the focus of the paper is on lagged causality, we assume independent priors for the layer parameters. Nevertheless, cross--layer contemporaneous dependence can be incorporated by assuming partial parameter pooling or hierarchical priors.}
  \begin{figure}[t]
		\centering
		\caption{Directed Acyclic Graph of the DSBMM for layer $\ell$. Observed variables (gray circles), parameters and latent variables (white circles) and conditional dependence between them (arrows). }
		\label{fig:DAG}
        		\resizebox{9.5cm}{!}{%
			\begin{tikzpicture}
				%[transform canvas={scale=0.6}]
				\node[latent] (zi) {$Z^{(l)}_{i1}$}; %
				\node[latent,left=of zi,yshift=0.3cm] (alpha) {$\alpha^{(l)}$};%
				\node[latent,right= 0.2 of zi] (zj) {$Z^{(l)}_{j1}$}; %
				{\tikzset{plate caption/.append style={above=10pt of #1.north west}}
					\plate {zizj} {(zi) (zj)} {$i,j \in \mathcal{V}$};} %
				\node[latent,below=of zizj,xshift=-0.5cm] (lambda) {$\theta^{(l)}_{Z^{(l)}_{i1}Z^{(l)}_{j1}}$};%
				\node[latent,right=0.2 of lambda] (nu) {$\nu^{(l)}_{Z^{(l)}_{i1}Z^{(l)}_{j1}}$};%
				\node[obs,below=0.3 of lambda,xshift=0.6cm] (y1) {$Y^{(l)}_{ij1}$};%
				\node[obs,left=0.5 of y1] (X1) {$X^{(l)}_{ij1}$};%
				\plate {lamnu} {(lambda) (nu) (y1) (X1)} {$(i,j) \in \mathcal{E}^{(l)}_1$};%
				\edge{alpha}{zizj}
				\edge{zi}{lambda}
				\edge{zj}{lambda}
				\edge{zi}{nu}
				\edge{zj}{nu}
				\edge{lambda}{y1}
				\edge{nu}{y1}
				\edge{X1}{y1}
				
				\node[latent,right=3.4 of zj] (zi2) {$Z^{(l)}_{i2}$}; %
				\node[latent,right= 0.2 of zi2] (zj2) {$ Z^{(l)}_{j2}$}; %
				{\tikzset{plate caption/.append style={above=10pt of #1.north west}} %####1.north in case of beamer
					\plate {zizj2} {(zi2) (zj2)} {$i,j \in \mathcal{V}$};}%
				\node[latent,below=of zizj2,xshift=-0.5cm] (lambda2) {$ \theta^{(l)}_{Z^{(l)}_{i2}Z^{(l)}_{j2}}$};%
				\node[latent,right=0.2 of lambda2] (nu2) {$\nu^{(l)}_{Z^{(l)}_{i2}Z^{(l)}_{j2}}$};%
				\node[obs,below=0.3 of lambda2,xshift=0.6cm] (y12) {$Y^{(l)}_{ij2}$};%
				\node[obs,left=0.5 of y12] (X12) {$X^{(l)}_{ij2}$};%
				\plate {lamnu2} {(lambda2) (nu2) (y12) (X12)} {$(i,j) \in \mathcal{E}^{(l)}_{2}$};%
				\edge{zi2}{lambda2};
				\edge{zj2}{lambda2};
				\edge{zi2}{nu2};
				\edge{zj2}{nu2};
				\edge{lambda2}{y12};
				\edge{nu2}{y12};
				\edge{X12}{y12};
				
				\plate {all1} {(lamnu) (alpha) (zizj)}{};
				\plate {allt} {(lamnu2) (zizj2)}{};
				\node[const,right=1.25 of zj,yshift=1.7cm](p){$P_{i2}^{(l)}$};
				\plate {p2}{(p)}{$i\in \mathcal{V}$};
				%\node[const,right=0.9 of zj,yshift=0.33cm](dots){\LARGE \vspace{0.1cm} ... \vspace{0.1cm}};
				%\edge{zizj}{zizj2};
				\draw [->] (zizj.east)-| (2,1.5) --(p2.west);
				\draw [->] (p2.east)-| (3.65,0.3) --(zizj2.west);
				\node[const,right=1.3 of zj2,yshift=1.7cm](pm2){$P_{i3}^{(l)}$};
				\plate {pt}{(pm2)}{$i\in \mathcal{V}$};
				\draw [->] (zizj2.east)-| (7.3,1.5) --(pt.west);
				\node[const,right=2.6 of zj2,yshift=0.33cm](dots2){\LARGE \vspace{0.1cm} ... \vspace{0.1cm}};
				\draw [->] (pt.east)-| (9.1,1.5) |- (dots2.west);
				\node[const,right=2.5 of zj2,yshift=0.33cm](blank){};
				
				\node[latent,left=of p2,xshift=-0.6cm,yshift=1cm](zmi){$Z_{i1}^{(-l)}$};
				\plate {zm}{(zmi)}{$i\in \mathcal{V}$};
				\node[latent,right=of p2,xshift=1.2cm,yshift=2.3cm](kap){$\kappa^{(l)}$};
				\node[latent,left=of pm2,xshift=-0.6cm,yshift=0.75cm](zmi2){$Z_{i2}^{(-l)}$};
				\plate {zm2}{(zmi2)}{$i\in \mathcal{V}$};
				
				\node[const,right=3 of kap,](dots3){\LARGE \vspace{0.1cm} ... \vspace{0.1cm}};
				\draw [->] (kap)to(dots3);
				
				\node[latent,left=of kap,xshift=-0.6cm,yshift=1cm](kbar){$\underline{\kappa}^{(\ell)}$};
				\node[latent,right=of kap,xshift=0.6cm,yshift=1cm](zet0){$\zeta_{0}^{(l)2}$};
				\node[latent,above=of kap,yshift=0.8cm,xshift=0.2cm](zet){$\zeta_{q\mathcal{U}}^{(l)2}$};
				\plate {zetg}{(zet)}{$\scriptstyle \mathcal{U}\in 2^{\mathcal{L}} \backslash \{\ell\}$};
				\plate {zetg2}{(zetg)}{$\scriptstyle q \in \mathcal{Q}\backslash \{Q^{(\ell)}\}$}
				
				\node[latent,above=of zet,yshift=-0.3cm,xshift=-0.2cm](rho){$\rho^{(l)}$};
				\node[latent,left=of rho,xshift=-0.6cm](i1){$\iota^{(l)}_{1}$};
				\node[latent,right=of rho,xshift=0.6cm](i2){$\iota^{(l)}_{2}$};
				
				\draw [->] (zm.east)to[bend left](p2.west);
				\draw [->] (zm2.east)to[bend left](pt.west);
				\draw [->] (kap.west)|-(3.05,3.8)--(p2.north);
				\draw [->] (kap.east)|-(8.477,3.79) --(pt.north);
				
				\draw [->] (kbar.east)to(kap.north west);
				\draw [->] (zet0.west)to(kap.north east);	
				\draw [->] (zetg2)to(kap);
				\draw [->] (rho)to(zetg2);
				\draw [->] (i1)to(rho);
				\draw [->] (i2)to(rho);			
			\end{tikzpicture}
		}	
    \end{figure}

Figure \ref{fig:DAG} summarizes the DAG of the DSBMM (only for a layer $\ell$ for the sake of simplicity). This representation underlines the influence of changes in community membership on the topological properties observed in the network, which in turn is affected by the community membership in the rest of the layers $Z^{(-\ell)}_{it}$. This illustration excludes the hyperparameter of the prior distribution of the connectivity parameters and the initial partition of the nodes $\alpha^{(\ell)}$ to focus on the transition parameters. The priors for these latter, that are the Normal and Multi--Laplacian distributions, are presented as a heteroschedastic Normal prior. 
	
	\subsection{Posterior approximation} \label{sec:postaprox}
Consistently with our application to trade flows, we assume $f^{(\ell)}(y| \theta^{(\ell)}_{qr},X^{(\ell)}_{ijt})$ is a log--normal distribution $\operatorname{LN}(\beta^{(\ell)}_{qr},\sigma^{(\ell)2}_{qr})$. Nevertheless, our modeling framework is general and can be easily modified to account for other distribution assumptions, such as a zero--truncated Poisson for count data \citep[e.g., see][]{matias2017statistical}. Since the likelihood function is not tractable, a Gibbs sampling procedure is derived to approximate the posterior distribution. 

Using the prior distributions presented in Section \ref{sec:prior}, the full conditional posterior distributions of the connectivity parameters and initial proportions are given in the following propositions. A full Gibbs sampling algorithm using the conditional distributions is then used to approximate the posterior distribution.

\begin{proposition}\label{prop:conne}
Under the prior distribution assumptions in \eqref{eq:priorcon}, the following full conditional distributions can be derived:
\begin{enumerate}[label=\textit{\arabic*)}]
 \setcounter{enumi}{0}
	\item $\beta^{(\ell)}_{qr}|\sigma_{qr}^{(\ell)2},\theta^{(-\ell)},\theta^{(\ell)}_{-qr},\nu, P, \alpha,Y,Z,D,X\sim\operatorname{N}\left(\overline{\beta}^{(\ell)}_{qr},\overline{\Sigma}^{(\ell)}_{qr}\right)$
	\item $\sigma_{qr}^{(\ell)2}|\beta^{(\ell)}_{qr},\theta^{(-\ell)},\theta^{(\ell)}_{-qr},\nu, P, \alpha, Y,Z,D,X\sim\operatorname{IG}\left(\overline{d}^{(\ell)}_{qr}/2,\overline{e}^{(\ell)}_{qr}/2\right)$
	\item $\nu^{(\ell)}_{qr}|\nu^{(-\ell)},\nu^{(\ell)}_{-qr},\theta,P, \alpha, Y,Z, D\sim \operatorname{Beta}\left(\overline{b}^{(\ell)}_{qr},\overline{c}^{(\ell)}_{qr}\right)$ 
	\item $\alpha^{(\ell)}\left|\alpha^{(-\ell)},\nu,\vartheta, P, Y,Z,D\right.\ \sim \operatorname{Dir}\left(\overline{\alpha}^{(\ell)}\right)$.
\end{enumerate}
\end{proposition} 
 
The multinomial representation of DSBMM does not allow for a conditional conjugate prior, and the use of the standard Metropolis--Hastings (MH) algorithm can be highly inefficient. As suggested by \citet{fruhwirth2010data}, data augmentation can be applied to avoid MHs steps. Therefore, in the following propositions we extend the Pólya gamma representation proposed by \citet{holsclaw2017bayesian} and \citet{polson2013bayesian} to the saturated multinomial logit and obtain tractable full conditional distributions for the transition parameters $\kappa_{r}^{(\ell)}$, the auxiliary variables of the two data augmentations, $\omega^{(\ell)}_{it,r}$ and $\zeta_{r\mathcal{U}}^{(l)2}$, and the hyperparameter related to the level of shrinkage $\rho^{(l)}$.

\begin{proposition}\label{prop:dataugLogit}
    Let $\eta^{(\ell)}_{it,r}=\widetilde{W}_{it-1}\kappa_r^{(\ell)}-R^{(\ell)}_{it,r}$, $R^{(\ell)}_{it,r}=\log(\sum_{k\neq r}^{Q^{(\ell)}}\exp(\widetilde{W}_{it-1}^{(\ell)}\kappa_k^{(\ell)}))$ and $\xi^{(\ell)}_{it,r}=W^{(\ell)}_{it,r}-1/2$. Under the Normal and Multi--Laplacian prior assumption $\pi(\kappa^{(\ell)}_{r})$, and from multinomial logistic probability assumption in \eqref{eq:multn1}, the full conditional distribution of $\kappa_{r}^{(\ell)}$ can be written as
\begin{equation}
		\begin{aligned}
			h(\kappa_r^{(\ell)}|Z,\omega^{(\ell)}_{1:N 1:T,r})\propto \pi\left(\kappa_r^{(\ell)}\right)\exp\left(-\frac{1}{2}\left(\tilde{\xi}^{(\ell)}_{r}-\eta^{(\ell)}_{r}\right)'\Omega_r^{(\ell)}\left(\tilde{\xi}^{(\ell)}_{r}-\eta^{(\ell)}_{r}\right)\right),
		\end{aligned}
	\end{equation}	
	where $\tilde{\xi}^{(\ell)}_{r}=(\xi^{(\ell)}_{12,r}/\omega^{(\ell)}_{12,r},\ldots,\xi^{(\ell)}_{NT,r}/\omega^{(\ell)}_{NT,r})'$, $\xi_r^{(\ell)}=(\xi_{11,r}^{(\ell)},\dots,\xi_{N(T-1),r}^{(\ell)})'$, $\eta^{(\ell)}_{r}=(\eta^{(\ell)}_{12,r},\ldots,\eta^{(\ell)}_{NT,r})'$, $\Omega_r^{(\ell)}=\operatorname{diag}(\omega^{(\ell)}_{12,r},\dots,\omega^{(\ell)}_{NT,r})$ and $\omega^{(\ell)}_{it,r}$ are auxiliary variables following a Pólya Gamma distribution, i.e. $\omega^{(\ell)}_{it,r}\sim \operatorname{PG}(1,0)$. 
\end{proposition}

\begin{proposition}\label{prop:PolyaLap}
    Let $\operatorname{GIG}(a,b,c)$, $a\in \mathbb{R}$ and $b,c>0$, be a Generalized Inverse--Gamma and $\operatorname{PG}(b,c)$, $b>0$ and $c\in \mathbb{R}$, a Pólya gamma distribution. From Proposition \ref{prop:dataugLogit} and the prior assumption \eqref{eq:rhoprior}, the following full conditional distributions can be derived:
\begin{enumerate}[label=\textit{\arabic*)}]
 \setcounter{enumi}{4}	\item $\kappa_r^{(\ell)}\left|Z,\omega^{(\ell)}_{1:N 1:T,r}\right.\sim\operatorname{N}\left(\overline{\kappa}_r^{(\ell)},\overline{K}_r^{(\ell)}\right)$
 	\item 		$\omega^{(\ell)}_{it,r}\left|Z,\kappa_r^{(\ell)}\right.\sim\operatorname{PG}\left(1,\eta^{(\ell)}_{it,r}\right)$
	\item $\zeta_{r\mathcal{U}}^{(l)2}\left|\kappa_r^{(\ell)},\rho^{(\ell)} \right.\sim\operatorname{GIG}\left(1/2,s(\mathcal{U})\rho^{(\ell)},||\kappa^{(\ell)}_{\mathcal{U},r} -\underline{\kappa}^{(\ell)}_{\mathcal{U},r}||^2_2\right)$
	\item $\rho^{(l)}\left|\zeta_{r\mathcal{U}}^{(l)2}\right.\sim \operatorname{G}(\overline{\iota}_1,\overline{\iota}_2)$

	%\item $\mathbb{P}\left(Z^{(\ell)}_{i,1:T}|Z^{(-\ell)}_{i,1:T},Z^{(\ell)}_{-i,1:T},\psi^{(\ell)}_{T}\right), 	\mathbb{P}\left(Z^{(\ell)}_{it}=q|Z^{^{(-\ell)}}_{i,1:t-1},Z^{^{(\ell)}}_{-i,1:t-1},\psi^{(\ell)}_{t-1}\right), \mathbb{P}\left(Z^{(\ell)}_{it}=q|Z^{^{(-\ell)}}_{i,1:t},Z^{^{(\ell)}}_{-i,1:t},\psi^{(\ell)}_{t}\right)$.
\end{enumerate}
    
\end{proposition}

Regarding the HMC, we compute iteratively the conditional prediction probability $\mathbb{P}(Z^{(\ell)}_{it}=r|Z^{^{(-\ell)}}_{i,1:t}, Z^{^{(\ell)}}_{-i,1:t-1},\psi^{(\ell)}_{t-1})$, the conditional filtered probability  $\mathbb{P}(Z^{(\ell)}_{it}=r|Z^{^{(-\ell)}}_{i,1:t+1},Z^{^{(\ell)}}_{-i,1:t},\psi^{(\ell)}_{t})$ and sample  $Z^{(\ell)}_{i,t}$ from the smoothed probability $\mathbb{P}(Z^{(\ell)}_{i,1:T}|Z^{(-\ell)}_{i,1:T},$ $Z^{(\ell)}_{-i,1:T},\psi^{(\ell)}_{T})$, where $\psi^{(\ell)}_t=[Y^{(\ell)}_{1:t},D^{(\ell)}_{1:t},X^{(\ell)}_{1:t}]$ \citep[see][]{fruhwirth2006finite,touloupou2020scalable}. See Section \ref{sec:ffbs} for further details. The last step in the Gibbs sampler generates samples for the allocation variables $Z^{(\ell)}_{{\color{black}  i \sdot}}$ by using a conditional forward filtering and backward sampling (FFBS). \textcolor{black}{For applications where the node set is time--varying, the membership dynamics can be reconstructed by assuming a node--centered sampling design and by following the missing handling strategy based on FFBS proposed in \citet{tabouy2020variational,hamaker2012regime}.} 

In mixture models and HMC models, state labels are not identifiable, and label changes may occur between Gibbs iterations, which is referred to as label switching. The DSBMs belong to the class HMC models, and the MCMC approximation can be affected by this issue  \citep{matias2017statistical}. Prior constraints that fix a specific cluster order can be used to achieve identification. We follow an alternative approach based on post--processing MCMC samples \citep{fruhwirth2006finite}.

	\subsection{Model selection and causality testing}\label{sec:modselcaus}

Regarding variable selection, a procedure based on the posterior estimates of the group LASSO is not straightforward because it does not lead to exact zeroes for the posterior modes or means. The alternative used in this paper is a credible interval criterion, that is, a parameter whose credible interval includes zero is considered to be null \citep[e.g.][]{van2019shrinkage}.

%Notice that sparsity is not always a desirable result (e.g., model averaging), but, in the DSBMM, our interest is in choosing a specific model to answer the question of whether a layer \textcolor{black}{NGB} causes another one, which means a criterion is needed to induce sparsity \citep[see for example][for a discussion on model averaging]{steel2020model}.

From a Bayesian perspective, a credible interval criterion has three main drawbacks: the results depend on the chosen credible level, it does not express the different sources of uncertainty, and does not account for the joint posterior distribution of the parameters. However, in a Bayesian LASSO framework, calculating Bayes factors or posterior variable inclusion probabilities is computationally intensive for a saturated multinomial logit \citep{hans2010model}. Alternatively, spike-and-slab prior distributions can be used to explore the model set \citep{george1997approaches}. Nevertheless, this space is very large, and the search procedure is either computationally intensive or inefficient. Additionally, a threshold is required to induce sparsity, and the results are typically sensitive to the choice of this threshold. This selective inference issue can be even more severe for highly correlated \textcolor{black}{lagged HMC variables} since it does not consider the joint distribution of the parameters \citep{bondell2012consistent}. Therefore, considering its computational convenience and given that its drawbacks are also challenging for alternative methods, the credible interval criterion appears to be a suitable option for large models, such as the DSBMM.

{\color{black}Regarding NGB causality testing, an advantage of our approach is that NGB causality can be tested through a few parameter restrictions and a theoretically valid procedure based on the Bayes factor (BF). In the multiple-layer case, blockwise testing is still feasible and can be applied to pairs of layers. Let us denote with
	$\kappa^{(\ell)}_{\mathcal{U}}=(\kappa^{(\ell)\prime}_{\mathcal{U},1},\ldots, \kappa^{(\ell)\prime}_{\mathcal{U}, Q^{(\ell)}-1})'$ and $\kappa^{(\ell)}_{\mathcal{U},r}=(\kappa^{(\ell)}_{\mathcal{U},r,1},\ldots,\kappa^{(\ell)}_{\mathcal{U},r,s(\mathcal{U})})'$  the collections of transition parameters in the membership probabilities of the nodes in layer $\ell$. Following Definition 1 and the Markov--chain conditionally independence assumption in \eqref{eq:membership}, the layer $\mathfrak{m}$ does NGB cause layer $\ell$, i.e. $Z^{(\mathfrak{m})} \stackrel{G}{\Rightarrow} Z^{(\ell)}$, if the null hypothesis $H_0:\kappa^{(\ell)}_{\mathcal{U}}={\bf 0}, \  \forall\ \mathcal{U}\in \mathcal{P}^{(\mathfrak{m})}$ is not satisfied, where $\mathcal{P}^{(\mathfrak{m})}$ denotes the parameters subscript indexes associated to the main effects and interactions of layer $\mathfrak{m}$ \textcolor{black}{lagged HMC variables}, i.e.   $\mathcal{P}^{(\mathfrak{m})}=\{ \mathcal{U}\in \mathcal{P}| \mathfrak{m}\in \mathcal{U}\}$. For example, in specifications with three blocks in layer $\ell$ and two blocks in layer $\mathfrak{m}$, layer $\mathfrak{m}$ does not NGB causes layer $\ell$ if the null $H_0:\kappa^{(\ell)}_{\{\mathfrak{m}\},1,1}=\kappa^{(\ell)}_{\{\mathfrak{m}\},2,1}=0\, (\text{main effects}),\ \kappa^{(\ell)}_{\{\ell,\mathfrak{m}\},1,1}=\kappa^{(\ell)}_{\{\ell,\mathfrak{m}\},1,2}=\kappa^{(\ell)}_{\{\ell,\mathfrak{m}\},2,1}=\kappa^{(\ell)}_{\{\ell,\mathfrak{m}\},2,2}=0\ (\text{interaction effects})$. The Bayes factor can be used for Bayesian testing, which involves the ratio of the posterior probability of the hypothesis to its prior probability. By assuming the same prior probabilities for the two hypotheses, the Bayes factor can be expressed as a ratio of the marginal likelihoods under the null and the alternative, i.e.  
	$$
	BF_{1,0}(Y)=\frac{m(Y|H_1)}{m(Y|H_0)},
	$$
	where $m(Y|H_i)=\int_{\Theta \times K} \sum_{\mathcal{Z}}h_{i}d\vartheta d\kappa$  for $i=0,1$ and $h_{i}=L(Y,Z|X,\vartheta,\kappa) \pi_{i}(\vartheta,\kappa)$. Notice that $\pi_{0}(\vartheta,\kappa)$ corresponds to the prior under the null hypothesis, where the prior of the restricted parameters $\kappa^{(\ell)}_{\mathcal{U}}={\bf 0}, \  \forall\ \mathcal{U}\in \mathcal{P}^{(\mathfrak{m})}$ reduces to a Dirac distribution at zero, while in the $\pi_{1}(\vartheta,\kappa)$  (alternative hypothesis) corresponds to the Multi--Laplacian prior.
	
	The marginal posterior can be approximated using MCMC draws (see \citet{llorente2023marginal} for a more extensive review). As in \citet{carallo2024generalized}, the method proposed by \citet{geyer1991estimating} can be used to approximate log--Bayes factor by expressing the posterior draws of the restricted and unrestricted models as a mixture, which can be written as a reverse logistic regression (RLR) in Proposition \ref{prop:reverselog}.
 \setcounter{proposition}{5}
    \begin{proposition}\label{prop:reverselog}
		Let $d=\log m(Y|H_1)-\log m(Y|H_0)$ be the log--Bayes factor, $h_{ij}=L(Y,Z(j),D|X,\vartheta(j),$ $\kappa(j)) \pi_{i}(\vartheta(j),\kappa(j))$ the full posteriors under $H_i$ $i=0,1$, and $\vartheta(j)$, $\kappa(j)$ and $Z(j)$ the $j$--th MCMC draw of the connectivity parameters, transition parameters and memberships, respectively, $j=1,\ldots,n$. Then, the RLR conditional likelihood is given by 
		$$
		L_Q(d)=\sum_{i\in \{0,1\}}\sum_{j=1}^{n}\log p_{i}(h_{ij},m(Y|H_i)),
		$$
		where $p_{i}(h_{ij},m(Y|H_i))$ is posterior probability of hypothesis $H_i$ given by 
		$$
		\begin{aligned}
			p_{i}(h_{ij},m(Y|H_i))%&=\frac{h_{ij}\exp(-\log m(Y|H_i)+\log(1/2))}{h_{0j}\exp(-\log m(Y|H_0)+\log(1/2))+h_{1j}\exp(-\log m(Y|H_1)+\log(1/2))}\\
			&=\frac{\exp(\mathbb{I}_{\{0\}}(i)(d+\log(h_{0j})-\log(h_{1j})))}{1+\exp(d+\log(h_{0j})-\log(h_{1j}))}.
		\end{aligned}
		$$
	\end{proposition}
	In other words, it is equivalent to maximizing the quasi--likelihood of the mixture with respect to $d$, which can be implemented using a standard logit regression with an intercept and a regressor that is the same for $i=\{0,1\}$ and whose slope parameter is restricted to be one.
}

\subsection{Simulation results}
\label{subsec:sim}
     % Summary paragraph of the simulation: ...
{\color{black} We study the effectiveness of the model under different DSBMM parameter prior assumptions (Normal and group LASSO priors), different non--linear Granger causality settings (no causality and unidirectional or bidirectional causality), and causality signs (coupling and decoupling layers). We also considered different network topologies (core--periphery and assortative structures) and sample size scenarios (a reference scenario, $N=50$ and $T=15$, a large $N$ scenario, $N=100$ and $T=15$, and a large $T$ scenario, $N=50$ and  $T=30$). In all causality settings and sample scenarios, the DSBMM combined with a Bayesian group LASSO prior (Multi--Laplacian prior) outperforms the Normal prior in correctly retrieving the NGB causality structure (see Table \ref{tab:propsynt} in Section \ref{sec:app_tab} of the Supplementary Material). We also provide some guidelines to choose the hyper--parameters of the Multi--Laplacian prior.}

{\color{black}  We also present a comparison between the DSBMM and alternative models from time--series analysis \citep[panel BVAR, see for example][]{canova2013panel,koop2013forecasting} and dynamic network literature: the Bayesian Learning for Dynamic network (BLDMN) introduced by \citet{durante2017bayesian} and the layer--independent DSBM introduced by \cite{matias2017statistical}. The results in Table \ref{tab:causBVAR} show that BVAR models have difficulties in detecting NGB causality. Additionally, the DSBMM outperformed other network models, especially out--of--sample (see Table \ref{tab:for_comp}). This outcome is due to the DSBMM's ability to incorporate information across layers, while capturing the topological change between an assortative and a core--periphery structure. More importantly, the DSBMM provides three modeling advantages: a conditional directed relationship by layer--pair (NGB causality), a general framework that jointly considers (un)directed and (un)weighted layers, and an intuitive interpretation in terms of clustering and classification of the nodes.}

  	\section{FTAs and trade networks}
	\label{sec:tradeFTA}
Our DSBMM model and inference provide a unified and coherent framework for international trade analysis, merging Gravity models and community detection algorithms. Since the General Agreement of Tariffs and Trade, FTAs have been an important tool conceived as a means to increase trade flows among countries. We use the DSBMM to infer the {\color{black}NGB} causal relationship between FTAs and trade flows.
 
	\subsection{Gravity model, communities and the DSBMM}\label{sec:gravDSBMM}
	Empirical studies have used the Gravity equation to explain bilateral trade and infer the effect of trade policies. This equation suggests that the economy's size (e.g., GDP, population) and the frictions (e.g., physical distance, language similarity, colonial links) of the countries involved determine trade flows. Although this gravity framework started as an intuitive empirical regularity, a theoretical foundation is possible using a general equilibrium model in a static framework à la Heckscher--Ohlin with monopolistic competition \citep[see for example][]{anderson2011gravity, kabir2017gravity}. These models extend the interpretation of trade flows beyond dyad relationships and focus on the structure of the trade network. In its reduced form, this family of models can be written as
\begin{equation}
	\label{eq:grav}
	\begin{cases}
		\log\left(Y^{(1)}_{ijt}\right)=\beta^{(1)}_0+\beta^{(1)}_1{\color{black}\log}\  GDP_{it}+\beta^{(1)}_2 {\color{black} \log} \ GDP_{jt} + (1-\varrho){\color{black}\log}\ t_{ijt}\\-{\color{black}(1-\varrho) \log}\ \Upsilon_{jt}-  {\color{black}(1-\varrho)\log}\ \Pi_{it}+ \varepsilon_{ijt}\\
		\left(\Pi_{it}\right)^{1-\varrho}=\sum_{j=1}^N\left(\frac{t_{i jt}}{\Upsilon_{jt}}\right)^{1-\varrho} \frac{GDP_{jt}}{\sum_{k=1}^N GDP_{kt}}\\
		\left(\Upsilon_{jt}\right)^{1-\varrho}=\sum_{i=1}^N\left(\frac{t_{i jt}}{\Pi_{it}}\right)^{1-\varrho} \frac{GDP_{it}}{\sum_{k=1}^N GDP_{kt}},\  \varrho>1,
	\end{cases}
\end{equation}
where the trade flow $Y^{(1)}_{ijt}$ increases with the node characteristics $GDP_{it}$ and $GDP_{jt}$, i.e. $\beta^{(1)}_{1},\beta^{(1)}_{2}>0$, and the frictions $t_{ijt}$ have a negative effect on trade. The two fully unobserved elements $\Upsilon_{jt}$ and $\Pi_{it}$ positively affecting the trade flow between $i$ and $j$ at time $t$, called inward and outward multi--resistance respectively, encompass an aggregated measure of the barriers between all the pairs of nodes in the trade network. {\color{black}The barriers are partially observed and are usually approximated by a set of observable variables ($T_{ijt}$), such as export subsidies and tariffs between countries or by unobserved effects ($\tilde{t}_{ijt}$), i.e., $t_{ijt}=\exp(T'_{ijt}\lambda+\tilde{t}_{ijt})$. To evaluate specific economic policies the main parameter of interest may include elements of the vector $(1-\varrho)\lambda$ and the reconstruction of the barrier network $(t_{ijt})^{1-\varrho}$, which allows to estimate general equilibrium impact of trade policies after solving for the multi--resistance terms in (\refeq{eq:grav}) \citep[for a more extensive discussion on trade policy evaluation and the structural Gravity model see for example][]{yotov2016advanced}. The parameter $\varrho$ can be interpreted as an elasticity of substitution and is usually estimated if the tariff data is available or calibrated based on previous studies \citep[e.g.,][]{anderson2020transitional}. While the DSBMM can be used to estimate the structural parameter of interest, our main objective is to address parameter heterogeneity and its relationship with FTAs through the multidimensional dependence and the NGB causality.}
	
In the DSBMM, the parameter heterogeneity can be present across the network and time. Previous works have underlined the importance of both dimensions. For instance, \citet{yotov2012simple} demonstrates the decreasing influence of geographical distance on trade over time due to globalization, but assumes the marginal effect of distance remains constant for all countries. On the other hand, based on firms' behavior, technology, market demand, and economic policies, the literature on trade has suggested specific network structures and potential dyad clustering. In particular, under some circumstances, the multi--resistance terms may not include information on all the network, but only on the barriers of the neighboring nodes of country $i$ and $j$, as in the Gravity model proposed by \citet{magerman2020pecking}\textcolor{black}{, where the neighboring countries of country $j$ importing from country $i$ refers to the set of active exporters $\{k\in \mathcal{V}:Y_{kj\textcolor{black}{t}}>0\}$}. This latter model predicts the existence of two communities in the trade network: a core and a periphery, resulting from firms' behavior, which determines their export destinations based on a pecking order. Other authors suggest that firms choose new destinations that are similar to previous ones, given the high cost of adapting products to heterogeneous markets, which may implicitly produce blocks of countries. Based on these considerations, a block structure in the network of barriers $t_{ijt}$ can be directly creating communities on $Y^{(1)}_{ij\textcolor{black}{t}}$, and/or indirectly through $\Upsilon_{j\textcolor{black}{t}}$ and $\Pi_{i\textcolor{black}{t}}$, while heterogeneous policy effects across the clusters may be also a possibility. For instance, this motivates the empirical work of \citet{sopranzetti2018overlapping} that studies the overlap between FTAs and trade networks from a dyad perspective.

Similarly, the empirical literature on community detection has provided evidence of clustering in trade and FTA networks, but it has mainly employed algorithmic methods, such as maximizing a modularity measure, to identify groups of countries with dense subgraphs, as in \citet{barigozzi2011identifying,bartesaghi2020communicability}. In this sense, DSBMM provides a statistical model, and subsequently, its inference includes a measure of uncertainty of the parameters and the latent membership. More importantly, it allows for dynamic membership of nodes and covariates that control for node characteristics, which are essential to link the community structure with the Gravity equation and parameter instability.  

 Specifically, we suggest that detecting communities in the trade flows after controlling for the observed covariates in (\ref{eq:grav}) concentrates the attention on the block structure in the unobserved components, that is, the multi--resistance terms and the unobserved barriers to trade. In order to focus on the multidimensional dependence, we assume only geographical distance is available to approximate the observed barriers ($dist_{ij}$), i.e.  $t_{ij\textcolor{black}{t}}=\exp(\lambda_{Z_{it}Z_{jt}} \textcolor{black}{\log} \ dist_{ij}+\tilde{t}_{Z_{it}Z_{jt}})$. The effect of distance ($\lambda_{Z_{it}Z_{jt}}$) and the unobserved barriers are block--dependent ($\tilde{t}_{Z_{it}Z_{jt}}$) and can vary across dyad and time, resulting in a partial pooling of the linear model, that is the first layer of a multidimensional network with $L=2$ is given by
 \begin{equation}
Y_{ijt}^{(1)}\left|X_{ijt},Z^{(1)}_{it}=q, Z^{(1)}_{jt}=r, \vartheta^{(1)}\right.\sim (1-\nu^{(1)}_{qr})\delta(y)+\nu^{(1)}_{qr} \operatorname{LN}\left(X'_{ijt}\beta_{qr}^{(1)},\sigma^2_{qr}\right),\ q,r\in \mathcal{Q}^{(1)}, \label{eq:lay1}
 \end{equation} 
where \textcolor{black}{the covariates are} $X_{ijt}=(1,\textcolor{black}{\log}\   GDP_{it},\textcolor{black}{\log} \ GDP_{jt},\textcolor{black}{\log} \ dist_{ij})'$, $\beta_{qr}^{(1)}=(\check{\beta}^{(1)}_{0qr},\beta^{(1)}_{1qr},\beta^{(1)}_{2qr},\beta^{(1)}_{3qr})'$, $\beta^{(1)}_{3qr}=(1-\varrho)\lambda_{qr}$, $\check{\beta}^{(1)}_{0qr}=\beta^{(1)}_{0qr}+(1-\varrho) \tilde{t}_{qr}-(1-\varrho)\textcolor{black}{\log}\ \Upsilon_r-(1-\varrho)\textcolor{black}{\log}\ \Pi_q$ and $q,r \in \mathcal{Q}^{(1)}$. Notice that zero--trade cases are allowed with probability $\nu^{(1)}_{qr}$ depending on the community structure. 

The second layer corresponds to the unweighted and directed FTA network, and it has its own community structure
\begin{eqnarray}
	&Y_{ijt}^{(2)}\left|Z^{(2)}_{it}=q, Z^{(2)}_{jt}=r, \vartheta^{(2)}\right.\sim \operatorname{Bern}\left( \nu^{(2)}_{qr}\right), q,r\in \mathcal{Q}^{(2)}. \label{eq:lay2}
\end{eqnarray}
	Since FTAs are part of the unobserved frictions $\tilde{t}_{qr}$, this specification (\ref{eq:lay1}) and (\ref{eq:lay2}) can be used to test if the community structure in the FTA network is {\color{black}NGB} causing the block membership in the trade network. For example, if $Q^{(1)}=Q^{(2)}=2$, (\ref{eq:dichmod}) becomes 
\begin{equation}
	\label{eq:dichmodap}
	\begin{aligned}
		\log\left(P^{(1)}_{it,Z^{(\ell)}_{it-1} 1} C^{(1)}_{it}\right)&=\kappa^{(1)}_{0,1}+\underbrace{\kappa^{(1)}_{\{1\},1,1}W^{(1)}_{it-1,1}+\kappa^{(1)}_{\{2\},1,1}W^{(2)}_{it-1,1}}_{\text{main effects}}+\underbrace{\kappa^{(1)}_{\{1,2\},1,1}W^{(1)}_{it-1,1}\cdot W^{(2)}_{it-1,1}}_{\text{first order effects}}\\
		\log\left(P^{(2)}_{it,Z^{(\ell)}_{it-1} 1} C^{(2)}_{it}\right)&=\kappa^{(2)}_{0,1}+\underbrace{\kappa^{(2)}_{\{1\},1,1}W^{(1)}_{it-1,1}+\kappa^{(2)}_{\{2\},1,1}W^{(2)}_{it-1,1}}_{\text{main effects}}+\underbrace{\kappa^{(2)}_{\{1,2\},1,1}W^{(1)}_{it-1,1}\cdot W^{(2)}_{it-1,1}}_{\text{first order effects}},
	\end{aligned}
\end{equation}
and the \textcolor{black}{NGB causality} tests can be stated as: FTAs does not \textcolor{black}{NGB} cause unobserved barriers and multi--resistance terms network, $H_0:\kappa^{(1)}_{\{2\},1,1}=\kappa^{(1)}_{\{1,2\},1,1}=0$; or, trade network does not \textcolor{black}{NGB} cause FTAs network, $H_0:\kappa^{(2)}_{\{1\},1,1}=\kappa^{(2)}_{\{1,2\},1,1}=0$. A third scenario would imply a bidirectional causality, $H_0:\kappa^{(1)}_{\{2\},1,1}=\kappa^{(1)}_{\{1,2\},1,1}=\kappa^{(2)}_{\{1\},1,1}=\kappa^{(2)}_{\{1,2\},1,1}=0$. 

 The FTA network is a trade policy instrument and may help predict the trade flows layer. However, this is not the only possible scenario because FTAs are not fully exogenous. As posed by \citet{baier2004economic}'s theoretical model, the welfare expectations of consumers and firms may depend on the pre--policy levels of trade flows and affect the likelihood of an FTA. Moreover, integration processes, domestic regulations, and production fragmentation are inducing changes in both networks and may \textcolor{black}{NGB} cause clustering patterns \citep[e.g.][]{baier2007free,baltagi2008estimating,ivlevs2010fdi,behrens2012dual}. Thus, it is not clear if FTAs and trade flows are unidirectional or bidirectional related in the \textcolor{black}{NGB} sense, and the empirical results obtained from panel analysis and impact evaluation methods on FTAs effectiveness at a dyad level show mixed evidence \citep[for a review see][]{baier2004economic}.

{\color{black}The DSBMM accommodates the existing challenges in estimating structural Gravity models. The latent block structure allows for heteroschedasticity ($\sigma_{qr}^{2}$) based on the pair block interacting and potential dependence between the probability of exporting and the volume of trade. Censoring in economic data has resulted in different approaches, such as sample selection, hurdle, and two--part models \citep[see for example the taxonomy proposed by][]{jones2000health}. In particular, in international trade, \citet{helpman2008estimating} assumes the zeros represent missing data, suggesting a Heckman procedure in which the selection and trade flows follow different mechanisms with unobserved correlated components. In the case of \citet{silva2006log}, the zeros are ``genuinely'' a no--trade situation between pairs of countries, and a single mechanism is assumed, that is, the covariates and the effect are the same for the level of trade and the cases of absence of trade. In this setting, the application of DSBMM to trade is equivalent to \citet{silva2006log} in assuming that zeros imply no trade. Still, it differs by considering a two--part specification with separate mechanisms for the excess of zeros and positive trade flows similar to the one in \citet{metulini2018spatial,egger2011trade}. More importantly, in the DSBMM both mechanisms are not independent, they share a set of common unobserved factors captured in the latent components that induce clustering in the network, i.e. membership variables $Z_{it}$ and $Z_{jt}$, that can capture positive or negative (nonlinear) relationship between the probability of exporting, the trade flow and the variance of the trade flows. Indeed, contrary to standard two--part models, the two mechanisms in the DSBMM cannot be independently estimated.

The simplified specification of the Gravity in \eqref{eq:lay1} can also be easily extended to relax some assumptions and focus on other objectives in the trade literature such as estimating the parameter of interest for trade policy, i.e. $(1-\varrho)\lambda$ and $(t_{ijt})^{1-\varrho}$. For instance, more covariates can be included to approximate trade barriers, explanatory variables can be incorporated into the probability of trade, or alternative distributions for trade flows can be used. The assumption of constant unobserved components within a pair of interacting blocks can be relaxed by introducing node--time and pair fixed effects or random effects. This latter option constitutes a mixed--effects alternative approach to deal with the multi--resistance terms, which can be interpreted as clustering the fixed effects, while allowing for the inclusion of time--invariant and node--time invariant covariates such as policy variables. To exemplify the flexibility of the DSBMM, some of these extensions are considered in Section \ref{sec:extenDSBMM} in Supplementary Material. Therefore, the application of DSBMM to the trade and FTA network provides a unified framework for two streams of literature: Gravity models and community detection.}

	\subsection{Data description}
	The dataset used in this application includes 159 countries detailed in Table \ref{tab:list}, for the period 1995--2017. The trade network corresponds to the value of all goods imported (CIF) between pairs of countries collected by the International Monetary Fund (IMF) in U.S. dollars, which are transformed into constant terms by using the GDP PPP deflator from the same source with base year 2010, as in \citet{helpman2008estimating}. The information reported to the IMF is not always complete due to missing reporting countries (nodes) or specific flows (edges) at a period $t$. We rely on the imputation techniques suggested by IMF \citep{dippelsman2018new}.
	
	Regarding the FTAs, the network is constructed using the database on Economic Integration Agreements maintained by the NSF-Kellogg Institute. The information provided on the type of FTA is dichotomized into the existence or non--existence of any trade agreement, from a non-reciprocal preferential trade arrangement (NR--PTA) to an Economic Union (EUN). The former type of agreement makes the adjacency matrix of the FTAs non--symmetric. This data is collected from various sources, including the CIA World Factbook and the WTO database, among others. If there is no information on any of the sources for a pair of countries, the database includes FTAs based on trade flows. For instance, if two countries have never traded and there is no international agreement between them in the databases reviewed, it is assumed that they have not signed an FTA. Although these imputations covered 40\% of the pairs between 1950--2012, the assumptions are reasonable and commonly accepted in the literature on FTA effectiveness \citep[e.g.][]{baier2019widely}. Finally, the covariates used for the first layer are from the dynamic gravity dataset of the U.S. International Trade Commission (USITC), which collects GDPs in real terms from the World Development Indicators (WDI), and the distance ($dist_{ij}$) refers to the population--weighted distance between each pair of countries  \citep{gurevich2018dynamic}.
	
	A general description of the resulting multi-layer FTAs--Trade Flows is presented in Table \ref{tab:netstat}, which confirms some of the empirical stylized facts in international trade. Since the beginning of the period, both networks have been dense, and without considering the direction of the edges, there is only one component (WCC), indicating that a path always exists to connect any pair of countries. Suppose the direction is accounted (SCC). In that case, the FTA has more than one component due to the presence of regional agreements and non--reciprocal agreements between developed and developing countries, which makes some nodes unilaterally unreachable; i.e., there exists a path from $i$ to $j$, but no path in the opposite direction. Over time, the FTAs network converges into a single component. In the trade network, following WCC and SCC, there is only one component during the period under investigation. It can be seen that the trade network is significantly more connected than the FTAs.   
	
	{
		\small	
		\setlength{\tabcolsep}{8pt}	
		% latex table generated in R 4.2.1 by xtable 1.8-4 package
% Sat Sep 10 12:37:05 2022
\begin{table}[t]
\centering
\caption{Connectivity in the Multiplex FTAs--Trade Flows} 
\label{tab:netstat}
\begin{tabular}{ccccccccc}
  \hline
Year & $AD$ & $AID$ & $De$ & $APL^{\dagger}$ & $WCC^{\dagger}$ & $SCC^{\dagger}$ & $AClueCoe^{\dagger}$ & $ABetCent^{\dagger}$ \\ 
  \hline \hline  \multicolumn{9}{l}{\textit{FTA Layer (unweighted and directed)}}  \\ 1995 & 51.69 & 25.84 & 0.16 & 2.68 & 1 & 31 & 0.36 & 153.75 \\ 
  2005 & 79.65 & 39.82 & 0.25 & 2.23 & 1 & 5 & 0.49 & 169.09 \\ 
  2015 & 97.38 & 48.69 & 0.31 & 1.88 & 1 & 1 & 0.53 & 139.03 \\ 
   \multicolumn{9}{l}{\textit{Trade Flow Layer (weighted and directed)}}   \\1995 & 169.84 & 84.92 & 0.54 & 8777.24 & 1 & 1 & 0.94 & 749.3 \\ 
  2005 & 215.85 & 107.92 & 0.68 & 1341.3 & 1 & 1 & 0.96 & 658.98 \\ 
  2015 & 244.42 & 122.21 & 0.77 & 1854.54 & 1 & 1 & 0.97 & 648 \\ 
   \hline
\end{tabular}
		\begin{minipage}[c]{0.95\textwidth}
			\centering
			\footnotesize\textit{Note}: Average Degree ($AD$), Average In--Degree ($AID$), Density ($De$), Average Path Length ($APL$), Number of Weakly and Strongly Connected Components ($WCC$ and $SCC$), Average Clustering Coefficients ($ACluCoe$), Average Betweenness Centrality ($ABetCen$). The indicators with $\dagger$ accounts for the weights of the Trade Flows, otherwise only the $D^{(1)}$ is used.
		\end{minipage}	
\end{table}

	}

	Both layers share common trends in most of the indicators in Table \ref{tab:netstat}. The average degrees and densities tend to increase over time, which can be the first indication of dependence. Consistent with this higher connectivity, the average path length for the unweighted layer has decreased; however, this is not the case for the weighted one. Despite its general decline, the trend experienced a reversal in the 2000s. The number of triangles in the FTA and trade networks has steadily increased, resulting in higher clustering coefficients. Moreover, the betweenness centrality is decreasing for both networks, which can be interpreted as a loss of influence of intermediary nodes given the presence of new one--step shortest paths and an average increase in the connectivity of the nodes. 
	
	The different topologies of the two layers and the temporal changes in connectivity call for the use of DSBM. Moreover, the simultaneous changes in the connectivity features may suggest a significant influence of trade policies (FTAs) on the trade network. To find evidence of this possibility a DSBMM framework can be applied. The DSBMM offers the possibility of testing for the existence of community structure in trade flows after controlling for GDP and distance, which, to the best of our knowledge, has not been directly examined in the literature. Moreover, the DSBMM can reveal whether changes in the community structure of the (unobserved) multi--resistance terms are related to changes in trade policy through their membership dynamics.

	\subsection{Results} \label{sec:results}
	
	To apply the DSBMM in (\ref{eq:lay1}) and (\ref{eq:lay2}) to the FTAs--Trade flows multidimensional network, the number of blocks $Q^{(\ell)}$, $\ell=1,2$, needs to be selected. \textcolor{black}{Standard criteria provide a large number of communities since either they are not designed for DSBMM or they are not consistent (see Section \ref{subsec:numb} in Supplementary Material for a discussion and further results). In addition, assuming time--varying number of communities can make inference computationally demanding. Thus, in this paper, we assume a constant number of communities and choose it following the results from previous studies. Some robustness checks of the main results are then discussed. We should note that assuming a constant number of blocks is not too restrictive, as some blocks can be empty in certain periods, which compensates for the reduced flexibility of the model. Our work can be extended by assuming a random number of communities and building on infinite Markov models \citep{dufays2016infinite}, or mixtures of finite mixture models combined with the telescopic sampling method recently introduced by \citet{fruhwirth2021generalized}.}
 
 Regarding the trade flows, based on \citet{magerman2020pecking}, a core--periphery structure is a parsimonious two-community representation of the network: a core, that is, a subset of highly connected nodes, and a periphery, the rest of the nodes that are more densely linked to the core than to the other nodes in the periphery. This can be interpreted using a pecking order scheme, firms in each country export following a destination list ordered according to market size and lower trade costs. Hence, countries with less productive firms (periphery) are only able to trade with the top of the list, while countries with more productive firms (core) can trade with large and smaller markets and/or less and more costly destinations. In order to verify the core--periphery structure, we assume $Q^{(1)}=2$. Regarding the FTAs layer, the literature provides little information on the number of communities. In this paper, following previous findings in \citet{barigozzi2011identifying}, we select four communities in the second layer, i.e., $Q^{(2)}=4$. Some robustness checks are then considered assuming $Q^{(1)}=Q^{(2)}=2$ and $Q^{(1)}=Q^{(2)}=4$ (see Figure \ref{fig:Gbcaus_app0} and \ref{fig:Gbcaus_app2} in the Supplementary Material).

\begin{figure}[t]
	\centering
	\caption{Joint posterior distribution of the average expected value and variance of trade flows (in log--scale), $\log \mathbb{E}(\nu^{(1)}_{qr}\exp((\beta_{qr}^{(1)})'X_{ijt}+\sigma_{qr}^{(1)2}/2)|Y,Z_{it}=q,Z_{jt}=r)$ and $\log \mathbb{E}(\nu^{(1)2}_{qr}\exp(2(\beta_{qr}^{(1)})'X_{ijt}+\sigma_{qr}^{(1)2})[\exp(\sigma^2)/\nu_{qr}-1]|Y,Z_{it}=q,Z_{jt}=r)$ respectively (top--left panel), the probability to trade ($\nu^{(1)}_{qr}$) and the conditional expected value of the trade flows (in log--scale, $\log \mathbb{E}(\nu^{(1)}_{qr}\exp(X^{(1)}_{ijt}\beta_{qr}^{(2)}+\sigma_{qr}^{(1)2}/2)|Y>0,Z_{it}=q,Z_{jt}=r)$,bottom--left panel) and the marginal posterior distribution of the probability to have a FTA ($\nu^{(2)}_{qr}$, right panel) by community pair (q,r)}	\label{fig:nusigma12}
	
	%($\sigma_{qr}^{(1)2}$) and the sparsity parameters of the Trade and FTA network ($\nu_{qr}^{(1)}$ and $\nu_{qr}^{(2)}$) under $Q^{(1)}=2$ and $Q^{(2)}=4$. For each block interaction $qr$ (x--axis): posterior distribution (colored-graph) and its quintiles (boxplot).
	\includegraphics[scale=0.5]{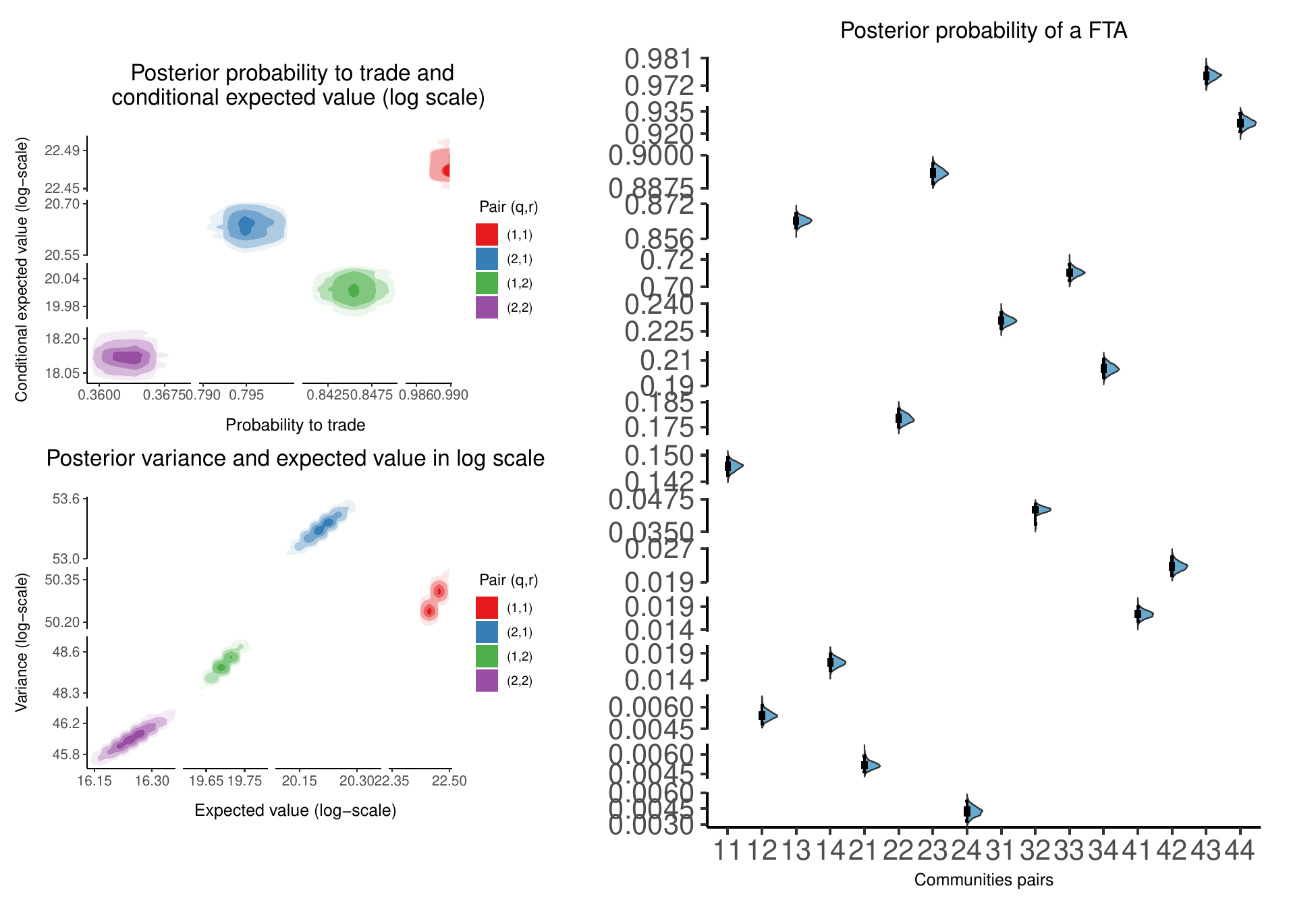}	
		\begin{minipage}{0.9\textwidth}
		\centering
		\footnotesize\textit{Note}: In the block interaction, $q$ and $r$ refer to the exporter and the importer community, respectively. The distributions are kernel density posterior estimates based on 30,000 MCMC samples obtained after 10,000 burn--in and thinning every 10 iterations.
	\end{minipage}	
	
		\end{figure}

The results under $Q^{(1)}=2$ and $Q^{(2)}=4$ confirm a core--periphery structure in the Trade layer (\textcolor{black}{top--left panel of} Figure \ref{fig:nusigma12}). Community 1 corresponds to the core with almost a fully connected sub--graph in terms of intra--block edges (value of $\nu_{11}^{(1)}$ close to 0.98 and is highly dense in the inter-block edges to community 2 (value of $\nu_{12}^{(1)}$ close to 0.84). Community 2 is the periphery since it is better connected to the core (value of $\nu_{21}^{(1)}$ close to 0.79) than to other nodes within its block (value of $\nu_{22}^{(1)}$ close to 0.36). \textcolor{black}{At the beginning of the sample, the community sizes are unbalanced and the core has a smaller membership probability compared to the periphery ($\hat{\alpha}^{(1)}_{1}=0.29$ and $\hat{\alpha}^{(1)}_{2}=0.71$, respectively). For a list of countries by block at the beginning and end of the period, see Figure \ref{fig:maps} and Table \ref{tab:listmem} in the Supplementary Material}. Additionally, \textcolor{black}{the flows with highest variance expressed as $ \mathbb{E}(\nu^{(1)2}_{qr}\exp(2(\beta_{qr}^{(1)})'X_{ijt}+\sigma_{qr}^{(1)2})[\exp(\sigma^2)/\nu_{qr}-1]|Y,Z_{it}=q,Z_{jt}=r)$, are those from the periphery to the core (bottom--left panel of Figure \ref{fig:nusigma12}) with a value close to 53.5 in log--scale. The latent membership variables in the DSBMM capture a non--linear and non--monotonic relationship between variance and the first moment of the trade flows, i.e. $ \mathbb{E}(\nu^{(1)}_{qr}\exp((\beta_{qr}^{(1)})'X_{ijt}+\sigma_{qr}^{(1)2}/2)|Y,Z_{it}=q,Z_{jt}=r)$. Although, within each community pair the relationship between mean and variance is positive, the variance in the intra--community trade flows of the core are lower than the variance in the flows from periphery to core. This result contradicts the models used in the literature that assume larger variance in larger trade flows \citep{silva2006log}. In the top--left panel of Figure \ref{fig:nusigma12}, a positive relationship, between the probability to trade and first moment of the trade volume, $\mathbb{E}(\nu^{(1)}_{qr}\exp(X^{(1)}_{ijt}\beta_{qr}^{(2)}+\sigma_{qr}^{(1)2}/2)|Y>0,Z_{it}=q,Z_{jt}=r)$, is observed suggesting a close dependence between both mechanisms, as underlined by previous studies \citep[e.g.,][]{helpman2008estimating,egger2011trade}.}. 

In the FTA layer, the structure seems more complex, except for community 3, where intra--block connectivity $\nu_{qq}^{(2)}$ is higher than its inter-block counterpart, the other blocks have a non--assortative feature, that is community 1, 2 and 4 are highly connected to block 3 and less densely linked to the other communities (see right panel of Figure \ref{fig:nusigma12}). This configuration is compatible with a core--periphery graph but with some particularities. \textcolor{black}{ The core is Community 3, which includes the countries with the largest number of treaties (high degree nodes) such as Western Europe and countries with a large number of indirect connections due to non--reciprocal agreements (highest eigen--centrality), such as the U.S. under the Generalized System of Preferences (GSP). At the beginning of the sample, the block sizes are unbalanced and two blocks, including the core, have a smaller initial membership probability ($\hat{\alpha}^{(2)}_{3}=0.14$ and  $\hat{\alpha}^{(2)}_{4}=0.11$) compared to the other two blocks ($\hat{\alpha}^{(2)}_{1}=0.41$ and $\hat{\alpha}^{(2)}_{2}=0.34$). See Table \ref{tab:listmem} for further details on the community composition. See Table \ref{tab:listmem} for further details.} For instance, the intra--edges of community 4 are almost as dense as their in--degree from the core, which is the case of countries from eastern Africa (see Figure \ref{fig:fta1995} and \ref{fig:nusigma12}). 

The block structure in the trade flows is consistent with the hypothesis of the picking order destination list. In particular, the most productive firms in the periphery (and in the core), which are those able to export, prioritize destinations with large market sizes and lower trade costs (the core) \citep{magerman2020pecking}. However, it is unclear from this theory why the flows from the periphery to the core are more unstable (higher variance). It could be due to supply factors such as productivity heterogeneity across countries in the periphery or demand factors from the core, since densely connected countries may have more alternative suppliers to consider.   
	
\begin{figure}[t]
		\centering
		\caption{Gravity equation parameters in the trade network ($\beta^{(1)}_{qr}$) under $Q^{(1)}=2$ and $Q^{(2)}=4$.  For each block interaction $qr$ (x--axis): posterior distribution (colored-graph) and its quintiles (boxplot).}\label{fig:parmsgrav}
		\includegraphics[scale=0.5]{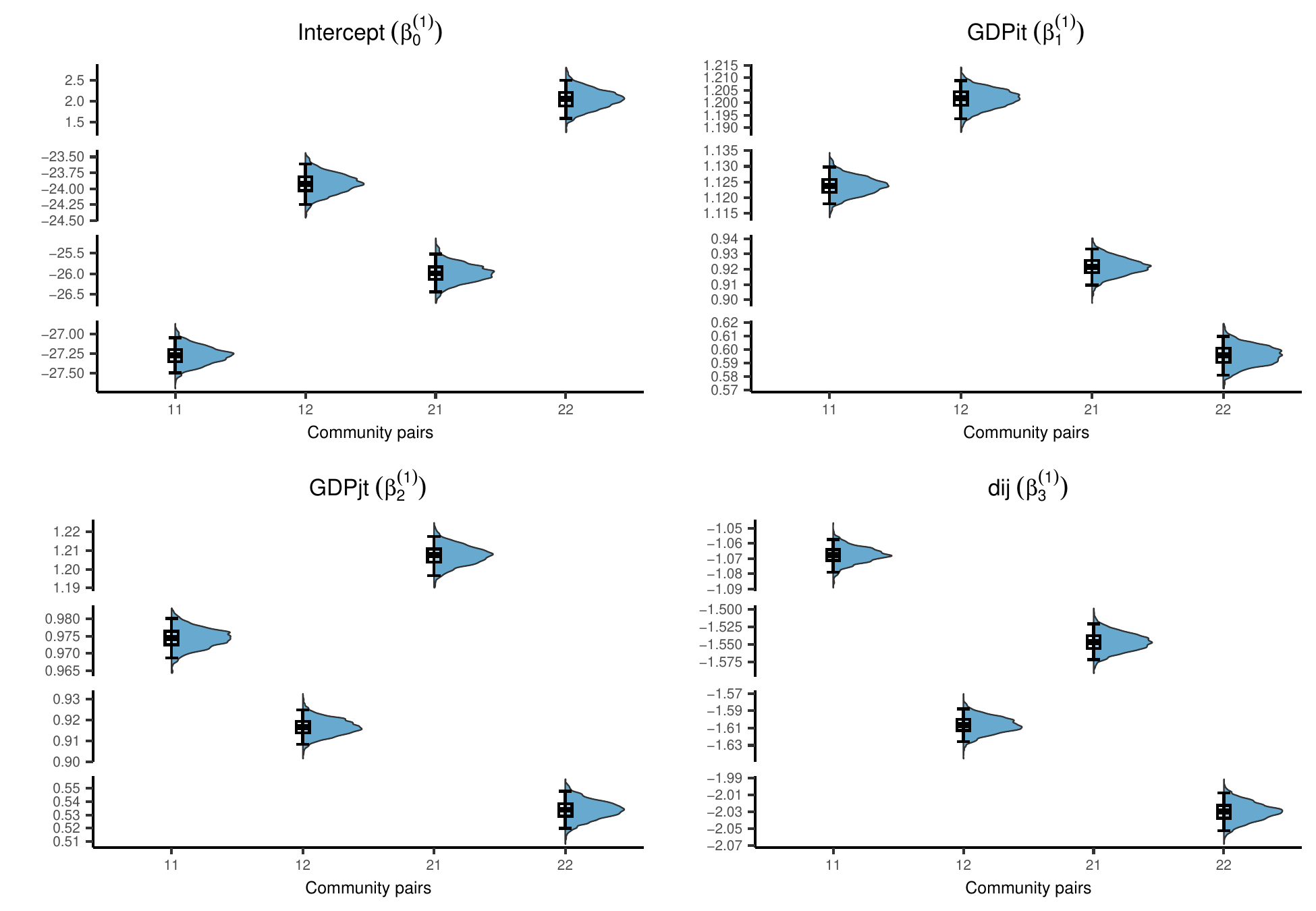}
		\begin{minipage}{0.9\textwidth}
			\centering
			\footnotesize\textit{Note}: Each distribution corresponds to a different block interaction $qr$, where $q$ and $r$ refer to the exporter and the importer community, respectively (x--axis). The distributions are kernel density posterior estimates based on 30,000 MCMC samples obtained after 10,000 burn--in and thinning every 10 iterations.
		\end{minipage}	
	\end{figure}

	Regarding the connectivity parameters in the Gravity equation, the results evidence significant differences between blocks, reflecting contrasting unobserved barriers and multi--resistance terms. \textcolor{black}{The lowest intercept is observed between the core countries, while the highest corresponds to the peripheral nodes (see Figure \ref{fig:parmsgrav})}. At the same time, this is compensated by a less negative distance parameter between the core countries and an almost twice as negative marginal effect within community 2. A similar pattern applies to the exporter and importer GDP elasticities, both are higher in the core. In the inter--block interaction, when the core (periphery) is the exporter, trade flows are more (less) responsive to the GDP, in contrast to its role as an importer. Hence, the heterogeneity in the Gravity equation parameters is coherent with the differences in the connectivity parameters of the unweighted part of the trade network ($\nu^{(1)}$) and the international trade literature studying intensive and extensive margins \citep[see for example][]{bernard2007firms}. The gaps in the distance parameter can be attributed to differences in technology and/or transportation costs. In the case of the GDP elasticities, the differences suggest the possibility of persistent trade deficits for the periphery, which is not fully accounted for by the standard trade theory that assumes the elasticities should be close to one for all countries. \textcolor{black}{The interpretation of the GDP coefficients is not straightforward because trade flows are measured in gross value, while GDP is measured in value added  \citep[e.g.,][]{anderson2004trade}. Nevertheless, these discrepancies in income elasticity have raised discussions on the influence of home market effects and oligopolistic market structures \citep{feenstra2001using,fujiwara2024firm} and they are also relevant in macroeconomic theory, in particular the literature on balance of payment constrained growth and structural change \citep[e.g.,][]{thirlwall2012balance,matsuyama2019engel} and consistent with the findings in  \citet{silva2006log}, based on Poisson Gravity models.}
	
	This network structure is not static, and countries may change membership through time and inherit the corresponding set of connectivity parameters, which to the extent of our review, is not taken into account in the standard partial or general equilibrium models used for ex-post and ex-ante trade policy evaluation. In this sense, for the trade network, 22.0\% of countries changed membership at least one time between 1995--2017 mostly from the periphery to core, while in the FTA layer, all countries deviate from their initial community (see Figure \ref{fig:changlay1} and \ref{fig:changlay2}). This new result complements existing community detection studies on trade flows, which deduce an increase in network density by identifying blocks at each point in time. However, given their static approach, they are not able to recover specific node changes.

 	\begin{figure}[t]
		\centering
		\caption{Transition probability parameters of the multinomial model ($\kappa^{(\ell)}$) for the Trade and FTAs network under $Q^{(1)}=2$ and $Q^{(2)}=4$, respectively. For each parameter $\kappa^{(\ell)}_{\mathcal{U},r,i}$ (x--axis): posterior distribution (colored-graph) and its quintiles (boxplot).}
		\label{fig:Gbcaus_app}
		\includegraphics[scale=0.45]{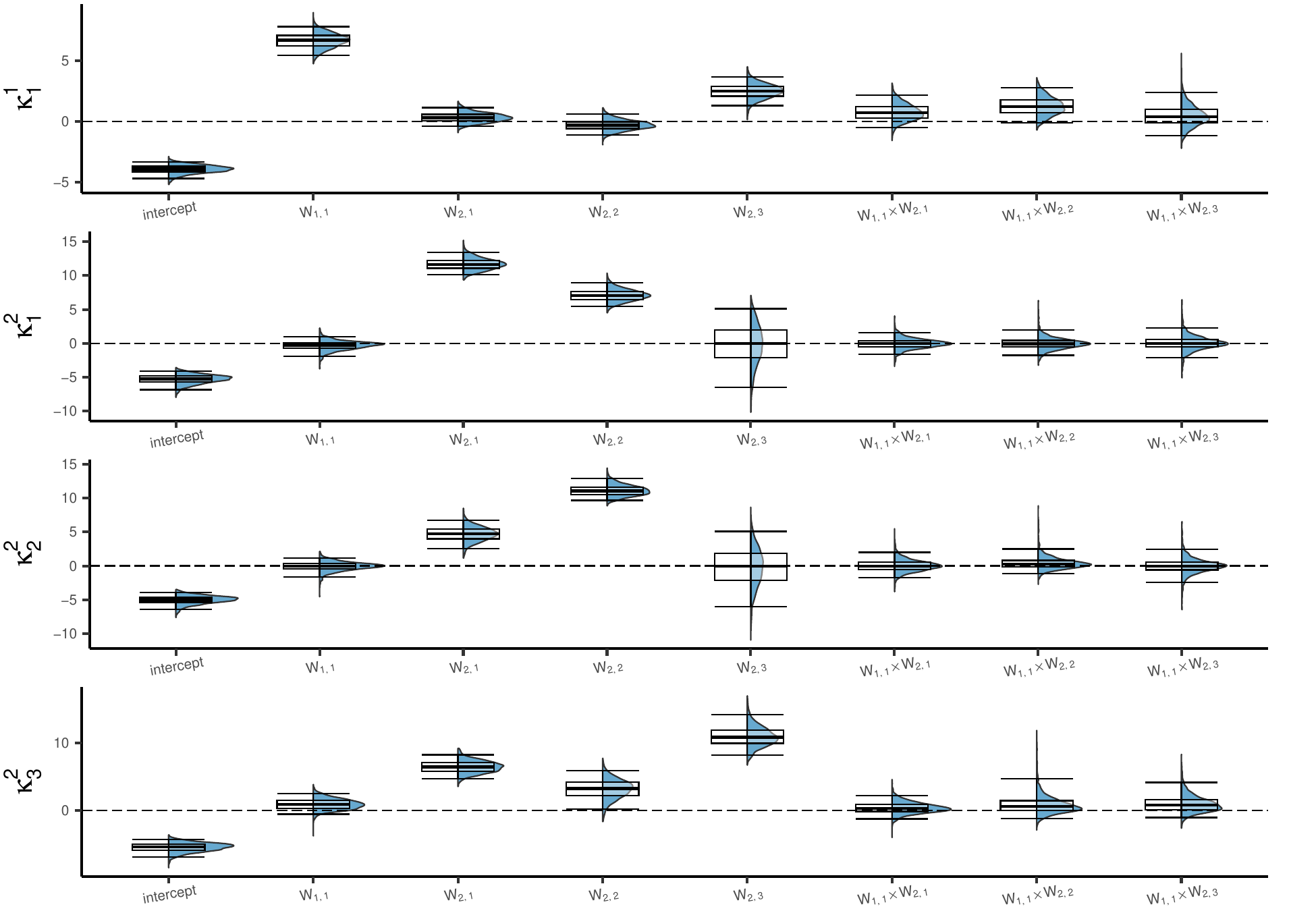}
		\begin{minipage}{0.9\textwidth}
			\centering
			\footnotesize\textit{Note}: Number of MCMC samples used for approximating the posterior: 30,000 and burn-in: 10,000. Thinning:10.
		\end{minipage}	
	\end{figure}
	The block dynamic allows us to test for \textcolor{black}{NGB causality} and check if the changes in the membership structure of the FTAs and in the memberships of the unobserved components of the Trade network are related. In Figure \ref{fig:Gbcaus_app}, the membership dependence between the two layers of the multidimensional network is summarized into four graphs, each of them associated with a vector $\kappa_{q}^{(\ell)}$. For instance, the first graph shows the transition probability of moving/staying in community 1 (core) of the trade network, and it can be noticed that its diagonal elements are significant, i.e. $W^{(1)}_{it-1,1}$ and the intercept, which denotes persistence in membership. More importantly, this persistence is attenuated by the influence of the country's membership in the FTA network. Essentially, a change in membership to the core of the FTA network induces a change in the trade network, potentially producing a decrease in the multi-resistance terms and unobserved barriers. The three graphs of the second layer describe the FTA network transition parameters and indicate that membership persistence is lower compared to the Trade network. From a variable selection approach, it appears not to be associated with the block structure of the latter. \textcolor{black}{In other words, there is evidence that FTA core membership $W_{2,3}$ influences the probability of moving to the Trade core (see Panel $\kappa^{1}_{1}$). The evidence of a predictive impact of Trade core membership $W_{1,1}$ on the transition to FTA core is weaker in all FTA communities (see panels $\kappa^{2}_{1}$,$\kappa^{2}_{2}$ and $\kappa^{2}_{3}$). Indeed, there are some cases of transitions to the Trade core (see boxes in Figure \ref{fig:changlay1}), which include countries in the FTA core from the beginning of the sample (e.g. HUN) and countries that entered the FTA core later in the sample (e.g. BLR, EST, LTU, LVA). See Figure \ref{fig:maps} and Table \ref{tab:listmem} for more details. However, a proper test of the NGB requires estimating the log--Bayes factor among the competing hypotheses. Following the reverse logistic approach, the log--factor in favor of no NGB from Trade to FTA, from FTA to Trade, and no NGB causality is -6136.21,-8853.52, and -5424.32, respectively. This suggests that all models incorporating some form of NGB causality perform better than models with no NGB causality. In particular, the posterior probability of the model allowing bilateral NGB causality is significantly higher than unilateral alternatives or no dependence---i.e., close to one.}  The same result applies for $Q^{(1)}=Q^{(2)}=4$, but for $Q^{(1)}=Q^{(2)}=2$ there is evidence of bidirectional causality (see Figure \ref{fig:Gbcaus_app0} and \ref{fig:Gbcaus_app2} in the Supplementary Material). 
	
In summary, the bidirectional block--causality suggests that the FTAs have effectively modified the trade flows network structure, favoring the transition of countries from the periphery to the core and increasing global trade integration and vice--versa. In other words, the level of trade for a specific pair of countries also helps predict their probability of signing or not a FTA, through the clustering structure. This is consistent with the endogeneity argument suggested by the trade literature, where expected gains from signing a treaty affect the trade policies. This result is robust to the number of blocks.

The core--periphery structure is also robust to the number of blocks. However, some countries are not part of the core of the FTA network and are densely connected in the trade network (e.g. China and Costa Rica). This could be due to the type of exported products, technological particularities, and the level of integration to the Global Value Chains (GVC), among other factors that can be easily included in the DSBMM as extra covariates or as extra layers (e.g. the global investment network). Moreover, the time--varying membership captured by our DSBMM implies evidence of parameter heterogeneity in the Gravity equation, which may be used by the trade literature to evaluate the impact of changes in tariffs. This heterogeneity is especially relevant in the evaluation of trade policies aiming at a structural change of the network beyond a specific pair of countries in the same community, for example, the theoretical scenarios of full autarky or in ambitious treaties such as the Transatlantic Trade and Investment Partnership (TTIP) or the Trans--Pacific Partnership (TPP), in these cases the counterfactual scenario may also imply different parameters.

    \section{Conclusion}
	\label{sec:con}
	This work has focused on extending the DSBM to a multidimensional setting with dependent Hidden Markov Chains (HMC). This task is achieved assuming node membership is driven by a different HMC in each layer. The HMC dependence across layers is specified as a saturated multinomial logit model. The resulting model, DSBMM, enables us to study the relationship between the block structure of layers and provides a framework for a \textcolor{black}{NGB causality} test. Compared to the literature on multidimensional SBM, our DSBMM is novel in many aspects. First, it can capture coupling trends between the membership structure of the layers, decoupling tendencies, or more complex relationships. Second, it provides information on the sign, magnitude, and direction of the dependence. Moreover, node- and dyad-specific covariates are included to capture heterogeneity that may affect the weights and the sparsity patterns of the networks. Finally, the dependence between the four types of layers --- weighted and unweighted, directed and undirected --- can be easily incorporated into our model. 
    
    To cope with over-paramerization, we propose a new Bayesian shrinkage approach for DSBMM based on Multi--Laplacian prior distributions. This prior class allows for different degrees of shrinkage between and within groups of \textcolor{black}{lagged HMC variables} and accounts for the correlation structure of the \textcolor{black}{hidden state variables}. A Pólya Gamma representation makes the posterior distribution more tractable, and an efficient Gibbs sampler is designed to approximate the posterior distribution. The efficiency of the sampler, the effectiveness of the Multi-Laplace prior in recovering the true {\color{black}NGB} causal structure, and the superior fitting ability of the DSBMM compared to benchmark models are studied through simulation experiments under different network configurations and parameter settings.
    
    We apply our DSBMM model to international trade networks and contribute to the debate on the effectiveness of FTAs beyond the country pairs level. Moreover, to the best of our knowledge, this is the first econometric model to integrate two commonly used approaches in international trade: community detection and gravity models. A temporal, multidimensional network of Trade Flows and FTAs is considered among 159 countries over 23 years. We found evidence of core--periphery structure in FTAs and trade networks, and of dyad and time heterogeneity of the Gravity equation parameters. The standard theoretical models do not account for this result, which may impact existing methods for FTA evaluations. Regarding the debate on the predictive power of FTAs, we find evidence of a \textcolor{black}{NGB} causal relationship between FTAs and trade. Moreover, the direction is sensitive to the choice of the number of communities. There is bidirectional block--causality between the FTAs and the unobserved barriers. 

    \section{Declaration of competing interests}
The authors declare that they have no known competing financial interests or personal relationships that could have appeared to influence the work reported in this paper.
    
\bibliographystyle{apalike}
\bibliography{biblio}

\begin{thebibliography}{}

\bibitem[Agudze et~al., 2022]{agudze2021markov}
Agudze, K.~M., Billio, M., Casarin, R., and Ravazzolo, F. (2022).
\newblock Markov switching panel with endogenous synchronization effects.
\newblock {\em Journal of Econometrics}, 230(2):281--298.

\bibitem[Airoldi et~al., 2008]{airoldi2008mixed}
Airoldi, E.~M., Blei, D.~M., Fienberg, S.~E., and Xing, E.~P. (2008).
\newblock Mixed membership stochastic blockmodels.
\newblock {\em Journal of Machine Learning Research}, 9(Sep):1981--2014.

\bibitem[Anderson, 2011]{anderson2011gravity}
Anderson, J.~E. (2011).
\newblock The {G}ravity model.
\newblock {\em Annual Review of Economics}, 3(1):133--160.

\bibitem[Anderson et~al., 2020]{anderson2020transitional}
Anderson, J.~E., Larch, M., and Yotov, Y.~V. (2020).
\newblock Transitional growth and trade with frictions: A structural estimation
  framework.
\newblock {\em The Economic Journal}, 130(630):1583--1607.

\bibitem[Anderson and Van~Wincoop, 2004]{anderson2004trade}
Anderson, J.~E. and Van~Wincoop, E. (2004).
\newblock Trade costs.
\newblock {\em Journal of Economic Literature}, 42(3):691--751.

\bibitem[Baier and Bergstrand, 2004]{baier2004economic}
Baier, S.~L. and Bergstrand, J.~H. (2004).
\newblock Economic determinants of free trade agreements.
\newblock {\em Journal of International Economics}, 64(1):29--63.

\bibitem[Baier and Bergstrand, 2007]{baier2007free}
Baier, S.~L. and Bergstrand, J.~H. (2007).
\newblock Do free trade agreements actually increase members' international
  trade?
\newblock {\em Journal of International Economics}, 71(1):72--95.

\bibitem[Baier et~al., 2019]{baier2019widely}
Baier, S.~L., Yotov, Y.~V., and Zylkin, T. (2019).
\newblock On the widely differing effects of free trade agreements: Lessons
  from twenty years of trade integration.
\newblock {\em Journal of International Economics}, 116:206--226.

\bibitem[Baltagi et~al., 2008]{baltagi2008estimating}
Baltagi, B.~H., Egger, P., and Pfaffermayr, M. (2008).
\newblock Estimating regional trade agreement effects on {FDI} in an
  interdependent world.
\newblock {\em Journal of Econometrics}, 145(1-2):194--208.

\bibitem[Barigozzi et~al., 2011]{barigozzi2011identifying}
Barigozzi, M., Fagiolo, G., and Mangioni, G. (2011).
\newblock Identifying the community structure of the international-trade
  multi-network.
\newblock {\em Physica A: Statistical Mechanics and Its Applications},
  390(11):2051--2066.

\bibitem[Bartesaghi et~al., 2020]{bartesaghi2020communicability}
Bartesaghi, P., Clemente, G.~P., and Grassi, R. (2020).
\newblock Communicability in the world trade network--a new perspective for
  community detection.
\newblock {\em Journal of Economic Interaction and Coordination}, 17(2):405 --
  441.

\bibitem[Behrens et~al., 2012]{behrens2012dual}
Behrens, K., Ertur, C., and Koch, W. (2012).
\newblock ‘{D}ual’ {G}ravity: {U}sing spatial econometrics to control for
  multilateral resistance.
\newblock {\em Journal of Applied Econometrics}, 27(5):773--794.

\bibitem[Bernard et~al., 2007]{bernard2007firms}
Bernard, A.~B., Jensen, J.~B., Redding, S.~J., and Schott, P.~K. (2007).
\newblock Firms in international trade.
\newblock {\em Journal of Economic Perspectives}, 21(3):105--130.

\bibitem[Billio et~al., 2016]{billio2016interconnections}
Billio, M., Casarin, R., Ravazzolo, F., and Van~Dijk, H.~K. (2016).
\newblock Interconnections between eurozone and {US} booms and busts using a
  {B}ayesian panel {M}arkov-switching {VAR} model.
\newblock {\em Journal of Applied Econometrics}, 31(7):1352--1370.

\bibitem[Bondell and Reich, 2012]{bondell2012consistent}
Bondell, H.~D. and Reich, B.~J. (2012).
\newblock Consistent high-dimensional {B}ayesian variable selection via
  penalized credible regions.
\newblock {\em Journal of the American Statistical Association},
  107(500):1610--1624.

\bibitem[Canova and Ciccarelli, 2013]{canova2013panel}
Canova, F. and Ciccarelli, M. (2013).
\newblock Panel vector autoregressive models: A survey.
\newblock In {\em VAR models in macroeconomics--new developments and
  applications: Essays in honor of Christopher A. Sims}, pages 205--246.
  Emerald Group Publishing Limited.

\bibitem[Carallo et~al., 2024]{carallo2024generalized}
Carallo, G., Casarin, R., and Robert, C.~P. (2024).
\newblock Generalized {P}oisson difference autoregressive processes.
\newblock {\em International Journal of Forecasting}, 40(4):1359--1390.

\bibitem[Casella et~al., 2010]{casella2010penalized}
Casella, G., Ghosh, M., Gill, J., and Kyung, M. (2010).
\newblock {Penalized regression, standard errors, and Bayesian lassos}.
\newblock {\em Bayesian Analysis}, 5(2):369 -- 411.

\bibitem[Csermely et~al., 2013]{csermely2013structure}
Csermely, P., London, A., Wu, L.-Y., and Uzzi, B. (2013).
\newblock Structure and dynamics of core/periphery networks.
\newblock {\em Journal of Complex Networks}, 1(2):93--123.

\bibitem[De~Paula, 2017]{de2017econometrics}
De~Paula, A. (2017).
\newblock Econometrics of network models.
\newblock In {\em Advances in Economics and Econometrics: Theory and
  Applications, Eleventh World Congress}, pages 268--323. Cambridge University
  Press Cambridge.

\bibitem[Dippelsman et~al., 2018]{dippelsman2018new}
Dippelsman, M.~R., Marini, M.~M., Stanger, M.~M., et~al. (2018).
\newblock New estimates for {D}irection of {T}rade {S}tatistics.
\newblock Technical report, International Monetary Fund.

\bibitem[Dufays, 2016]{dufays2016infinite}
Dufays, A. (2016).
\newblock Infinite-state markov-switching for dynamic volatility.
\newblock {\em Journal of Financial Econometrics}, 14(2):418--460.

\bibitem[Dufour and Taamouti, 2010]{dufour2010short}
Dufour, J.-M. and Taamouti, A. (2010).
\newblock Short and long run causality measures: Theory and inference.
\newblock {\em Journal of Econometrics}, 154(1):42--58.

\bibitem[Durante et~al., 2017]{durante2017bayesian}
Durante, D., Mukherjee, N., and Steorts, R.~C. (2017).
\newblock Bayesian learning of dynamic multilayer networks.
\newblock {\em Journal of Machine Learning Research}, 18(43):1--29.

\bibitem[Egger et~al., 2011]{egger2011trade}
Egger, P., Larch, M., Staub, K.~E., and Winkelmann, R. (2011).
\newblock The trade effects of endogenous preferential trade agreements.
\newblock {\em American Economic Journal: Economic Policy}, 3(3):113--143.

\bibitem[Fang et~al., 2011]{fang2011double}
Fang, Z., Wang, J., Liu, B., and Gong, W. (2011).
\newblock Double pareto lognormal distributions in complex networks.
\newblock In {\em Handbook of Optimization in Complex Networks: Theory and
  Applications}, pages 55--80. Springer.

\bibitem[Feenstra et~al., 2001]{feenstra2001using}
Feenstra, R.~C., Markusen, J.~R., and Rose, A.~K. (2001).
\newblock Using the {G}ravity equation to differentiate among alternative
  theories of trade.
\newblock {\em Canadian Journal of Economics/Revue canadienne
  d'{\'e}conomique}, 34(2):430--447.

\bibitem[Fr{\"u}hwirth-Schnatter, 2006]{fruhwirth2006finite}
Fr{\"u}hwirth-Schnatter, S. (2006).
\newblock {\em Finite mixture and {M}arkov switching models}, volume 425.
\newblock Springer.

\bibitem[Fr{\"u}hwirth-Schnatter and Fr{\"u}hwirth, 2010]{fruhwirth2010data}
Fr{\"u}hwirth-Schnatter, S. and Fr{\"u}hwirth, R. (2010).
\newblock Data augmentation and {MCMC} for binary and multinomial logit models.
\newblock In {\em Statistical Modelling and Regression Structures}, pages
  111--132. Springer.

\bibitem[Fr{\"u}hwirth-Schnatter et~al., 2021]{fruhwirth2021generalized}
Fr{\"u}hwirth-Schnatter, S., Malsiner-Walli, G., and Gr{\"u}n, B. (2021).
\newblock Generalized mixtures of finite mixtures and telescoping sampling.
\newblock {\em Bayesian Analysis}, 16(4):1279--1307.

\bibitem[Fujiwara, 2024]{fujiwara2024firm}
Fujiwara, K. (2024).
\newblock Firm heterogeneity, home market effect, and {G}ravity equation in an
  oligopoly.
\newblock {\em Open Economies Review}, 35(5):1115--1131.

\bibitem[George and McCulloch, 1997]{george1997approaches}
George, E.~I. and McCulloch, R.~E. (1997).
\newblock Approaches for {B}ayesian variable selection.
\newblock {\em Statistica Sinica}, 7(2):339--373.

\bibitem[Geyer, 1991]{geyer1991estimating}
Geyer, C.~J. (1991).
\newblock Estimating normalizing constants and reweighting mixtures in {M}arkov
  chain {M}onte {C}arlo: Technical report 568.
\newblock {\em School of Statistics, University of Minnesota}.

\bibitem[Gurevich and Herman, 2018]{gurevich2018dynamic}
Gurevich, T. and Herman, P. (2018).
\newblock The dynamic {G}ravity dataset: Technical documentation.

\bibitem[Hamaker and Grasman, 2012]{hamaker2012regime}
Hamaker, E. and Grasman, R. (2012).
\newblock Regime switching state-space models applied to psychological
  processes: Handling missing data and making inferences.
\newblock {\em Psychometrika}, 77:400--422.

\bibitem[Han et~al., 2015]{han2015consistent}
Han, Q., Xu, K., and Airoldi, E. (2015).
\newblock Consistent estimation of dynamic and multi-layer block models.
\newblock In {\em International Conference on Machine Learning}, pages
  1511--1520.

\bibitem[Hans, 2010]{hans2010model}
Hans, C. (2010).
\newblock Model uncertainty and variable selection in {B}ayesian lasso
  regression.
\newblock {\em Statistics and Computing}, 20:221--229.

\bibitem[Held and Holmes, 2006]{held2006bayesian}
Held, L. and Holmes, C.~C. (2006).
\newblock Bayesian auxiliary variable models for binary and multinomial
  regression.
\newblock {\em Bayesian Analysis}, 1(1):145--168.

\bibitem[Helpman et~al., 2008]{helpman2008estimating}
Helpman, E., Melitz, M., and Rubinstein, Y. (2008).
\newblock Estimating trade flows: {T}rading partners and trading volumes.
\newblock {\em The Quarterly Journal of Economics}, 123(2):441--487.

\bibitem[Hoff, 2021]{hoff2018}
Hoff, P. (2021).
\newblock Additive and multiplicative effects network models.
\newblock {\em Statistical Science}, 36(1):34--50.

\bibitem[Holsclaw et~al., 2017]{holsclaw2017bayesian}
Holsclaw, T., Greene, A.~M., Robertson, A.~W., and Smyth, P. (2017).
\newblock Bayesian nonhomogeneous {M}arkov models via {P}{\'o}lya-{G}amma data
  augmentation with applications to rainfall modeling.
\newblock {\em The Annals of Applied Statistics}, 11(1):393--426.

\bibitem[Hu et~al., 2015]{hu2015shortcomings}
Hu, S., Jia, X., Zhang, J., Kong, W., and Cao, Y. (2015).
\newblock Shortcomings/limitations of blockwise {G}ranger causality and
  advances of blockwise new causality.
\newblock {\em IEEE Transactions on Neural Networks and Learning Systems},
  27(12):2588--2601.

\bibitem[Ivlevs and De~Melo, 2010]{ivlevs2010fdi}
Ivlevs, A. and De~Melo, J. (2010).
\newblock {FDI}, the brain drain and trade: channels and evidence.
\newblock {\em Annals of Economics and Statistics/Annales d’{\'E}conomie et
  de Statistique}, pages 103--121.

\bibitem[Jones, 2000]{jones2000health}
Jones, A.~M. (2000).
\newblock Health econometrics.
\newblock In {\em Handbook of health economics}, volume~1, pages 265--344.
  Elsevier.

\bibitem[Jovanovski and Kocarev, 2019]{jovanovski2019bayesian}
Jovanovski, P. and Kocarev, L. (2019).
\newblock Bayesian consensus clustering in multiplex networks.
\newblock {\em Chaos: An Interdisciplinary Journal of Nonlinear Science},
  29(10):103142.

\bibitem[Kabir et~al., 2017]{kabir2017gravity}
Kabir, M., Salim, R., and Al-Mawali, N. (2017).
\newblock {The {G}ravity model and trade flows: Recent developments in
  econometric modeling and empirical evidence}.
\newblock {\em Economic Analysis and Policy}, 56:60--71.

\bibitem[Kaufmann, 2015]{kaufmann2015k}
Kaufmann, S. (2015).
\newblock K-state switching models with time-varying transition
  distributions—{D}oes loan growth signal stronger effects of variables on
  inflation?
\newblock {\em Journal of Econometrics}, 187(1):82--94.

\bibitem[Kim et~al., 2018]{kim2018review}
Kim, B., Lee, K.~H., Xue, L., and Niu, X. (2018).
\newblock A review of dynamic network models with latent variables.
\newblock {\em Statistics Surveys}, 12:105.

\bibitem[Kivel\"{a} et~al., 2014]{Kiv14}
Kivel\"{a}, M., Arenas, A., Barthelemy, M., Gleeson, J.~P., Moreno, Y., and
  Porter, M.~A. (2014).
\newblock Multilayer networks.
\newblock {\em Journal of Complex Networks}, 2(3):203--271.

\bibitem[Koop, 2013]{koop2013forecasting}
Koop, G.~M. (2013).
\newblock Forecasting with medium and large {B}ayesian {VAR}s.
\newblock {\em Journal of Applied Econometrics}, 28(2):177--203.

\bibitem[Lee and Wilkinson, 2019]{lee2019review}
Lee, C. and Wilkinson, D.~J. (2019).
\newblock A review of stochastic block models and extensions for graph
  clustering.
\newblock {\em Applied Network Science}, 4(1):1--50.

\bibitem[Lei and Lin, 2022]{lei2022bias}
Lei, J. and Lin, K.~Z. (2022).
\newblock Bias-adjusted spectral clustering in multi-layer stochastic block
  models.
\newblock {\em Journal of the American Statistical Association}, pages 1--13.

\bibitem[Little, 1993]{little1993pattern}
Little, R.~J. (1993).
\newblock Pattern-mixture models for multivariate incomplete data.
\newblock {\em Journal of the American Statistical Association},
  88(421):125--134.

\bibitem[Llorente et~al., 2023]{llorente2023marginal}
Llorente, F., Martino, L., Delgado, D., and Lopez-Santiago, J. (2023).
\newblock Marginal likelihood computation for model selection and hypothesis
  testing: an extensive review.
\newblock {\em SIAM Review}, 65(1):3--58.

\bibitem[Magerman et~al., 2020]{magerman2020pecking}
Magerman, G., De~Bruyne, K., and Van~Hove, J. (2020).
\newblock Pecking order and core-periphery in international trade.
\newblock {\em Review of International Economics}, 28(4):1113--1141.

\bibitem[Mariadassou et~al., 2010]{mariadassou2010uncovering}
Mariadassou, M., Robin, S., and Vacher, C. (2010).
\newblock Uncovering latent structure in valued graphs: a variational approach.
\newblock {\em The Annals of Applied Statistics}, 4(2):715--742.

\bibitem[Matias and Miele, 2017]{matias2017statistical}
Matias, C. and Miele, V. (2017).
\newblock Statistical clustering of temporal networks through a dynamic
  stochastic block model.
\newblock {\em Journal of the Royal Statistical Society: Series B (Statistical
  Methodology)}, 79(4):1119--1141.

\bibitem[Matsuyama, 2019]{matsuyama2019engel}
Matsuyama, K. (2019).
\newblock Engel's law in the global economy: Demand-induced patterns of
  structural change, innovation, and trade.
\newblock {\em Econometrica}, 87(2):497--528.

\bibitem[Metulini et~al., 2018]{metulini2018spatial}
Metulini, R., Patuelli, R., and Griffith, D.~A. (2018).
\newblock A spatial-filtering zero-inflated approach to the estimation of the
  {G}ravity model of trade.
\newblock {\em Econometrics}, 6(1):9.

\bibitem[Mosconi and Seri, 2006]{mosconi2006non}
Mosconi, R. and Seri, R. (2006).
\newblock Non-causality in bivariate binary time series.
\newblock {\em Journal of Econometrics}, 132(2):379--407.

\bibitem[Nigai, 2017]{nigai2017tale}
Nigai, S. (2017).
\newblock A tale of two tails: Productivity distribution and the gains from
  trade.
\newblock {\em Journal of International Economics}, 104:44--62.

\bibitem[Olivella et~al., 2022]{olivella2022dynamic}
Olivella, S., Pratt, T., and Imai, K. (2022).
\newblock Dynamic stochastic blockmodel regression for network data:
  Application to international militarized conflicts.
\newblock {\em Journal of the American Statistical Association},
  117(539):1068--1081.

\bibitem[Otranto, 2005]{otranto2005multi}
Otranto, E. (2005).
\newblock The multi-chain {M}arkov switching model.
\newblock {\em Journal of Forecasting}, 24(7):523--537.

\bibitem[Park and Casella, 2008]{park2008bayesian}
Park, T. and Casella, G. (2008).
\newblock The {B}ayesian lasso.
\newblock {\em Journal of the American Statistical Association},
  103(482):681--686.

\bibitem[Paul and Chen, 2020]{paul2020random}
Paul, S. and Chen, Y. (2020).
\newblock A random effects stochastic block model for joint community detection
  in multiple networks with applications to neuroimaging.
\newblock {\em The Annals of Applied Statistics}, 14(2):993--1029.

\bibitem[Piermartini and Yotov, 2016]{piermartini2016estimating}
Piermartini, R. and Yotov, Y. (2016).
\newblock Estimating trade policy effects with structural {G}ravity.
\newblock Technical report, WTO Staff Working Paper.

\bibitem[Polson et~al., 2013]{polson2013bayesian}
Polson, N.~G., Scott, J.~G., and Windle, J. (2013).
\newblock Bayesian inference for logistic models using {P}{\'o}lya--{G}amma
  latent variables.
\newblock {\em Journal of the American Statistical Association},
  108(504):1339--1349.

\bibitem[Raman et~al., 2009]{raman2009bayesian}
Raman, S., Fuchs, T.~J., Wild, P.~J., Dahl, E., and Roth, V. (2009).
\newblock The {B}ayesian group-lasso for analyzing contingency tables.
\newblock In {\em Proceedings of the 26th Annual International Conference on
  Machine Learning}, pages 881--888.

\bibitem[Salter-Townshend and McCormick, 2017]{salter2017latent}
Salter-Townshend, M. and McCormick, T.~H. (2017).
\newblock Latent space models for multiview network data.
\newblock {\em The Annals of Applied Statistics}, 11(3):1217.

\bibitem[Silva and Tenreyro, 2006]{silva2006log}
Silva, J.~S. and Tenreyro, S. (2006).
\newblock The log of {G}ravity.
\newblock {\em The Review of Economics and Statistics}, 88(4):641--658.

\bibitem[Song and Belin, 2004]{song2004imputation}
Song, J. and Belin, T.~R. (2004).
\newblock Imputation for incomplete high-dimensional multivariate normal data
  using a common factor model.
\newblock {\em Statistics in Medicine}, 23(18):2827--2843.

\bibitem[Sopranzetti, 2018]{sopranzetti2018overlapping}
Sopranzetti, S. (2018).
\newblock Overlapping free trade agreements and international trade: {A}
  network approach.
\newblock {\em The World Economy}, 41(6):1549--1566.

\bibitem[Stanley et~al., 2016]{stanley2016clustering}
Stanley, N., Shai, S., Taylor, D., and Mucha, P.~J. (2016).
\newblock Clustering network layers with the strata multilayer stochastic block
  model.
\newblock {\em IEEE Transactions on Network Science and Engineering},
  3(2):95--105.

\bibitem[Tabouy et~al., 2020]{tabouy2020variational}
Tabouy, T., Barbillon, P., and Chiquet, J. (2020).
\newblock Variational inference for stochastic block models from sampled data.
\newblock {\em Journal of the American Statistical Association},
  115(529):455--466.

\bibitem[Thirlwall, 2012]{thirlwall2012balance}
Thirlwall, A.~P. (2012).
\newblock Balance of payments constrained growth models: history and overview.
\newblock {\em Models of balance of payments constrained growth: History,
  theory and empirical evidence}, pages 11--49.

\bibitem[Touloupou et~al., 2020]{touloupou2020scalable}
Touloupou, P., Finkenst{\"a}dt, B., and Spencer, S.~E. (2020).
\newblock Scalable bayesian inference for coupled hidden markov and semi-markov
  models.
\newblock {\em Journal of Computational and Graphical Statistics},
  29(2):238--249.

\bibitem[Van~Erp et~al., 2019]{van2019shrinkage}
Van~Erp, S., Oberski, D.~L., and Mulder, J. (2019).
\newblock Shrinkage priors for {B}ayesian penalized regression.
\newblock {\em Journal of Mathematical Psychology}, 89:31--50.

\bibitem[Van~Hasselt, 2011]{van2011bayesian}
Van~Hasselt, M. (2011).
\newblock Bayesian inference in a sample selection model.
\newblock {\em Journal of Econometrics}, 165(2):221--232.

\bibitem[Yang et~al., 2011]{yang2011detecting}
Yang, T., Chi, Y., Zhu, S., Gong, Y., and Jin, R. (2011).
\newblock Detecting communities and their evolutions in dynamic social
  networks—a {B}ayesian approach.
\newblock {\em Machine Learning}, 82(2):157--189.

\bibitem[Yotov, 2012]{yotov2012simple}
Yotov, Y.~V. (2012).
\newblock A simple solution to the distance puzzle in international trade.
\newblock {\em Economics Letters}, 117(3):794--798.

\bibitem[Yotov et~al., 2016]{yotov2016advanced}
Yotov, Y.~V., Piermartini, R., Larch, M., et~al. (2016).
\newblock {\em An advanced guide to trade policy analysis: the structural
  {G}ravity model}.
\newblock WTO iLibrary.

\bibitem[Zhao et~al., 2012]{zhao2012consistency}
Zhao, Y., Levina, E., and Zhu, J. (2012).
\newblock Consistency of community detection in networks under degree-corrected
  stochastic block models.
\newblock {\em The Annals of Statistics}, 40(4):2266--2292.

\end{thebibliography}
	
	%\newpage
	\appendix
	\renewcommand{\theequation}{\thesection.\arabic{equation}}

	\section{Proofs and further results}
	\label{sec:fullcond}
{\color{black}
\subsection{Proof of Proposition \ref{prop:modelprop} and Corollary \ref{col:1}}

	\begin{enumerate}[leftmargin=0.2cm]
		\item $\mathbb{E}(Y^{(\ell)}_{ijt}|\vartheta^{(\ell)},Z^{(\ell)}_{it}=q,Z^{(\ell)}_{jt}=r)=\mathbb{E}(\mathbb{E}(Y^{(\ell)}_{ijt}|D^{(\ell)}_{ijt},\vartheta^{(\ell)},Z^{(\ell)}_{it}=q,Z^{(\ell)}_{jt}=r)|\vartheta^{(\ell)},Z^{(\ell)}_{it}=q,Z^{(\ell)}_{jt}=r)=\mathbb{E}(D^{(\ell)}_{ijt}\exp(\beta^{(\ell)}_{0qr}+\sigma_{qr}^{(\ell)2}/2)|\vartheta^{(\ell)},Z^{(\ell)}_{it}=q,Z^{(\ell)}_{jt}=r)=\nu^{(\ell)}_{qr}\exp(\beta^{(\ell)}_{0qr}+\sigma_{qr}^{(\ell)2}/2)$;
		\item 				$
		\mathbb{V}(Y^{(\ell)}_{ijt}|\vartheta^{(\ell)},Z^{(\ell)}_{it}=q,Z^{(\ell)}_{jt}=r)=\mathbb{E}(Y^{(\ell)2}_{ijt}|\vartheta^{(\ell)},Z^{(\ell)}_{it}=q,Z^{(\ell)}_{jt}=r)-\mathbb{E}(Y^{(\ell)}_{ijt}|\vartheta^{(\ell)},Z^{(\ell)}_{it}=q,Z^{(\ell)}_{jt}=r)^2=\nu^{(\ell)}_{qr}\exp(\sigma_{qr}^{(\ell)2})\exp(2\beta^{(\ell)}_{0qr}+\sigma_{qr}^{(\ell)2})-\nu^{(\ell)2}_{qr}\exp(2\beta^{(\ell)}_{0qr}+\sigma_{qr}^{(\ell)2})=\mathbb{E}(Y^{(\ell)}_{ijt}|\vartheta^{(\ell)},Z^{(\ell)}_{it}=q,Z^{(\ell)}_{jt}=r)^2 \left(\frac{\exp(\sigma_{qr}^{(\ell)2})}{\nu^{(\ell)}_{qr}}-1\right)
		$;
		\item $\mathbb{C}ov(Y^{(\ell)}_{i_1j_1t},Y^{(\ell)}_{i_2j_2t}|Z^{(\ell)}_{i_1t},Z^{(\ell)}_{j_1t},Z^{(\ell)}_{i_2t},Z^{(\ell)}_{j_2t})=\mathbb{C}ov(\mathbb{E}(Y^{(\ell)}_{i_1j_1t}|\vartheta^{(\ell)},Z^{(\ell)}_{i_1t},Z^{(\ell)}_{j_1t}),\mathbb{E}(Y^{(\ell)}_{i_2j_2t}|\vartheta^{(\ell)},Z^{(\ell)}_{i_2t},Z^{(\ell)}_{j_2t})|Z^{(\ell)}_{i_1t},Z^{(\ell)}_{j_1t},Z^{(\ell)}_{i_2t},Z^{(\ell)}_{j_2t})=(\mathbb{E}(\nu_{Z_{i_{1}t}Z_{j_{1}t}}^{(\ell)2})\mathbb{E}(\exp(2\beta^{(\ell)}_{0Z_{i_{1}t}Z_{j_{1}t}}+\sigma_{Z_{i_{1}t}Z_{j_{1}t}}^{(\ell)2}))-\mathbb{E}(\nu_{Z_{i_{1}t}Z_{j_{1}t}}^{(\ell)})^2\mathbb{E}(\exp(\beta^{(\ell)}_{0Z_{i_{1}t}Z_{j_{1}t}}+\sigma_{Z_{i_{1}t}Z_{j_{1}t}}^{(\ell)2}/2))^2)\mathbb{I}_{\{Z_{i_{2}t}\}}(Z_{i_{1}t})\mathbb{I}_{\{Z_{j_{2}t}\}}(Z_{j_{1}t})=\left(\frac{\underline{b}^{(\ell)}(\underline{b}^{(\ell)}+1)\exp(2(\underline{\beta}^{(\ell)}_{0}+\underline{\Sigma}^{(\ell)}))\underline{e}^{(\ell)}}{(\underline{b}^{(\ell)}+\underline{c}^{(\ell)})(\underline{b}^{(\ell)}+\underline{c}^{(\ell)}+1)(\underline{e}^{(\ell)}-1)}-\frac{4\underline{e}^{(\ell)2}\underline{b}^{(\ell)2}\exp(2\underline{\beta}^{(\ell)}_{0}+\underline{\Sigma}^{(\ell)})}{(\underline{b}^{(\ell)}+\underline{c}^{(\ell)})^2(2\underline{e}^{(\ell)}-1)^2}\right)\mathbb{I}_{\{Z_{i_{2}t}\}}(Z_{i_{1}t})\mathbb{I}_{\{Z_{j_{2}t}\}}(Z_{j_{1}t})$,
	\end{enumerate}
    where the last equality in the property 3 considers that the exponential transformation of the a Gamma variable follows a Pareto distribution with scale 1 and shape $\underline{e}^{(\ell)}$, i.e. $\exp(\sigma^{(\ell)2}_{Z_{i_{1}t}Z_{j_{1}t}})\sim \operatorname{Par}(1,\underline{e}^{(\ell)})$, where $\sigma_{Z_{i_{1}t}Z_{j_{1}t}}^{(\ell)2}\sim \operatorname{G}(1,1/\underline{e}^{(\ell)})$, while the exponential transformation of a Normal variable is log--normal distributed, i.e. $\exp(\beta^{(\ell)}_{0Z_{i_{1}t}Z_{j_{1}t}})\sim \operatorname{LN}(\underline{\beta}^{(\ell)}_{0},\underline{\Sigma}^{(\ell)})$. 
  
  Following the previous results we obtain
   $\mathbb{E}(Y^{(\ell)}_{ijt}|Z^{(\ell)}_{it}=q,Z^{(\ell)}_{jt}=r)=\frac{2\underline{b}^{(\ell)}\underline{e}^{(\ell)}\exp(\underline{\beta}^{(\ell)}_{0}+\underline{\Sigma}^{(\ell)}/2)}{(\underline{b}^{(\ell)}+\underline{c}^{(\ell)})(2\underline{e}^{(\ell)}-1)}, \ \underline{e}^{(\ell)}>1/2$ and $\mathbb{V}(Y^{(\ell)}_{ijt}|Z^{(\ell)}_{it}=q,Z^{(\ell)}_{jt}=r)=\frac{\underline{b}^{(\ell)}\underline{e}^{(\ell)}\exp(\underline{\beta}^{(\ell)}_0+\underline{\Sigma}^{(\ell)}/2)^2}{(\underline{b}^{(\ell)}+\underline{c}^{(\ell)})}\left(\frac{\exp(\underline{\Sigma}^{(\ell)})}{\underline{e}^{(\ell)}-2}-\frac{4\underline{e}^{(\ell)}\underline{b}^{(\ell)}}{(\underline{b}^{(\ell)}+\underline{c}^{(\ell)})(2\underline{e}^{(\ell)}-1)^2}\right), \ \underline{e}^{(\ell)}>2$. The result in Corollary \ref{col:1} then follows from the definition of $VM=\mathbb{V}(Y^{(\ell)}_{ijt}|Z^{(\ell)}_{it}=q,Z^{(\ell)}_{jt}=r)/\mathbb{E}(Y^{(\ell)}_{ijt}|Z^{(\ell)}_{it}=q,Z^{(\ell)}_{jt}=r)$.
}	        
  \subsection{Proof of Proposition \ref{prop:multiL}}

  Applying the representation in equation (7) of \citet{raman2009bayesian} to the distribution $\text{M-Laplace}(\kappa^{(\ell)}_{\mathcal{U},r}|\mathbf{0},(c(\mathcal{U})^{(\ell)})^{-1})$, for  $\kappa^{(\ell)}_{\mathcal{U},r}$ with $\mathcal{U}\neq\{\ell\}$, one obtains $\kappa^{(\ell)}_{\mathcal{U},r}|\zeta^{(\ell)}_{\mathcal{U},r} \sim \operatorname{N}(\mathbf{0},\operatorname{diag}(\zeta^{(\ell)2}_{\mathcal{U},r}\mathbf{1}_{s(\mathcal{U})}'))$ and $\zeta^{(\ell)2}_{\mathcal{U},r}|\rho^{(\ell)}\sim \operatorname{G}(\frac{s(\mathcal{U})+1}{2},\frac{2}{\rho^{(l)} s(\mathcal{U})})$. The results follows stacking and reordering $\kappa^{(\ell)}_{\ell,r}$ and $\kappa^{(\ell)}_{0r}$ into one vector $\kappa^{(\ell)}_{r}$ that follows a conditional Normal distribution.

 \subsection{Proof of Proposition \ref{prop:conne}}

	Using (\ref{eq:2}), the complete likelihood can be written as
	\begin{equation}
		\begin{split}
			L(Y,Z,D|X,\vartheta,\kappa)=
			\prod_{\ell=1}^{L}&\Bigg( \prod_{t=1}^{T}\left(\prod_{i\neq j}^N \left( (1-D^{(\ell)}_{ijt})\delta(Y^{(\ell)}_{ijt})+D^{(\ell)}_{ijt} f^{(\ell)}\left(Y^{(\ell)}_{ijt}| \theta^{(\ell)}_{Z^{(\ell)}_{it} Z^{(\ell)}_{jt}} \right)\right)\right.\Bigg. \\
			&\Bigg.\left.\left(\nu^{(\ell)}_{Z^{(\ell)}_{it} Z^{(\ell)}_{jt}}\right)^{D^{(\ell)}_{ijt}}(1-\nu^{(\ell)}_{Z^{(\ell)}_{it} Z^{(\ell)}_{jt}})^{1-D^{(\ell)}_{ijt}}\right)\Bigg)\left(\prod_{t=2}^{T}\prod_{i=1}^N \mathbb{P}(Z_{it}|Z_{it-1})\mathbb{P}(Z_{i1}|\alpha)\right),
		\end{split}
	\end{equation}	
where the time and layer dependence are driven by the HMCs, conditionally to $X$. 

\bigskip

\par\noindent $1)-2)$ The full conditional distribution of the parameters $\theta_{qr}^{(\ell)}$ for  $q, r \in \mathcal{Q}^{(\ell)}$ is:
	\begin{equation}
		\label{eq:16}
		h\left(\theta^{(\ell)}_{qr}|\theta^{(-\ell)},\theta^{(\ell)}_{-qr},\nu, P, \alpha, Y,Z,D,X\right)\propto \ \prod_{t=1}^{T} \prod_{(i, j,t) \in \mathcal{D}^{(\ell),qr1}} f^{(\ell)}\left(Y^{(\ell)}_{ijt}|\theta^{(\ell)}_{qr},X^{(\ell)}_{ijt}\right)\pi(\theta^{(\ell)}_{qr}),
	\end{equation}
	with $ \mathcal{D}^{(\ell),qr1}=\left\{(i,j,t)\in \mathcal{E}^{(\ell)}\times \mathcal{T}|D^{(\ell)}_{ijt}=1, Z^{(\ell)}_{it}=q,Z^{(\ell)}_{it}=r \right\}$. If the layer is weighted and $Y^{(\ell)}_{ijt}=a\in \mathbb{R}_{+}$  for $(i,j,t)\in \mathcal{E}_t^{(\ell)}$, $a\sim\operatorname{LN}(X^{(\ell)'}_{ijt}\beta^{(\ell)}_{qr},\sigma_{qr}^{(\ell)2})$, where $\theta^{(\ell)}_{qr}=(\beta^{(\ell)'}_{qr},\sigma_{qr}^{(\ell)2})'$. Since we assumed independent conditionally conjugate priors $\operatorname{N}_{p}(\underline{\beta}^{(\ell)}_{qr},\underline{\Sigma}^{(\ell)}_{qr})$ and $\operatorname{IG}(\underline{d}^{(\ell)}_{qr}/2,\underline{e}^{(\ell)}_{qr}/2)$, then \eqref{eq:16} becomes
 	\begin{equation}
	\begin{aligned}	h\left(\theta^{(\ell)}_{qr}|\theta^{(-\ell)},\theta^{(\ell)}_{-qr},\nu, P, \alpha, Y,Z,D,X\right)&\propto \left(\sigma_{qr}^{(\ell)2}\right)^{-\left(\overline{d}^{(\ell)}_{qr} / 2+1\right)}\exp \left(-\frac{1}{2 \sigma_{qr}^{(\ell)2}}\left[(\varepsilon_{qr}^{(\ell)})^{\prime} (\overline{\Sigma}^{(\ell)}_{qr})^{-1}(\varepsilon^{(\ell)}_{qr})+e^{(\ell)}_{qr}\right]\right),
 \end{aligned}
	\end{equation}
where $\varepsilon_{qr}^{(\ell)}=(\beta^{(\ell)}_{qr}-\overline{\beta}^{(\ell)}_{qr})$, which leads to 
	\begin{eqnarray}
\beta^{(\ell)}_{qr}|\sigma_{qr}^{(\ell)2},\theta^{(-\ell)},\theta^{(\ell)}_{-qr},\nu, P, \alpha, Y,Z,D,X&\sim\operatorname{N}_{p}\left(\overline{\beta}^{(\ell)}_{qr},\overline{\Sigma}^{(\ell)}_{qr}\right)\\	\sigma_{qr}^{(\ell)2}|\beta^{(\ell)}_{qr},\theta^{(-\ell)},\theta^{(\ell)}_{-qr},\nu, P, \alpha, Y,Z,D,X&\sim\operatorname{IG}\left(\overline{d}^{(\ell)}_{qr}/2,\overline{e}^{(\ell)}_{qr}/2\right)
	\end{eqnarray}
	with 
	\begin{eqnarray}	\overline{\beta}^{(\ell)}_{qr}&=&\left(\underline{\Sigma}^{(\ell)-1}_{qr}+\frac{1}{\sigma_{qr}^{(\ell)2}} (X^{(\ell)}_{qr1})^{\prime} X_{qr1}^{(\ell)}\right)^{-1}\left(\underline{\Sigma}_{qr}^{(\ell)-1} \underline{\beta}^{(\ell)}_{qr}+\frac{1}{\sigma_{qr}^{(\ell)2}} (X_{qr1}^{(\ell)})^{\prime} \log(Y_{qr1})\right)\\
		\overline{\Sigma}^{(\ell)}_{qr}&=&\left(\underline{\Sigma}_{qr}^{(\ell)-1}+\frac{1}{\sigma_{qr}^{(\ell)2}} (X_{qr1}^{(\ell)})^{\prime} X_{qr1}^{(\ell)}\right)^{-1}\\
		\overline{d}^{(\ell)}_{qr}&=&\underline{d}^{(\ell)}_{qr}+\sum_t \left|\mathcal{D}^{(\ell),qr1}_t\right|\\
		\overline{e}^{(\ell)}_{qr}&=&\underline{e}^{(\ell)}_{qr}+\left(\log(Y^{(\ell)}_{qr1})- (X^{(\ell)}_{qr1})'\beta^{(\ell)}_{qr}\right)'\left(\log(Y^{(\ell)}_{qr1})- (X^{(\ell)}_{qr1})'\beta^{(\ell)}_{qr}\right),
	\end{eqnarray}
 where $X_{qr1}$ and $Y_{qr1}$ collect respectively $(i,j)\in \mathcal{D}^{(\ell),qr1}\ \forall \ t\in \mathcal{T}$ and $|A|$ the cardinality of the set $A$.

\bigskip
 
\par\noindent $3)$ The full conditional distribution of the probability of having an active edge between two nodes is
	\begin{equation}
		\label{eq:22}
		\begin{aligned}	h\left(\nu^{(\ell)}_{qr}|\nu^{(-\ell)},\nu^{(\ell)}_{-qr},\theta,P, \alpha, Y,Z, D\right)&\propto \ \prod_{(i,j,t)\in \mathcal{D}^{(\ell),qr}}\left(\nu^{(\ell)}_{qr}\right)^{D^{(\ell)}_{ijt}} \left(1-\nu^{(\ell)}_{qr}\right)^{1-D^{(\ell)}_{ijt}}\pi(\nu^{(\ell)}_{qr})\\
			& \propto \left(\nu_{qr}\right)^{\left|\mathcal{D}^{(\ell),qr1}\right|} \left(1-\nu^{(\ell)}_{qr}\right)^{\left|\mathcal{D}^{(\ell),qr}\right|-\left|\mathcal{D}^{(\ell),qr1}\right|}\pi(\nu^{(\ell)}_{qr}),
		\end{aligned}
	\end{equation}
where $ \mathcal{D}^{(\ell),qr}=\left\{(i,j,t)\in \mathcal{E}^{(\ell)}\times \mathcal{T}|Z^{(\ell)}_{it}=q,Z^{(\ell)}_{jt}=r \right\}$,  $ \mathcal{D}^{(\ell),qr1}=\left\{(i,j,t)\in \mathcal{E}^{(\ell)}\times \mathcal{T}|Z^{(\ell)}_{it}=q,\right.$ \\ $\left.Z^{(\ell)}_{jt}=r, D^{(\ell)}_{ijt}=1 \right\}$. From the conjugate prior assumption $\pi(\nu^{(\ell)}_{qr})=\operatorname{Beta}(\underline{b}^{(\ell)}_{qr},\underline{c}^{(\ell)}_{qr})$ and (\ref{eq:22}) it follows:
 	\begin{equation}
		\begin{aligned}	h\left(\nu^{(\ell)}_{qr}|\nu^{(-\ell)},\nu^{(\ell)}_{-qr},\theta,P, \alpha, Y,Z, D\right)&\propto \left(\nu_{qr}\right)^{\left|\mathcal{D}^{(\ell),qr1}\right|+\underline{b}^{(\ell)}_{qr}-1} \left(1-\nu^{(\ell)}_{qr}\right)^{\left|\mathcal{D}^{(\ell),qr}\right|-\left|\mathcal{D}^{(\ell),qr1}\right|+\underline{c}^{(\ell)}_{qr}-1}\\
  & \propto \operatorname{Beta}\left(\overline{b}^{(\ell)}_{qr},\overline{c}^{(\ell)}_{qr}\right),
		\end{aligned}
	\end{equation}
	where $\overline{b}^{(\ell)}_{qr}=\left|\mathcal{D}^{(\ell),qr1}\right|+\underline{b}^{(\ell)}_{qr}$ and $\overline{c}^{(\ell)}_{qr}=\left|\mathcal{D}^{(\ell),qr}\right|-\left|\mathcal{D}^{(\ell),qr1}\right|+\underline{c}^{(\ell)}_{qr}$. 
	
	\bigskip
\par\noindent $4)$ The full conditional for the transition probabilities and  the initial allocation 
	\begin{equation}
		\begin{aligned}	h\left(\alpha^{(\ell)}|\alpha^{(-\ell)},\nu,\theta, P, Y,Z,D\right)&\propto \ \prod_{i=1}^N \prod_{k^{(\ell)}=1}^{Q^{(\ell)}} (\alpha^{(\ell)}_{k^{(\ell)}})^{\mathbb{I}_{\{k^{(\ell)}\}}(Z^{(\ell)}_{i1})} \pi\left(\alpha^{(\ell)}\right)
			\propto \prod_{k=1}^{Q^{(\ell)}} (\alpha^{(\ell)}_{k})^{\sum_{i=1}^N\mathbb{I}_{\{k\}}(Z^{(\ell)}_{i1})} \pi\left(\alpha^{(\ell)}\right).
		\end{aligned}
	\end{equation}
If a priori $\alpha^{(\ell)}\sim\operatorname{Dir}(\underline{\alpha}^{(\ell)})$, then
	\begin{equation}
		\begin{aligned}	h\left(\alpha^{(\ell)}|\alpha^{(-\ell)},\nu,\theta, P, Y,Z,D\right)&\propto \prod_{k=1}^{Q^{(\ell)}} (\alpha^{(\ell)}_{k})^{\sum_{i=1}^N\mathbb{I}_{\{k\}}(Z^{(\ell)}_{i1})+\underline{\alpha}^{(\ell)}_{k}-1}&\propto \operatorname{Dir}\left(\overline{\alpha}^{(\ell)}\right),
		\end{aligned}
	\end{equation}
 where $\overline{\alpha}^{(\ell)}=(\sum_{i=1}^N\mathbb{I}_{\{1\}}(Z^{(\ell)}_{i1})+\underline{\alpha}^{(\ell)}_{1},\ldots, \sum_{i=1}^N\mathbb{I}_{\{Q^{(\ell)}\}}(Z^{(\ell)}_{i1})+\underline{\alpha}^{(\ell)}_{Q^{(\ell)}})'$.

\subsection{Proof of Proposition \ref{prop:dataugLogit}}
	Regarding the transition parameters of the Markov chains, let us define $\varphi(\eta^{(\ell)}_{it,r})=(\exp(\eta^{(\ell)}_{it,r}))/(1+\exp(\eta^{(\ell)}_{it,r}))$. The full conditional of the transition parameters is
\begin{equation}
	\begin{aligned}
		\label{eq:hkappa2}
		h(\kappa_{r}^{(\ell)}&|Z)\propto \ \pi\left(\kappa_r^{(\ell)}\right)\prod_{t=2}^T\prod_{i=1}^{N} \left(\varphi(\eta^{(\ell)}_{it,r})\right)^{W^{(\ell)}_{it,r}}\left(\sum_{k=1}^{Q^{(\ell)}}\exp\left(\widetilde{W}_{it-1}\kappa_{k}^{(\ell)}\right)\right)^{W^{(\ell)}_{it,r}-1}\prod_{k\neq r}\exp\left(\widetilde{W}_{it-1}\kappa_k^{(\ell)}\right)^{W^{(\ell)}_{it,r}}\\
		&\propto \pi\left(\kappa_{r}^{(\ell)}\right)\prod_{t=2}^T\prod_{i=1}^{N} \left(\varphi(\eta^{(\ell)}_{it,r})\right)^{W^{(\ell)}_{it,r}}\left(1+\exp\left(\eta^{(\ell)}_{it,r}\right)\right)^{W^{(\ell)}_{it,r}-1}\\
		%&\propto \pi\left(\kappa_r^{(\ell)}\right)\prod_{t=2}^T\prod_{i=1}^{N} \exp\left(\eta^{(\ell)}_{it,r}W^{(\ell)}_{it,r}\right)\left(1+\exp\left(\eta^{(\ell)}_{it,r}\right)\right)^{-1} \\
		&\propto \pi\left(\kappa_r^{(\ell)}\right)\prod_{t=2}^T\prod_{i= 1}^{N} \exp\left(\xi^{(\ell)}_{it,r}\eta^{(\ell)}_{it,r}\right)\int_{0}^{+\infty}\exp\left(-\omega^{(\ell)}_{it,r}\left(\eta^{(\ell)}_{it,r}\right)^2/2\right)h(\omega^{(\ell)}_{it,r})d\omega^{(\ell)}_{it,r}\\
		&\propto \int_{\mathbb{R}^{N(T-1)}}\underbrace{\pi\left(\kappa_r^{(\ell)}\right)\prod_{t=2}^T\prod_{i=1}^{N}\mathbb{I}_{\{\mathbb{R}^{+}\}}\left(\omega^{(\ell)}_{it,r}\right)\exp\left(\xi^{(\ell)}_{it,r}\eta^{(\ell)}_{it,r}-\omega^{(\ell)}_{it,r}\left(\eta^{(\ell)}_{it,r}\right)^2/2\right)h(\omega^{(\ell)}_{it,r})}_{h(\kappa_r^{(\ell)},\omega^{(\ell)}_{1:N 1:T,r}|Z)}d\omega^{(\ell)}_{it,r},
	\end{aligned}
\end{equation}
where $\eta^{(\ell)}_{it,r}=\widetilde{W}_{it-1}\kappa_r^{(\ell)}-R^{(\ell)}_{it,r}$, $R^{(\ell)}_{it,r}=\log(\sum_{k\neq r}^{Q^{(\ell)}}\exp(\widetilde{W}_{it-1}\kappa_k^{(\ell)}))$, $\xi^{(\ell)}_{it,r}=W^{(\ell)}_{it,r}-1/2$ and $\omega^{(\ell)}_{it,r}$ are auxiliary variables following a Pólya Gamma distribution, i.e. $\omega^{(\ell)}_{it,r}\sim \operatorname{PG}(1,0)$. This can be rewritten as:
\begin{equation}
		\begin{aligned}		h(\kappa_r^{(\ell)}|Z,\omega^{(\ell)}_{1:N 1:T,r})\propto \pi\left(\kappa_r^{(\ell)}\right)\exp\left(-\frac{1}{2}\left(\tilde{\xi}^{(\ell)}_{r}-\eta^{(\ell)}_{r}\right)'\Omega_r^{(\ell)}\left(\tilde{\xi}^{(\ell)}_{r}-\eta^{(\ell)}_{r}\right)\right),
		\end{aligned}
	\end{equation}	
	where $\tilde{\xi}^{(\ell)}_{r}=(\xi^{(\ell)}_{12,r}/\omega^{(\ell)}_{12,r},\ldots,\xi^{(\ell)}_{NT,r}/\omega^{(\ell)}_{NT,r})'$, $\xi_r^{(\ell)}=(\xi_{11,r}^{(\ell)},\dots,\xi_{N(T-1),r}^{(\ell)})'$, $\eta^{(\ell)}_{r}=(\eta^{(\ell)}_{12,r},\ldots,\eta^{(\ell)}_{NT,r})'$ and $\Omega_r^{(\ell)}=\operatorname{diag}(\omega^{(\ell)}_{12,r},\dots,\omega^{(\ell)}_{NT,r})$.

\subsection{Proof of Proposition \ref{prop:PolyaLap}}
\par\noindent $5)$ The full conditional distributions for $\kappa_r^{(\ell)}$ under $\kappa^{(\ell)}_{r}\sim \operatorname{N}(\underline{\kappa}^{(\ell)}_{r},\underline{K}^{(\ell)}_{r})$ is given by:
\begin{equation}
	\begin{aligned}	&\!\!\!h(\kappa_{r}^{(\ell)}|Z,\omega^{(\ell)}_{1:N 1:T,r})\propto \exp\left(-\frac{1}{2}\left(\varepsilon_r^{(\ell)}\right)'(\underline{K}^{(\ell)}_{r})^{-1}\left(\varepsilon_r^{(\ell)}\right)-\frac{1}{2}\left(\nu^{(\ell)}_{r}\right)'\Omega_r^{(\ell)}\left(\nu^{(\ell)}_{r}\right)\right)\propto\operatorname{N}\left(\overline{\kappa}_r^{(\ell)},\overline{K}_r^{(\ell)}\right),
\end{aligned}
\end{equation}
where $\varepsilon_r^{(\ell)}=(\kappa_r^{(\ell)}-\underline{\kappa}_r^{(\ell)})$,  $\nu^{(\ell)}_{r}=(\tilde{\xi}^{(\ell)}_{r}+R_{r}^{(\ell)}-\widetilde{W}\kappa^{(\ell)}_{r})$, and
\begin{eqnarray} \overline{\kappa}_r^{(\ell)}=\overline{K}_r^{(\ell)}\left(\widetilde{W}'\underbrace{\Omega_r^{(\ell)}\tilde{\xi}_r^{(\ell)}}_{\xi_r^{(\ell)}}+\widetilde{W}'\Omega_r^{(\ell)}R_r^{(\ell)}+\left(\underline{K}_r^{(\ell)}\right)^{-1}\underline{\kappa}_r^{(\ell)}\right),	\quad \overline{K}_r^{(\ell)}=\left(\widetilde{W}'\Omega_r^{(\ell)} \widetilde{W}+\left(\underline{K}_r^{(\ell)}\right)^{-1}\right)^{-1}.
	\end{eqnarray}
Stacking in one vector all variables $\kappa_r^{(\ell)}$ $r=1,\ldots, Q^{(\ell)}-1$ of the Pólya-Gamma representation one gets	
	\begin{equation}
		\begin{aligned}		h(\kappa^{(\ell)}|Z,\omega^{(\ell)}_{1:N 1:T})&\propto \pi\left(\kappa^{(\ell)}\right)\exp\left(-\frac{1}{2}\left(\tilde{\xi}^{(\ell)}-\eta^{(\ell)}\right)'\Omega^{(\ell)}\left(\tilde{\xi}^{(\ell)}-\eta^{(\ell)}\right)\right)\propto\operatorname{N}\left(\overline{\kappa}^{(\ell)},\overline{K}^{(\ell)}\right),
  \end{aligned}
  \end{equation}
where $\overline{\kappa}^{(\ell)}=\overline{K}^{(\ell)}\left(\widetilde{W}_{Q,\ell}'\xi^{(\ell)}+\widetilde{W}_{Q,\ell}'\Omega^{(\ell)}R^{(\ell)}+\left(\underline{K}^{(\ell)}\right)^{-1}\underline{\kappa}^{(\ell)}\right)$ and $\overline{K}^{(\ell)}=\left(\widetilde{W}_{Q,\ell}'\Omega^{(\ell)} \widetilde{W}_{Q,\ell}+\left(\underline{K}^{(\ell)}\right)^{-1}\right)^{-1}$ with $\widetilde{W}_{Q,\ell}=({\bf I}_{Q^{(\ell)}-1}\otimes\widetilde{W})$.

 \par\noindent $6)$ Let $\operatorname{PG}\left(1,a\right)$ denote a general Pólya Gamma distribution. The full conditional for  $\omega^{(\ell)}_{it,r}$ is given by  

 \begin{equation}
	\begin{aligned}	
h(\omega^{(\ell)}_{i t,r}|Z,\kappa_{r}^{(\ell)})&\propto \exp\left(-\omega^{(\ell)}_{it,r}\left(\eta^{(\ell)}_{it,r}\right)^2/2\right)h(\omega^{(\ell)}_{it,r})\propto \operatorname{PG}\left(1,\eta^{(\ell)}_{it,r}\right),
\end{aligned}
\end{equation}
%which is proportional to a general Pólya Gamma distribution \citep{polson2013bayesian}. 
%Thus, the full posterior is equivalent to $\omega^{(\ell)}_{it,r}\left|Z,\kappa_r^{(\ell)}\right.\sim\operatorname{PG}\left(1,\eta^{(\ell)}_{it,r}\right).$

\par\noindent $7)-8)$ The full conditional of the group LASSO parameters in (\ref{eq:grouplass}) under $\rho^{\ell}\sim \operatorname{G}(\underline{\iota}^{(\ell)}_1,\underline{\iota}^{(\ell)}_2)$ is given by
 \begin{equation}
	\begin{aligned}	
h(\zeta_{r\mathcal{U}}^{(l)2}\left|\kappa_r^{(\ell)},\rho^{(\ell)} \right.)&\propto (\zeta_{r\mathcal{U}}^{(l)2})^{-1/2}\exp\left(-\frac{1}{2}\left(\frac{||\kappa^{(\ell)}_{\mathcal{U},r} -\underline{\kappa}^{(\ell)}_{\mathcal{U},r}||^2_2}{\zeta_{r\mathcal{U}}^{(l)2}}+\zeta_{r\mathcal{U}}^{(l)2}s(\mathcal{U})\rho^{(\ell)}\right)\right)\\
h(\rho^{(l)}\left|\zeta_{r\mathcal{U}}^{(l)2}\right.)&\propto (\rho^{\ell})^{\underline{\iota}^{(l)}_{1}+\phi^{(\ell)}}\exp\left(-\rho^{(\ell)}\left(1/\underline{\iota}^{(l)}_{2}+\frac{\sum_{r=1}^{Q^{(\ell)}-1}\sum_{\mathcal{U}}\zeta^{(l)2}_{r\mathcal{U}}s(\mathcal{U})}{2}\right)\right),
\end{aligned}
\end{equation}
where $\phi^{(\ell)}=((Q^{(\ell)}-1)(p-Q^{(\ell)}+1)+(2^{L}-2)(Q^{(\ell)}-1))/2$, which leads to 
	\begin{equation}		\zeta_{r\mathcal{U}}^{(l)2}\left|\kappa_r^{(\ell)},\rho^{(\ell)} \right.\sim\operatorname{GIG}\left(1/2,s(\mathcal{U})\rho^{(\ell)},||\kappa^{(\ell)}_{\mathcal{U},r} -\underline{\kappa}^{(\ell)}_{\mathcal{U},r}||^2_2\right)
	\end{equation}	
	\begin{equation}	
		\rho^{(l)}\left|\zeta_{r\mathcal{U}}^{(l)2}\right.\sim \operatorname{G}\left(\underline{\iota}^{(l)}_{1}+\phi^{(\ell)} ,\left(1/\underline{\iota}^{(l)}_{2}+(\sum_{r=1}^{Q^{(\ell)}-1}\sum_{\mathcal{U}}\zeta^{(l)2}_{r\mathcal{U}}s(\mathcal{U}))/2\right)^{-1}\right),
	\end{equation}	
	respectively, where $\operatorname{GIG}$ denotes the Generalized Inverse Gaussian distribution \citep{raman2009bayesian}. 
%\vspace{-0.5cm}
\textcolor{black}{\subsection{Proof of Proposition \ref{prop:reverselog}}
Let $m_i=m(Y|H_i)$, plugging the decomposition $h_{ij}=\exp(d+(\log (h_{0j})-m_0+m_1)\mathbb{I}_{0}(i)+(\log(h_{1j})-m_0+m_1)\mathbb{I}_{1}(i))$, into the RLR conditional likelihood of \citet{geyer1991estimating}, with $n+n$ terms, one gets $p_{i}(h_{ij},m_i)=h_{ij}\exp(-\log m_i+\log(n/(2n))))/$ $(h_{0j}\exp(-\log m_0+\log(n/(2n))))+h_{1j}\exp(-\log m_1+\log(n/(2n)))))=\exp(\mathbb{I}_{\{0\}}(i)(d+\log(h_{0j})-\log(h_{1j}))))/(1+\exp(d+\log(h_{0j})-\log(h_{1j})))$.}

\subsection{Forward filtering backward sampling}\label{sec:ffbs}
Let $h(Z_{it}^{(\ell)},Z_{jt}^{(\ell)})=f^{(\ell)}(Y^{(\ell)}_{jit}| \theta^{(\ell)}_{Z^{(\ell)}_{jt}Z^{(\ell)}_{it}},X^{(\ell)}_{jit}  )\nu^{(\ell)}_{Z^{(\ell)}_{jt} Z^{(\ell)}_{it} }$, then the full conditional of $Z^{(\ell)}_{i,1:T}$ is 
	\begin{equation*}
		\label{eq:zli}
		\begin{aligned}
			\mathbb{P}(Z^{(\ell)}_{i,1:T}&|Z^{^{(-\ell)}}_{i,1:T},Z^{^{(1:L)}}_{-i,1:T},Y,D,X,\vartheta_t)\propto 
   %\  \left(\prod_{t=2}^{T} \mathbb{P}(Z^{(\ell)}_{it}|Z_{it-1})\mathbb{P}(Z^{(\ell)}_{i1}|\alpha)\right)\\& \prod_{t=1}^{T}\left\{\prod_{j=1}^N \left(\left[(1-D^{(\ell)}_{ijt})\delta(Y^{(\ell)}_{ijt})+D^{(\ell)}_{ijt} f^{(\ell)}\left(Y^{(\ell)}_{ijt}| \theta^{(\ell)}_{Z^{(\ell)}_{it} Z^{(\ell)}_{jt}},X^{(\ell)}_{ijt} \right)\right]\left(\nu_{Z^{(\ell)}_{it} Z^{(\ell)}_{jt}}^{(\ell)}\right)^{D^{(\ell)}_{ijt}}\left(1-\nu^{(\ell)}_{Z^{(\ell)}_{it} Z^{(\ell)}_{jt}}\right)^{1-D^{(\ell)}_{ijt}} \right)\right.\\
%			&\left.\left(\left[(1-D^{(\ell)}_{jit})\delta(Y^{(\ell)}_{jit})+ D^{(\ell)}_{jit}f^{(\ell)}\left(Y^{(\ell)}_{jit}| \theta^{(\ell)}_{Z^{(\ell)}_{jt}Z^{(\ell)}_{it} },X^{(\ell)}_{jit} \right)\right]\left(\nu_{Z^{(\ell)}_{jt}Z^{(\ell)}_{it} }^{(\ell)}\right)^{D^{(\ell)}_{jit}}\left(1-\nu^{(\ell)}_{Z^{(\ell)}_{jt}Z^{(\ell)}_{it}} \right)^{1-D^{(\ell)}_{jit}}\right)\right\} \\ 
%			\propto\ &
   \left(\prod_{t=2}^{T} \mathbb{P}(Z^{(\ell)}_{it}|Z_{it-1})\mathbb{P}(Z^{(\ell)}_{i1}|\alpha)\right)\left(\prod_{t=2}^{T} \mathbb{P}(Z^{(-\ell)}_{it}|Z_{it-1})\right)\\
   &\prod_{t=1}^{T}\left[\left(\prod_{j\in \mathcal{D}_{it}^{(\ell),out}}  h(Z_{it}^{(\ell)},Z_{jt}^{(\ell)})\right) \left(\prod_{j\in \mathcal{D}_{it}^{(\ell),in}}  h(Z_{jt}^{(\ell)},Z_{it}^{(\ell)})\right) \prod_{j=1}^{N}\left(1-\nu_{Z_{it} Z_{jt}}^{(\ell)}\right)^{1-D^{(\ell)}_{ijt}} \left(1-\nu_{Z^{(\ell)}_{jt} Z^{(\ell)}_{it} }^{(\ell)}\right)^{1-D^{(\ell)}_{jit}}\right]\\
			&\propto \left(\prod_{t=2}^{T} \mathbb{P}(Z^{(\ell)}_{it}|Z_{it-1})\mathbb{P}(Z^{(\ell)}_{i1}|\alpha)\right) \left(\prod_{t=2}^{T} \mathbb{P}(Z^{(-\ell)}_{it}|Z_{it-1})\right) \prod_{t=1}^{T}\left[g^{(\ell)}(Y^{(\ell)}_{i\sdot,t},Y^{(\ell)}_{\sdot i, t},D^{(\ell)}_{i\sdot,t},D^{(\ell)}_{\sdot i, t}|Z,X^{(\ell)}_{i\sdot,t},X^{(\ell)}_{\sdot i, t},\vartheta_t)\right],
		\end{aligned}
	\end{equation*}
	where $Z^{(1:L)}_{-i,1:T}$ is the set of HMC of the other nodes in all layers, $Z^{(-\ell)}_{i,1:T}$ the HMC of the same node except $\ell$, $Y^{(\ell)}_{i\sdot,t}$ is the $i$--th row of adjacency matrix omitting main diagonal and $\mathcal{D}_{it}^{(\ell),out}=\{j \in \mathcal{V} |D^{(\ell)}_{ijt}=1 \}$ and $\mathcal{D}_{it}^{(\ell),in}=\{j \in \mathcal{V} |D^{(\ell)}_{jit}=1 \}$, the set of neighbours. 
From  (\ref{eq:zli}), prediction and filtering are	
\begin{equation*}
	\begin{aligned}
		p_{r,t|t-1}^{(i,\ell)}=&\sum_{q=1}^{Q^{(\ell)}}\mathbb{P}\left(Z^{(\ell)}_{it}=r|Z^{(\ell)}_{it-1}=q,Z^{(-\ell)}_{i,1:t-1},Z^{(\ell)}_{-i,1:t-1}\right)p_{q,t-1|t-1}^{(i,\ell)}\\
		p_{r,t|t}^{(i,\ell)}\propto p_{r,t|t-1}^{(i,\ell)}g^{(\ell)}(Y^{(\ell)}_{i\sdot,t},Y^{(\ell)}_{\sdot i, t},&D^{(\ell)}_{i\sdot,t},D^{(\ell)}_{\sdot i, t}|\vartheta,Z,X^{(\ell)}_{i\sdot,t},X^{(\ell)}_{\sdot i, t})\prod_{\mathfrak{m}\neq \ell}\mathbb{P}\left(Z^{(\mathfrak{m})}_{it+1}=r|Z^{(\mathfrak{m})}_{it}=q,Z^{(-\mathfrak{m})}_{i,1:t},Z^{(\mathfrak{m})}_{-i,1:t}\right),
	\end{aligned}
\end{equation*}
where $p_{r,t|s}^{(i,\ell)}=\mathbb{P}(Z^{(\ell)}_{it}=r|Z^{(-\ell)}_{i,1:s+1},Z^{(\ell)}_{-i,1:s},\psi^{(\ell)}_{s})$ $\psi^{(\ell)}_t=(Y^{(\ell)}_{1:t},D^{(\ell)}_{1:t},X^{(\ell)}_{1:t})$. We apply Forward-Filtering Backward-Sampling (FFBS):  
\begin{equation*}
	\mathbb{P}\left(Z^{(\ell)}_{i,1:T}|Z^{(-\ell)}_{i,1:T},Z^{(\ell)}_{-i,1:T},\psi^{(\ell)}_{T}\right)=\mathbb{P}\left(Z^{(\ell)}_{iT}|Z^{(-\ell)}_{i,1:T},Z^{(\ell)}_{-i,1:T},\psi^{(\ell)}_T\right)\prod_{t=1}^{T-1}\mathbb{P}\left(Z^{(\ell)}_{it}|Z^{(\ell)}_{it+1},Z^{(-\ell)}_{i,1:t+1},Z^{(\ell)}_{-i,1:T},\psi^{(\ell)}_t\right)
\end{equation*}	
where $\mathbb{P}(Z^{(\ell)}_{it}|Z^{(\ell)}_{it+1},Z^{(-\ell)}_{i,1:T},Z^{(\ell)}_{-i,1:T},\psi^{(\ell)}_t)\propto \mathbb{P}(Z^{(\ell)}_{it+1}|Z^{(\ell)}_{it},Z^{(-\ell)}_{i,1:t},Z^{(\ell)}_{-i,1:T},\psi^{(\ell)}_t) \mathbb{P}(Z^{(\ell)}_{it}|Z^{(-\ell)}_{i,1:t+1},Z^{(\ell)}_{-i,1:T},\psi^{(\ell)}_t)$.
\end{document}